\documentclass[useAMS,usenatbib]{mn2e}
\usepackage{graphicx}
\usepackage{rotating}
\usepackage{float}
\usepackage{aecompl}
\usepackage{color}
\usepackage{pdflscape}

\newcommand\aj{AJ}
\newcommand\apj{ApJ}
\newcommand\apjl{ApJL}
\newcommand\apjs{ApJS}
\newcommand\aap{A\&A}
\newcommand\araa{ARA\&A}
\newcommand\mnras{MNRAS}
\newcommand\pasj{PASJ}
\newcommand\pasp{PASP}
\newcommand\msun{M$_{\odot}$}

\title[Submillimeter Array Observations of NGC 2264-C]
{Submillimeter Array Observations of NGC 2264-C: Molecular Outflows and Driving Sources}

\author[N. Cunningham, S.L ~Lumsden, C.J ~Cyganowski, L.T ~Maud, C. Purcell]
  {Nichol Cunningham$^{1,2}${\thanks{E-mail:ncunning@nrao.edu}}, Stuart L. Lumsden$^1$, Claudia J. Cyganowski$^{3,4}$, Luke T. Maud$^{1,5}$, \newauthor Cormac Purcell$^6$\\
$^{1}$School of Physics and Astronomy, University of Leeds, LS2 9JT, UK \\
$^2$National Radio Astronomy Observatory, PO Box 2, Green Bank, WV, 24944, USA\\
$^3$Scottish Universities Physics Alliance (SUPA), School of Physics and Astronomy, University of St Andrews, St Andrews, Fife KY16 9SS, UK\\
$^4$Harvard-Smithsonian Center for Astrophysics, Cambridge, MA 02138, USA \\
$^{5}$Leiden Observatory, Leiden University, PO Box 9513, 2300 RA Leiden, The Netherlands\\
$^6$Sydney Institute for Astronomy, University of Sydney, NSW 2006, Australia}

\date{Accepted 2016 February 12. Received 2016 February 11; in original form 2014 October 17}
\pagerange{\pageref{firstpage}--\pageref{lastpage}} \pubyear{2016}
\def\LaTeX{L\kern-.36em\raise.3ex\hbox{a}\kern-.15em
    T\kern-.1667em\lower.7ex\hbox{E}\kern-.125emX}

\begin{document}
\label{firstpage}
\maketitle

\begin{abstract}
We present 1.3\,mm Submillimeter Array (SMA) observations at $\sim$3$\arcsec$ resolution towards the brightest section of the intermediate/massive star forming cluster NGC\,2264-C. The millimetre continuum emission reveals ten 1.3\,mm continuum peaks, of which four are new detections. The observed frequency range includes the known molecular jet/outflow tracer SiO (5-4), thus providing the first high resolution observations of SiO towards NGC\,2264-C. We also detect molecular lines of twelve additional species towards this region, including CH$_3$CN, CH$_3$OH, SO, H$_2$CO, DCN, HC$_3$N, and $^{12}$CO. The SiO (5-4) emission reveals the presence of two collimated, high velocity (up to 30\,km\,s$^{-1}$ with respect to the systemic velocity) bi-polar outflows in NGC\,2264-C. In addition, the outflows
are traced by emission from $^{12}$CO, SO, H$_2$CO, and CH$_3$OH. We find an evolutionary spread between cores residing in the same parent cloud.  The two unambiguous outflows are driven by the brightest mm continuum cores, which are IR-dark, molecular line weak, and likely the youngest cores in the region.  Furthermore, towards the RMS source AFGL\,989-IRS1, the IR-bright and most evolved source in NGC\,2264-C, we observe no molecular outflow emission. A molecular line rich ridge feature, with no obvious directly associated continuum source, lies on the edge of a low density cavity and may be formed from a wind driven by AFGL\,989-IRS1. In addition, 229\,GHz class I maser emission is detected towards this feature. 

\end{abstract}

\begin{keywords}
individual objects: NGC 2264-C – ISM: jets and outflows – ISM:
molecules – stars: formation
\end{keywords}

\section{Introduction}

Can high-mass star formation be described as a scaled-up version of
low-mass star formation?  Protostellar jets and outflows are a
ubiquitous feature of the star formation process; as such, their study
provides an important means of addressing this fundamental question.
Observationally, young stellar objects (YSOs) of all masses are known
to drive bipolar molecular outflows.  If a single driving mechanism
operates in both the low- and high-mass regimes
\citep[e.g.][]{Richer2000}, this would be evidence for a common
formation process. The initial evidence for a single mechanism was an
observed correlation, over five orders of magnitude, between
bolometric luminosity and outflow force, power and mass flow rate
\citep[e.g.][]{CabritandBertout1992,ShepherdandChurchwell1996}.  Outflows
observed towards massive young stellar objects (MYSOs) can contain
momentum, mass and energy up to a few orders of magnitude larger
(e.g. \citealt{Beuther2002}; \citealt{Wu2004}; \citealt{Zhang2005})
than outflows observed toward lower mass YSOs \citep[e.g.][]{KimandKurtz2006}.

A single outflow mechanism would imply that all stars acquire mass by
a similar disc-accretion process that scales with source
luminosity. However, outflows observed towards MYSOs were initially
suggested to be less collimated than those identified towards low mass
YSOs (e.g. \citealt{Wu2004}). In contrast, both \citet{Beuther2004}
and \citet{Zhang2005} identified similar degrees of collimation
towards the jets/outflows driven by young MYSOs as found in
jets/outflows from low mass YSOs. To account for both low and high
degrees of collimation of massive outflows, \citet{BeutherandShepherd2005} proposed a
scenario in which an initially well collimated jet/outflow
de-collimates with time, eventually forming a wide angle wind. In the
early formation stages outflows driven by MYSOs are potentially
magnetically dominated and therefore highly collimated (e.g. HH80/81,
\citealt{CarrascoGonzalez2010}).  For more evolved MYSOs, a
radiatively driven stellar wind naturally gives rise to a less
collimated outflow (e.g. \citealt{Vaidya2011}). 

Intermediate-mass YSOs (2 \msun$\le$ M$_{*}\le$8 \msun) link the low- and high-mass regimes, and so
have the potential to provide key insights into whether outflow
properties scale smoothly with the mass of the driving (proto)star.  A
complication is that intermediate and high-mass stars generally form
in clustered environments \citep{LadaandLada2003}; as a result, it is
not always unambiguously clear from observations which core(s) in a
region are powering the jet/outflow(s), particularly at lower spatial
resolution.  High angular resolution studies that can resolve
individual cores and outflows are thus crucial to understanding
jets/outflows from intermediate and high-mass stars.

We targeted NGC 2264-C, one of the nearest intermediate/high-mass star
forming regions in the Red \emph{MSX} Source (RMS) survey
\citep{Lumsden2013}, with the Submillimeter Array (SMA).  The aim of
these observations was to resolve the SiO emission seen with the JCMT
(Section~\ref{obs:jcmt}), and so identify
which source(s) in the region are driving \emph{active} outflows.  A
particular goal was to assess the relationship of the RMS source to
the outflows seen at low angular resolution, and so to aid in the
interpretation of a large, low-resolution outflow survey of RMS
objects \citep{Maud2015outflows}.

\begin{figure*}
\includegraphics[width=0.95\textwidth]{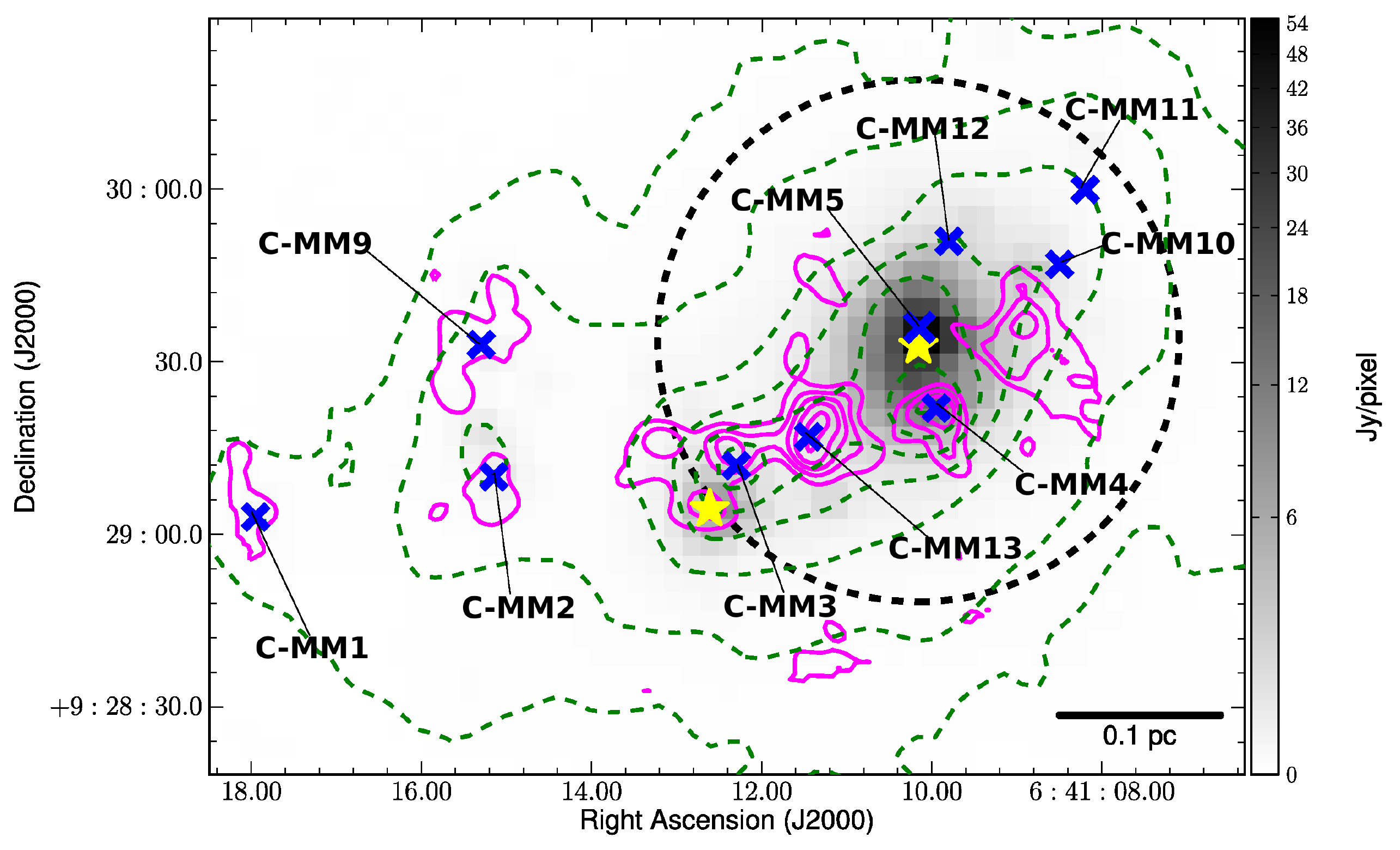}
\caption{Herschel PACS 70\,$\mu$m emission (greyscale, the emission is plotted on a log scale) overlaid with contours of SCUBA 450\,$\mu$m emission (dashed green) and N$_2$H$^+$ (1-0) integrated intensity (PdBI, magenta). The N$_2$H$^+$ (1-0) integrated intensity map was presented in \citet{Peretto2007} and provided to us by N. Peretto, the contours are the same as given in that paper and are from 1\,Jy\,beam$^{-1}$ to 5\,Jy\,beam$^{-1}$ in steps of 1\,Jy\,beam$^{-1}$. The SCUBA contours are given by 1$\sigma$=0.9\,Jy\,beam$^{-1}$ $\times$5,10,20,30,40,50,60. Blue crosses (x) mark the positions of millimetre sources reported by \citet{Peretto2006,Peretto2007}; yellow stars mark the positions of the brightest 24\,$\mu$m \emph{Spitzer} point sources in the region (taken directly from the archive mosaic).  The section of NGC\,2264-C presented in this work is shown by the dashed black circle which represents the 10\% power level of the SMA primary beam centred on R.A. (J2000) 06$^h$41$^m$10.13$^s$, Dec. (J2000) 09$^{\circ}$29$^{'}$34.0$^{''}$. \label{image:infrared1}}
\end{figure*}

Located in the Mon OB1 giant molecular cloud
complex at a distance of 738$^{+57}_{-50}$ pc
\citep{Kamezaki2014}, NGC-2264-C is a comparatively well-studied
region, with a wealth of ancillary data and known CO outflows. 
AFGL 989-IRS1 \citep{Allen1972}, a
9.5\,M$_{\odot}$ B2 star, dominates the region in the IR. 
Thirteen
millimetre continuum sources have been identified \citep[][see Figure
\ref{image:infrared1}]{WardThompson2000,Peretto2006,Peretto2007},
which have typical diameters of $\sim$0.04\,pc and masses ranging
from $\sim$\,2$-$40\,M$_{\odot}$ (\citealt{Peretto2006};
\citealt{Peretto2007}).  From comparing observations and
SPH simulations, \citet{Peretto2007} suggest that NGC 2264-C is in a global state
of collapse onto the central, most massive millimetre core, C-MM3
($\sim$\,40\,M$_{\odot}$).
\citet{Maury2009} observed $^{12}$CO\,(2-1) emission with the IRAM 30m
($\sim$11$\arcsec$ resolution), detecting a network of 11 outflow lobes with projected velocities ranging from 10-30\,km\,s$^{-1}$,
lengths of 0.2-0.8 pc and momentum fluxes in the range
0.5\,$\times$\,10\,$^{-5}$ to
50\,$\times$\,10\,$^{-5}$\,M$_{\odot}$\,km\,s$^{-1}$yr$^{-1}$. However,
the limited angular resolution meant that only a small minority of the
detected outflow lobes could be unambiguously identified with driving
sources (millimetre continuum cores).  Subsequent high resolution SMA observations by
\citet{Saruwatari2011} focusing on the central class 0 protostellar
core, C-MM3, identified a compact, young north-south bipolar outflow
driven by this source in both $^{12}$CO and CH$_3$OH emission,
emphasising the power of high-resolution, multi-line observations.

The ambiguity in identifying outflow driving sources
based on low-resolution single-dish observations calls for high angular resolution observations of a
reliable jet/outflow tracer. Silicon monoxide (SiO) emission is an
effective tracer of jets/outflows from both low mass YSOs
(e.g. \citealt{Codella2014}; \citealt{Tafalla2010};
\citealt{LopezSepulcre2011}; \citealt{Gibb2004}; \citealt{Sakai2010})
and MYSOs (e.g. \citealt{Gibb2007}; \citealt{Codella2013};
\citealt{SanchezMonge2013}; \citealt{Leurini2013}). Importantly, SiO
emission, unlike CO, does not suffer from confusion with easily
excited ambient material. 
CO emission can also be easily excited in outflows, which may obscure the underlying effects; it is
possible for weak outflows to effectively be ``fossil'' momentum
driven remnants even after the central driving engine declines
(e.g. \citealt{Klaassen2006}, \citealt{Hunter2008}). 
In contrast, SiO requires the passage of
fast shocks to release it into the gas phase
(e.g. \citealt{Gusdorf2008b}; \citealt{Guillet2009}) and thus is an
excellent tracer of the fast shocks associated with an active outflow
near the stellar driving source (e.g.\ \citealt{Schilke1997}).

We present the first high angular resolution study ($\sim$3$\arcsec$ resolution
with the Submillimeter Array) of SiO (5-4) emission towards NGC
2264-C.  The main goals of the SMA observations are to identify
\emph{active} outflows and their central
driving sources, and to determine the relationship of these active
outflows to the multiple outflows previously reported based on
single-dish observations of
the region.  In Section 2 we summarise the
observations presented in this paper.  We present our results in
Section 3, discuss the physical properties of the detected cores
and outflows in Section 4, and summarise our conclusions in Section 5.

\begin{figure*}
\includegraphics[width=14cm]{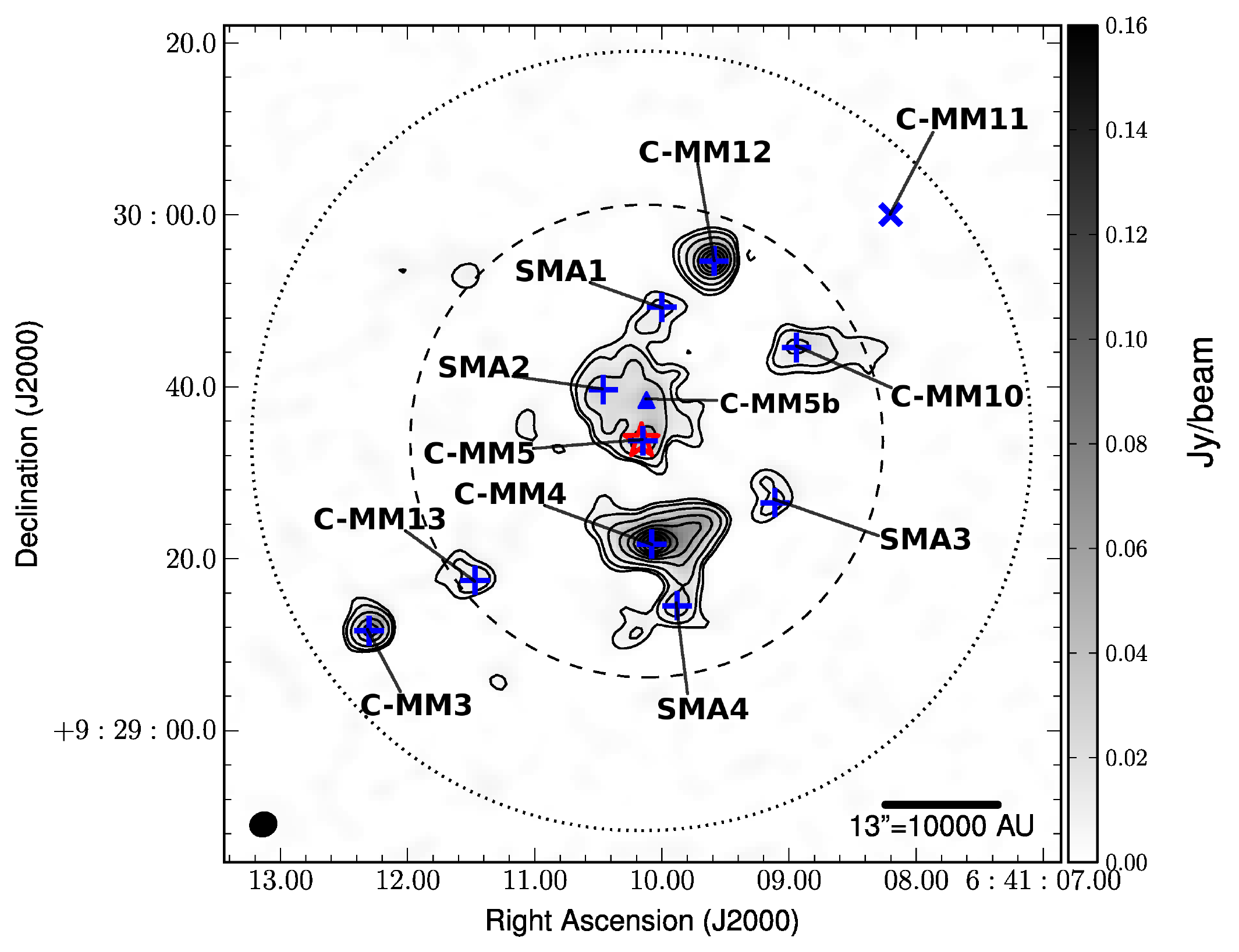}
\caption{Map of the 1.3\,mm continuum emission towards the  section of NGC 2264-C observed with the SMA. The greyscale shows the SMA 1.3\,mm continuum emission from 0 to 0.16 Jy beam$^{-1}$; the peak flux in the field is 0.159 Jy beam$^{-1}$. The black contours represent (3, 5, 10, 20, 30, 40, 50, 60, 70, 80)$\times$ $\sigma$$=$\,2\,mJy beam$^{-1}$.  The peak continuum positions for the 10 leaves (millimetre continuum peaks) identified in the dendrogram are marked with blue pluses (+) and labelled SMA- for new detections and C-MM- for previously detected sources. The blue triangle marks the peak position of the possible millimetre continuum peak C-MM5b. The red star represents the position of AFGL 989-IRS1, taken from the 2MASS data  in the RMS Survey, and the blue cross (x) marks the position of C-MM11 from \citet{Peretto2007}. The black dashed circle  represents the FWHP of the SMA primary beam, and the outer black dotted circle represents the 10\% power level of the SMA primary beam. The SMA synthesised beam (3.06\,$\times$\,2.69$''$, PA= -69.3) is shown at lower left. }
\label{image:continuum}
\end{figure*}

\section{Observations and Data Reduction}
\subsection{Submillimeter Array Observations}
Our Submillimeter Array (SMA)\footnote{The Submillimeter Array is a
  joint project between the Smithsonian Astrophysical Observatory and
  the Academia Sinica Institute of Astronomy and Astrophysics and is
  funded by the Smithsonian Institution and the Academia Sinica.}
observations were made with 8 antennas in the compact configuration on
2010 December 12. The pointing centre for the observations is
R.A. (J2000) 06$^h$41$^m$10.13$^s$, Dec. (J2000)
09$^{\circ}$29$^{'}$34.0$^{''}$. We observed on-source for a total of
$\sim$4 hours, spread over an 8 hour track to improve UV
coverage.  The system temperatures ranged from
$\sim$\,100\,-\,180\,K depending on source elevation. A typical value
of $\tau_{(225GHz)}$ $\sim$\,0.1\,-\,0.15 was obtained during the
observations.
At 1.3\,mm, the SMA primary beam
is $\sim$\,55\,$\arcsec$ (FWHP), and the largest recoverable scale for the array in the compact configuration is
$\sim$\,20${\arcsec}$. 
The total observed bandwidth is $\sim$\,8\,GHz, covering $\sim$216.8\,-\,220.8\,
GHz in the lower sideband and $\sim$228.8\,-\,232.8\,GHz in the upper
sideband. 

Initial calibration was accomplished in MIRIAD
\citep{Sault1995}, with further processing undertaken in
CASA\footnote{http://casa.nrao.edu}.
The bandpass calibration was derived from observations
of the quasar 3C454.3. The gain calibrators were J0530+135, J0532+075,
and J0739+016, and the absolute flux calibration was derived from Uranus.  The fluxes
derived for the quasars were found to be within 20\% of the SMA
monitoring values, which suggests that the absolute flux calibration is good
to within 20\%.  The upper and lower sidebands
were treated individually during calibration. The line and continuum
emission were separated using the command \emph{uvcontsub} in CASA: only
line-free channels were used to estimate the
continuum. Self-calibration was performed on the continuum data in
each sideband, with the solutions applied to the line data. 
The SMA correlator was configured to provide a uniform spectral
resolution of 0.8125 MHz; the line data were resampled to a velocity
resolution of 1.2\,km\,s$^{-1}$, then Hanning smoothed.

The continuum data from the lower and upper sidebands were combined to produce the final continuum image. 
The continuum image and line image cubes were cleaned using a robust
weighting of 0.5. This results in a synthesised beam size of
$\sim$\,3.06\,$\times$\,2.69\,$\arcsec$ with a P.A.\ of approximately
$-67$\,deg for the final 1.3\,mm continuum image, which has a 1$\sigma$ rms
noise level of $\sim$\,2\,mJy beam$^{-1}$.
In the Hanning-smoothed spectral line image cubes, the typical 1$\sigma$ rms noise (per channel) is
$\sim$40\,mJy beam$^{-1}$.
Unless otherwise noted, images displayed in figures have not been corrected for the primary beam response of the SMA. 
All reported measurements were made from images corrected for the primary beam response.

\subsection{JCMT Observations\label{obs:jcmt}}

The SiO (J\,=\,8-7) line (347.3305\,GHz) was observed with the
Heterodyne Array Receiver Program (HARP) and Auto-Correlation Spectral
Imaging System \citep[ACSIS;][]{Buckle2009jcmt} at the James Clerk
Maxwell Telescope\footnote{The James Clerk Maxwell Telescope has
historically been operated by the Joint Astronomy Centre
on behalf of the Science and Technology Facilities Council
of the United Kingdom, the National Research Council
of Canada and the Netherlands Organisation for Scientific
Research.} (JCMT) on 2009 April 26.  The SiO (J\,=\,8-7)
data shown in this paper are part of a larger project (M09AU18), which
will be presented elsewhere. A position-switched jiggle chop map with
the tracking centred on R.A. (J2000) 06$^h$41$^m$10.10$^s$,
Dec. (J2000) 09$^{\circ}$29$^{'}$34$^{''}$ was made with the HARP
array. The array is made up of 16 receiver elements, however, at the
time of observing receiver H14 was not operational and is therefore
missing from our jiggle map. The reference position for the sky
subtractions was R.A. (J2000) 06$^h$46$^m$35.50$^s$, Dec. (J2000)
10$^{\circ}$10$^{'}$30$^{''}$. The standard pointing check was made
before the start of each $\sim$\,30 minute observing block on a
point-like spectral line standard, which was also used to check the
overall flux scale. At 345 GHz the JCMT has a beam size of
$\sim$\,15\,$\arcsec$ and a main beam efficiency $\eta_{mb}$ of 0.61
\citep{Buckle2009jcmt}.  During the observations,
$\tau_{(225GHz)}$\,$\sim$\,0.042.  The SiO data were smoothed to a
velocity resolution of 1.6\,km\,s$^{-1}$ to improve the
signal-to-noise. A 1$\sigma$ rms noise level of
T$_{\rm MB}$\,$\sim$\,0.04\,K was achieved in the final map (totalling 66 minutes of
observation time). The data were processed using Starlink\footnote{http://starlink.eao.hawaii.edu/starlink} packages and
converted to the main-beam temperature scale (T$_{\rm MB}$).

\subsection{Archival Data}

To complement our SMA and JCMT HARP observations,
we utilise archival mid/far-infrared data.  
These archival data include PACS \citep{Poglitsch2010} 70\,$\mu$m
observations (observational ID 1342205056, P.I. F.\ Motte) taken with
the ESA \emph{Herschel Space Observatory}\footnote{{\it Herschel} is
  an ESA space observatory with science instruments provided by
  European-led Principal Investigator consortia and with important
  participation from NASA.} \citep{Pilbratt2010}, \emph{Spitzer}
MIPSGAL 24\,$\mu$m observations \citep{Rieke2004} and SCUBA
450\,$\mu$m data \citep{DiFrancesco2008}.

The Herschel PACS 70\,$\mu$m and SCUBA 450\,$\mu$m maps are presented in Figure~\ref{image:infrared1}, overlaid with the positions of the brightest \emph{Spitzer} MIPSGAL 24\,$\mu$m point sources in our SMA field of view{\color{blue}\footnote{The positions were measured directly from the downloaded archive level 2 data, the data is saturated at the position of AFGL989-IRS1}} . We have corrected the astrometry for the PACS archival data using the \emph{Spitzer} MIPSGAL 24\,$\mu$m images, which were calibrated against 2MASS point source positions \citep{Skrutskie2006}.

\section{Results}

\begin{figure*} 
\includegraphics[width=0.46\textwidth]{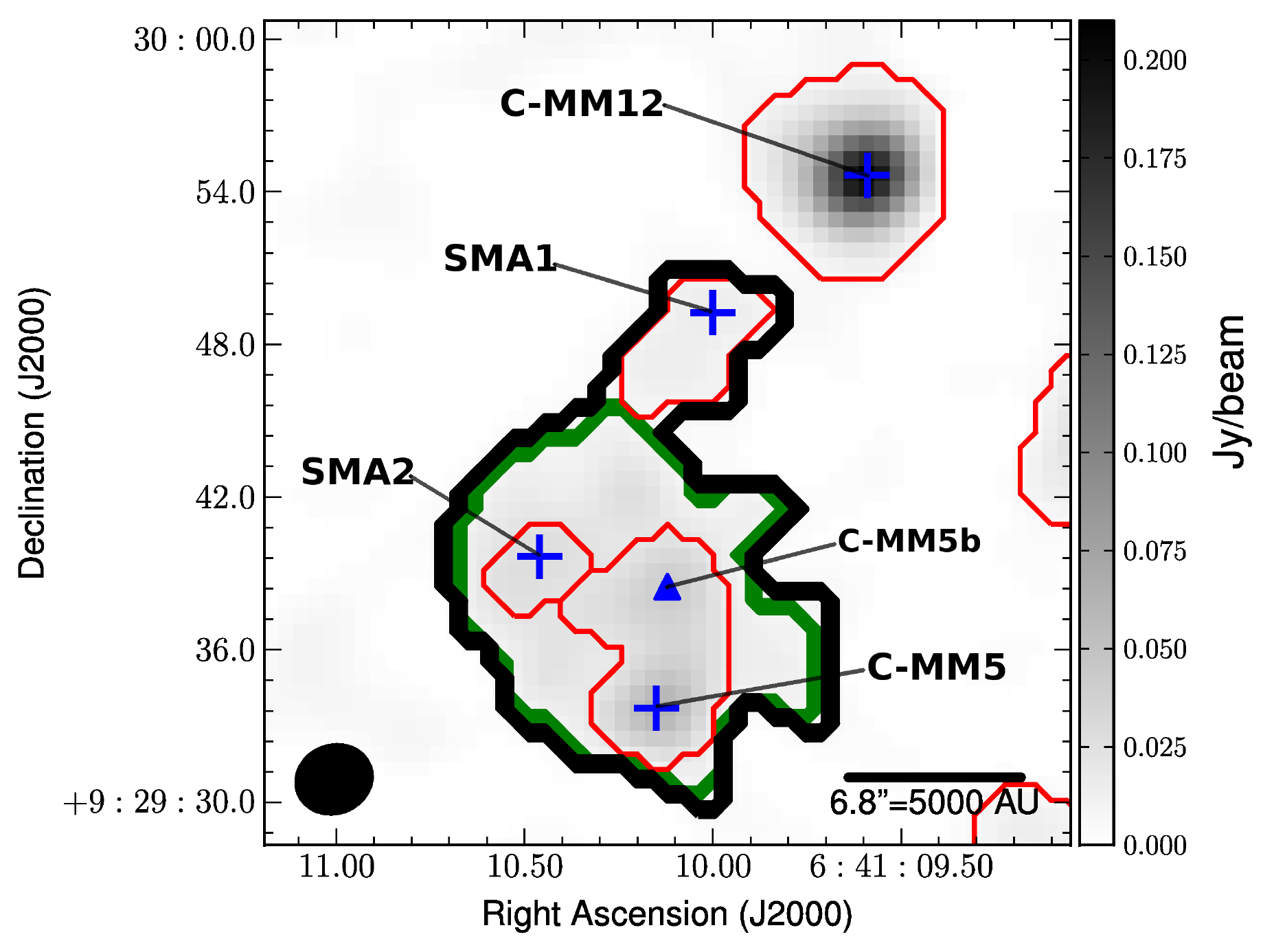}
\includegraphics[width=0.46\textwidth]{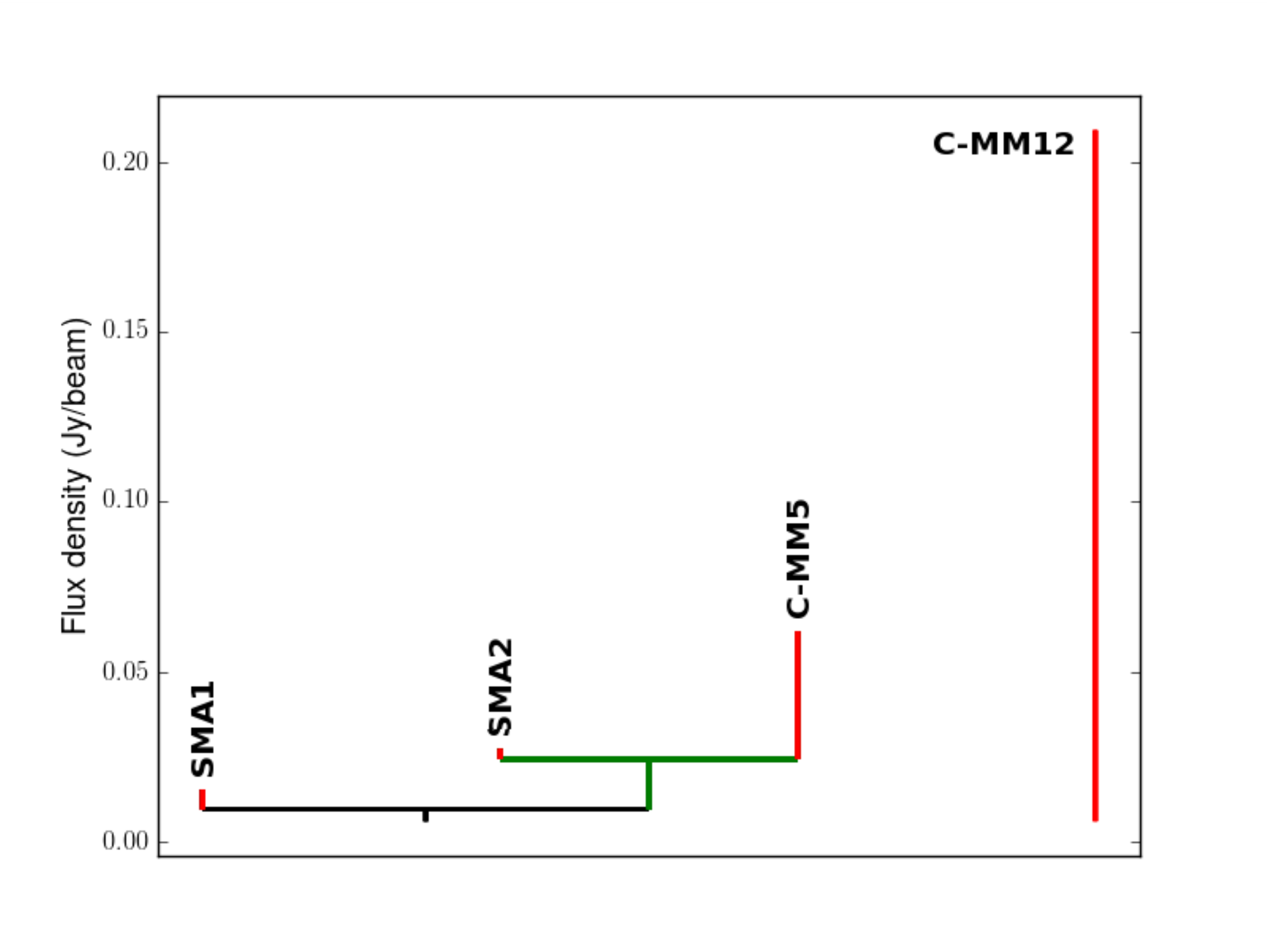}
\caption{Examples of dendrogram tree structure for C-MM5, C-MM12, SMA1 and SMA2. The left panel shows the 1.3 mm continuum image, corrected for the primary beam response, in greyscale, overlaid with 3$\sigma$ contours from the 1.3\,mm continuum image prior to correction for the primary beam.  These contours correspond to structures identified in the dendrogram analysis (Section~\ref{section:continuum}). Structures are identified using a minimum threshold of 4\,$\sigma$ (1$\sigma$\,$=$\,2mJy/beam) and a minimum number of 18 contiguous pixels, with a step increase of 1\,$\sigma$ for further substructure to be identified. The black contours (left panel) and corresponding black lines (tree diagram, right panel) show an example of a parent tree structure with further substructure (in this case the structure containing C-MM5, SMA1, and SMA2). The green contour represents an additional branch within this structure, which hosts two cores, C-MM5 and SMA2. The red contours (left panel) and corresponding red lines (tree diagram, right panel) denote the highest level of a parent structure or of an independent structure with no further substructure if present. C-MM12 shows only parent structure and has no further substructure. A colour version of this figure is available online. \label{image:dendrogram}}
\end{figure*}

\begin{table*}
\begin{minipage}{180mm}
\begin{center}
\caption{Properties of Millimetre Continuum Sources \label{table:Continuum}}
\centering
\begin{tabular}{c c c c c c c c c}
\hline\hline
Source $^a$&\multicolumn{2}{c}{J2000.0 Coordinates$^b$}&\multicolumn{2}{c}{J2000.0 Coordinates$^c$} & R$_{eff}$$^d$ &I$_{\rm peak}$$^{e}$& S$_{\nu}$ $^{f}$\\
\cline{2-3}
\cline{4-5}
           &  $\alpha$($^{hms}$)   & $\delta$($^{\circ}\,'\,''$)  &  $\alpha$($^{hms}$)   &  $\delta$($^{\circ}\,'\,''$)    &  (pc)  & (mJy beam$^{-1}$) &   (mJy)   \\
\hline
C-MM3  &  06 41 12.31&  +\,09 29 11.50 &   06 41 12.28 & +\,09 29 11.96 &    0.011   &     370  &    395   \\
C-MM4  &  06 41 10.08&  +\,09 29 21.70 &   06 41 09.95 & +\,09 29 21.40 &    0.017   &     185  &    496   \\
C-MM5  &  06 41 10.15&  +\,09 29 33.70 &   06 41 10.14 & +\,09 29 35.87 &    0.012   &     62   &    131   \\
C-MM10 &  06 41 08.94&  +\,09 29 44.56 &   06 41 08.75 & +\,09 29 44.24 &    0.015   &     37   &    113   \\
C-MM12 &  06 41 09.59&  +\,09 29 54.64 &   06 41 09.63 & +\,09 29 54.62 &    0.013   &     209  &    274   \\
C-MM13 &  06 41 11.47&  +\,09 29 17.47 &   06 41 11.51 & +\,09 29 18.00 &    0.0099  &     34   &    53    \\
SMA1   &  06 41 10.00&  +\,09 29 49.25 &   06 41 10.12 & +\,09 29 47.97 &    0.0079   &     15   &    21    \\
SMA2   &  06 41 10.46&  +\,09 29 39.66 &   06 41 10.46 & +\,09 29 39.16 &    0.0057   &     27   &    26    \\
SMA3   &  06 41 09.11&  +\,09 29 26.50 &   06 41 09.16 & +\,09 29 27.38 &    0.0093   &     17   &    31    \\
SMA4   &  06 41 09.88&  +\,09 29 14.53 &   06 41 09.88 & +\,09 29 14.85 &    0.0053   &     48   &    36    \\

\hline

\end{tabular}
\end{center}
{\bf Notes.}\\
$^{a}$ {Sources identified in the dendrogram analysis: C-MM-
  denotes previously known/named millimetre sources \citep{Peretto2006,Peretto2007}; SMA- denotes new 1.3\,mm detections.}\\
$^{b}$ {Position of millimetre continuum peak.}\\
$^{c}$ {Position of intensity-weighted centroid, computed using \textit{Computing Dendrogram Statistics} \citep{Astropy2013}.}\\
$^{d}$ {Effective radius, computed from the total leaf area using \textit{Computing Dendrogram Statistics} \citep{Astropy2013}, through R$_{eff}=\sqrt{Area/\pi}$ and assuming the distance of 738pc to NGC 2264-C.}\\
$^{e}$ {Peak intensity of 1.3 mm continuum emission, corrected for the primary beam response.}\\
$^{f}$ {Integrated 1.3 mm flux density, measured from the primary-beam-corrected continuum image using {\sc CASAVIEWER} and
  the 3\,$\sigma$ mask from the dendrogram computed from the uncorrected image (see Section~\ref{section:continuum}). For sources which are nested within parent structures (e.g. C-MM5, SMA1 and SMA2), only the flux from within the highest level of that structure (e.g. the leaves) is extracted (see the red contours in Figure \ref{image:dendrogram}).}\\
\end{minipage}
\end{table*}

\subsection{1.3\,mm Continuum Emission \label{section:continuum}}
\begin{figure*}
\centering
\includegraphics[width=0.95\textwidth]{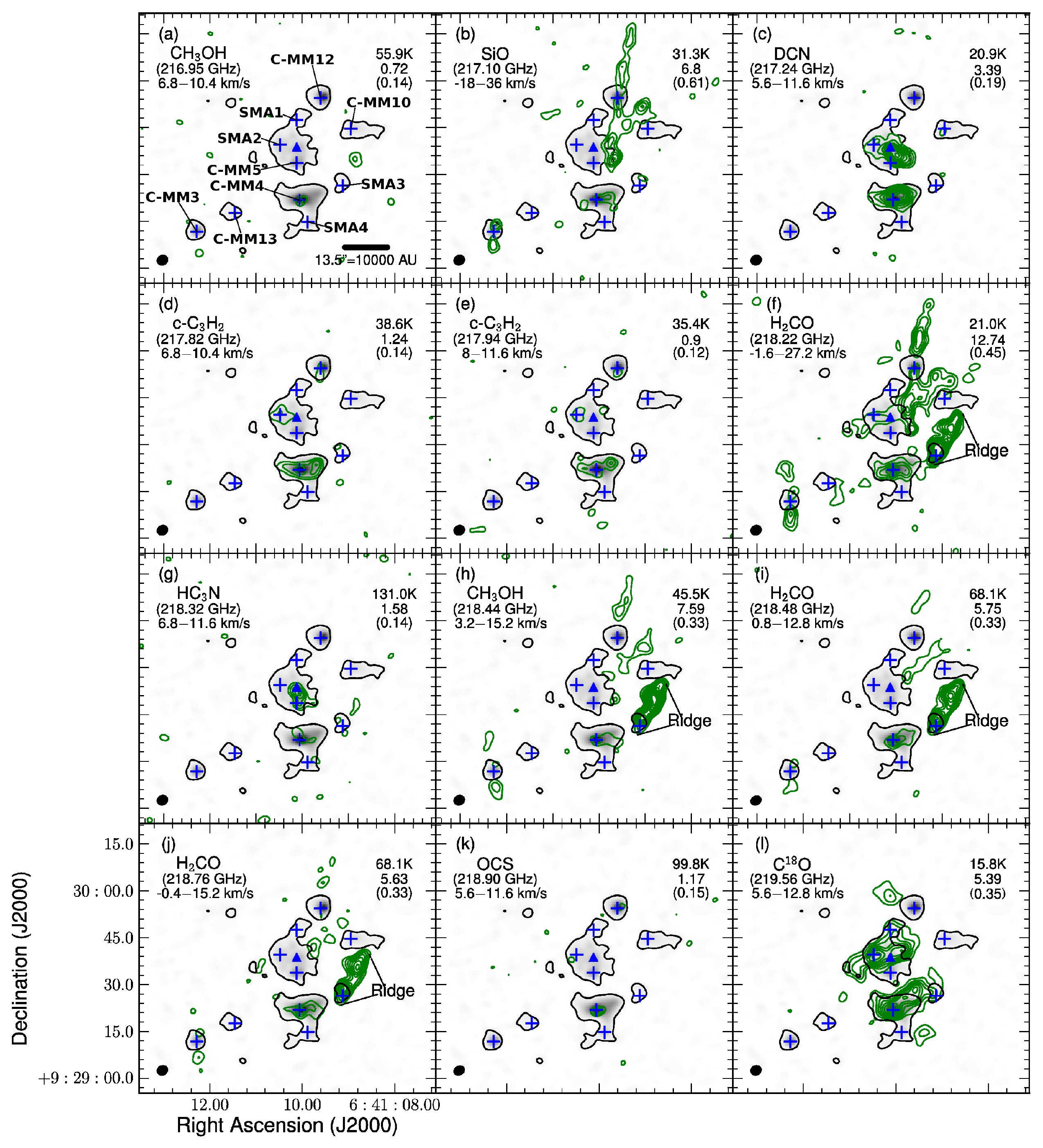}
\caption{SMA integrated intensity maps (green contours) of detected
  molecular lines, overlaid on the 1.3\,mm continuum emission
  (greyscale and black contour, contour level 3$\sigma$$=$6\,mJy
  beam$^{-1}$).  The integrated velocity range is given in the top
  left corner of each panel, along with the molecule and line rest
  frequency.  E$_{\rm upper}$ for the transition and the peak and rms
  of each map (both in Jy beam$^{-1}$ * km\,s$^{-1}$) are given at
  upper right in each panel.  Blue pluses (+) mark the positions of
  the millimetre continuum peaks from Table~{\ref{table:Continuum}},
  and the blue triangle marks the position of C-MM5b (see
  Section~\ref{section:continuum}). Contour levels for the integrated
  intensity maps are from 3\,$\sigma$ to peak in steps of 2\,$\sigma$.
  Panel (p) shows the CH$_{3}$CN emission integrated over the k=0-4
  components and thus the velocity range and frequency is not given.  \label{image:chemistry}}
\end{figure*}

\begin{figure*}
\centering
\includegraphics[width=0.95\textwidth]{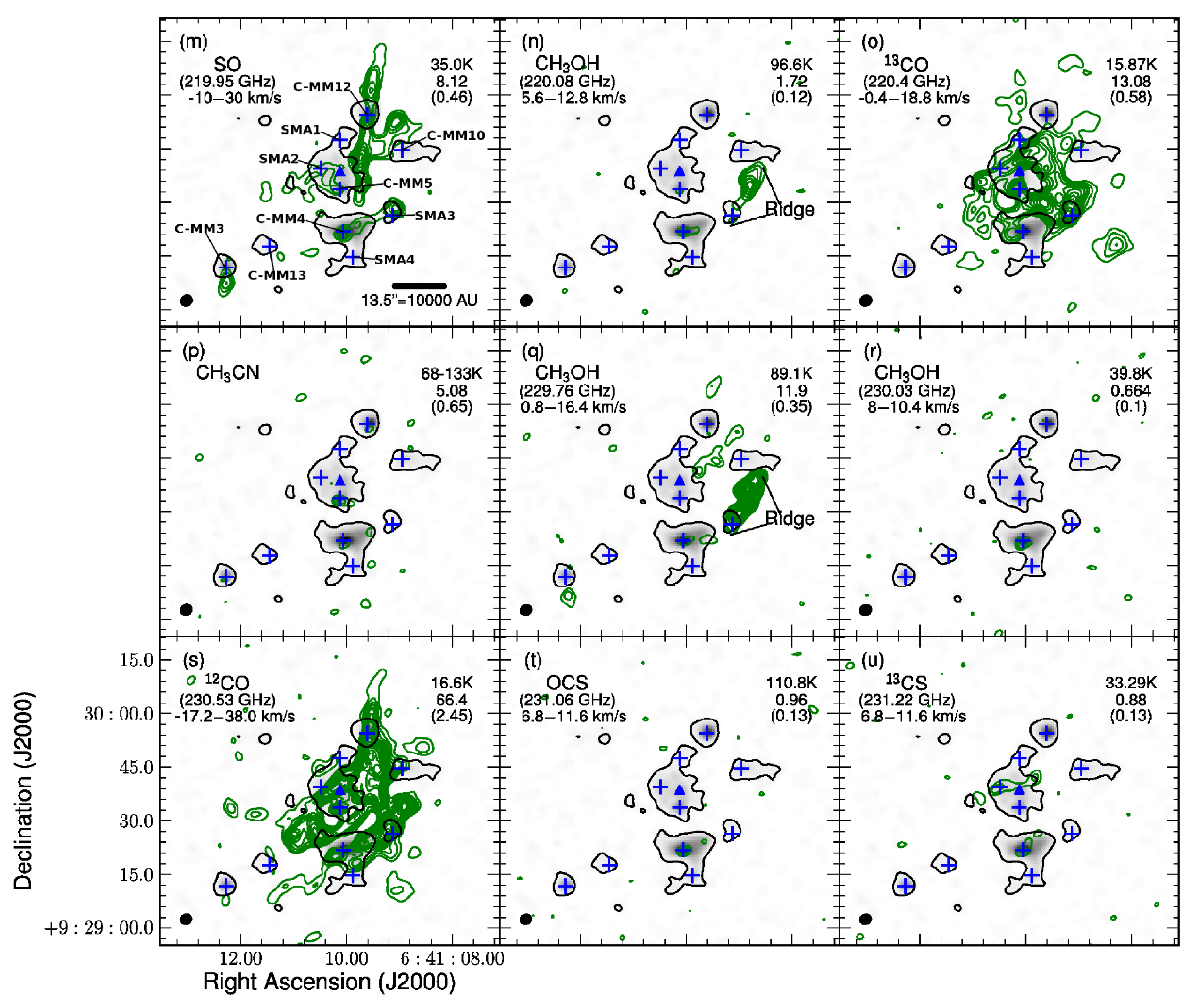}
\contcaption{}
\end{figure*}

The SMA 1.3\,mm continuum image is presented in Figure \ref{image:continuum}. The structure of the 1.3\,mm emission was characterised through the implementation of dendrograms \citep{Rosolowsky2008}. Dendrograms or structure tree diagrams can be used to define hierarchical structure in clouds. The dendrogram tree structures are comprised of three main components: a trunk, branches and leaves, which correspond to increasing levels of intensity in the dataset. We set the minimum threshold intensity required to identify a parent tree structure (e.g. trunk) to be 4\,$\sigma$ (in the continuum image prior to correction for the primary beam response; $\sigma$\,$\sim$\,2\,mJy beam$^{-1}$) and a minimum area of 18 contiguous pixels (the equivalent area of the synthesised beam). Further nested substructures (i.e. branches and leaves) require an additional 1\,$\sigma$ increase in intensity, again over a minimum of 18 contiguous pixels. To compute the dendrogram we have used the python implementation astrodendro\footnote{This research made use of astrodendro, a Python package to compute dendrograms of Astronomical data (http://www.dendrograms.org/).}.

We identify 10 millimetre continuum sources in our SMA 1.3 mm image with the dendrogram analysis, of which four are new detections (labelled SMA1-SMA4 in Figure~\ref{image:continuum}). The other six 1.3\,mm continuum sources have been previously reported (\citealt{Peretto2006}, \citealt{Peretto2007}, \citealt{WardThompson2000}) and are labelled in Figure~\ref{image:continuum} according to the naming scheme of Peretto et al.\ (C-MM-).
Of the ten 1.3\,mm continuum sources identified in the dendrogram analysis, five are found as separate, independent parent structures with no substructure, and five are found nested into two tree structures. The two nested structures are comprised of C-MM4 and SMA4, and of C-MM5, SMA1, and SMA2, respectively (see Figure \ref{image:dendrogram} for an example of nested structure). 
The parameters of the millimetre continuum sources (e.g.\ peak intensity, integrated flux density, and effective radius) were extracted using the analysis code \textit{Computing Dendrogram Statistics}\footnote{This research made use of Astropy, a community-developed core Python package for Astronomy \citep{Astropy2013}.} and the {\sc CASAVIEWER} tool, and are presented in Table~\ref{table:Continuum}.  All properties are estimated from the primary beam corrected 1.3 mm continuum image using a 3\,$\sigma$ mask derived from the continuum image prior to correction for the primary beam response. This approach was taken to avoid including noise in the parameter estimates for millimetre continuum sources far from the pointing centre. 

We detect all of the previously identified millimetre continuum peaks within the FWHP of our SMA primary beam.  Of the sources reported by \citet{Peretto2006} and \citet{Peretto2007} that fall within the field of view shown in Figure~\ref{image:continuum}, only C-MM11 --which lies outside the 20\% power level of the SMA primary beam-- is not detected in our SMA image.  Our nondetection of C-MM11 is also consistent with the relative intensities of the millimetre continuum sources reported by \citet[][1.2 mm IRAM 30-m observations]{Peretto2006}: C-MM11 is the weakest Peretto et al.\ source within our SMA field (Figure~\ref{image:infrared1}).

Our four new detections SMA1, SMA2, SMA3, and SMA4 have the lowest integrated flux densities of the ten millimetre continuum
sources: the brightest new detections (SMA3 and SMA4) have integrated flux densities that are approximately half that of the weakest
previously reported source (C-MM13) in our SMA image (Table~\ref{table:Continuum}).  This is consistent with SMA1-SMA4 not having been reported based on previous
observations.  We note that SMA4 is visible as a 3$\sigma$ contour in Figure 1 of \citet[][PdBI 3.2 mm observations]{Peretto2007}, but was considered a marginal detection and not studied further.

Inspecting the continuum map by eye reveals an additional millimetre continuum peak not identified by the dendrogram analysis. We separate C-MM5 into two cores: one at the peak position for C-MM5 tabulated in Table~{\ref{table:Continuum}}, and a second $\sim$4.5$\arcsec$ north (see  Figure \ref{image:continuum}). The intensity of this second peak surpasses the dendrogram threshold for substructure, however the number of contiguous pixels is below the set limit, so the leaf is merged into C-MM5. If the number of contiguous pixels required is reduced to 77\% of the area of the synthesised beam, C-MM5 is separated into two leaves by the dendrogram analysis. Inspection of the molecular line emission also indicates these two leaves may be separate objects (Section~\ref{section:results_lines}). We therefore suggest that the northern leaf (hereafter C-MM5b) and C-MM5 are distinct millimetre continuum peaks, which we are unable to unequivocally separate at the resolution of our SMA data.

\subsection{Molecular Line Emission \label{section:results_lines}}

We detect molecular line emission in 23 transitions of 13 species above 5\,$\sigma$ with the SMA,
including SiO, CO and its isotopologues ($^{13}$CO and C$^{18}$O), SO,
OCS, H$_2$CO, CH$_3$OH, DCN, c-C$_3$H$_2$, $^{13}$CS, HC$_3$N and
CH$_3$CN.  All lines detected at $>$5$\sigma$ have E$_{\rm upper}$ between $\sim$\,16 and 131\,K. 
Table 2 lists the species, transition,
rest frequency, E$_{\rm upper}$, and whether the line is detected at
$>$5\,$\sigma$ towards the millimetre continuum peak of each of the
11 cores (including C-MM5b, see Section~\ref{section:continuum}).
Table 3 presents the parameters (peak line
intensity, line centre velocity, linewidth, and integrated line
intensity) obtained from single Gaussian fits to lines detected at
$>$5\,$\sigma$, at each millimetre continuum peak.  No fits are
presented for $^{12}$CO and $^{13}$CO due to the complexity of the
line profiles, which are strongly affected both by self absorption and
by artefacts from poorly imaged large-scale emission; $^{12}$CO and $^{13}$CO are similarly excluded from Table 2.

Figure~\ref{image:chemistry} presents integrated intensity maps for
all detected transitions, including $^{12}$CO and $^{13}$CO.  The velocity range over which emission is
integrated is determined for each transition, to encompass all
channels with emission above 3$\sigma$.
As shown in Figure~\ref{image:chemistry}, the morphology of the
detected line emission is complex.  For many transitions, the spatial
morphology of the line emission differs markedly from that of the
millimetre continuum. The emission in the integrated
  intensity maps falls into three main categories: (i) ``compact''
  molecular line emission that directly traces millimetre continuum
  peaks, which includes emission from CH$_3$CN, OCS (18-17) and OCS (19-18), HC$_3$N, and the CH$_3$OH
  5(1 , 4)- 4(2 , 2), CH$_3$OH 8(0 , 8)- 7(1 , 6) E and CH$_3$OH 3(
  -2, 2)- 4( -1, 4) E transitions; (ii) ``diffuse'' molecular line emission that
  overlaps but may not be clearly associated with millimetre continuum
  emission (e.g.\ may differ in morphology), which includes emission from c-C$_3$H$_2$, DCN and C$^{18}$O; and
  (iii) spatially extended molecular line emission that is more likely
  associated with outflows, which includes emission from
  SiO, SO, H$_2$CO, and the CH$_3$OH 4(2 , 2)- 3(1 ,
  2) and CH$_3$OH 8( -1, 8)- 7(0 , 7) E transitions.  Emission from $^{13}$CS appears to fall into both the ``compact'' (i)
  and ``diffuse'' (ii) categories.  As shown in Figure~\ref{image:chemistry}, the velocity-integrated $^{12}$CO and
  $^{13}$CO emission extends across much of the field, $^{12}$CO and
  $^{13}$CO are therefore not included in these categories; high-velocity $^{12}$CO, however, traces outflows, as discussed in Section~\ref{sec:outflows}.  
A distinct feature, which we have labelled the ``ridge'',
is also evident in Figure~\ref{image:chemistry}.  The ridge is
associated with the strongest H$_2$CO emission in the field, as well
as the strongest emission in several CH$_3$OH transitions.  While the
southern edge of the ridge overlaps SMA3, the majority of the ridge
structure does not appear to be associated with a millimetre continuum
source.  The nature of this intriguing feature is discussed in
Section~\ref{discussion:ridge}.
\subsubsection{Line Emission Associated with Millimetre Continuum Sources: Systemic Velocity Estimates \label{results:vlsr}}

As shown in Figure~\ref{image:chemistry}, compact molecular line emission that peaks at the millimetre continuum peaks is seen in only a few lines (OCS, CH$_3$CN, HC$_3$N and the 216.946 GHz, 220.078 and 230.027 GHz CH$_3$OH transitions), and only towards C-MM4, C-MM5, and C-MM5b (not all of these lines are detected towards all three sources; see Table 2).
For these sources, we present estimates of the $v_{\rm LSR}$'s of the individual millimetre continuum cores in Table~\ref{table:vlsr}, based on the fits (Table 3) to lines that exhibit spatially compact emission. We note that for CH$_3$CN, the ladder transitions with a detection $<$5$\sigma$ are excluded from the $v_{\rm LSR}$ estimates. 

DCN and c-C$_3$H$_2$ exhibit more extended emission (which we refer to as ``diffuse'') that is coincident with, and morphologically similar to, the millimetre continuum emission of C-MM4 and SMA2.  For C-MM4, in particular, the similar morphologies of the line and millimetre continuum emission suggest that this more extended molecular line emission is associated with the millimetre continuum source. Strong DCN emission is also detected in the vicinity of C-MM5 and C-MM5b, but in this case the morphology of the line emission differs from that of the millimetre continuum, so it is not clear that the DCN emission is directly associated with the continuum source(s). Table~\ref{table:vlsr} presents estimates of core $v_{\rm LSR}$'s based on fits to lines that exhibit ``diffuse'' emission (including C$^{18}$O as well as DCN and c-C$_3$H$_2$) and are detected at the position of a given core's millimetre continuum peak. We emphasise that while this approach allows us to estimate $v_{\rm LSR}$'s for more sources, the results must be treated with greater caution than the estimates based on compact molecular line emission. The morphology of the C$^{18}$O emission, for example, does not match that of the millimetre continuum, suggesting that the detected C$^{18}$O may not arise from the dense gas of the millimetre continuum cores. Our $v_{\rm LSR}$ estimates exclude lines that likely trace outflows and exhibit clearly
extended emission (e.g.\ SiO, SO, H$_2$CO and the remaining CH$_3$OH lines).

The $v_{\rm LSR}$'s of the millimetre continuum sources, estimated as described above, range from $\sim$7.8 to 9.9\,km\,s$^{-1}$. For C-MM4 and C-MM5 (for which we have the most data from line fits, see Tables 3 and \ref{table:vlsr}), the $v_{\rm LSR}$ estimates based on compact and ``diffuse'' tracers differ by up to $\sim$1\,km\,s$^{-1}$ (for C-MM5), and the standard deviation of the $v_{\rm LSR}$'s measured from individual lines is $\sim$0.5\,km\,s$^{-1}$ (Table~\ref{table:vlsr}).  
\setcounter{table}{2}
\begin{landscape}
\begin{small} 
\begin{center}

{{\bf Table 2}: Molecular Line Transitions Detected towards Millimetre Continuum Peaks in NGC 2264-C \label{table:molecularlineproperties}}
\centering
\begin{tabular}{c c c c c c c c c c c c c c c}
\hline\hline
Species        & Transition         & Frequency$^a$ & E$_{upper}$$^a$ &   &   &   &   & Detected$^b$  &   &   &   	&   &   &      \\[0.5ex]
\cline{5-15}
               &                   &    (GHz) &    (K)         & CMM3 & CMM4 & CMM5 & CMM10 & CMM12 & CMM13 & CMM5b & SMA1 & SMA2 & SMA3 & SMA4 \\  
\hline
\multicolumn{14}{c}{LSB}\\  
\hline
CH$_3$OH        &5(1 , 4)- 4(2 , 2) & 216.946  &   55.9           & N & Y & N & N & N & N & N & N & N & N & N \\
SiO             &   (5 - 4)         & 217.105  &  31.3            & Y & Y & N & N & Y & N & Y & N & N & N & N \\ 
DCN             &   (3 - 2)         & 217.239  &  20.9            & N & Y & Y & Y & N & N & Y & N & Y & N & N  \\
{\it c}-C$_3$H$_2$ $^c$& 6$_{16}$ - 5$_{05}$  & 217.822   &  38.6 & N & Y & N & N & Y & N & N & N & Y & N & N  \\
{\it c}-C$_3$H$_2$$^c$&  6$_{06}$ - 5$_{15}$  & 217.822   &  38.6 & N & Y & N & N & Y & N & N & N & Y & N & N \\
{\it c}-C$_3$H$_2$&  5$_{14}$ - 5$_{23}$  & 217.940   &   35.4    & N & Y & N & N & N & N & N & N & Y & N & N  \\
H$_2$CO         & 3( 0, 3)- 2( 0, 2)& 218.222  &   21.0           & Y & Y & N & Y & Y & N & Y & N & Y & Y & N  \\
HC$_3$N         & (24 - 23)         & 218.324  &   131.0          & N & Y & Y & N & N & N & Y & N & N & N & N  \\ 
CH$_3$OH        &4(2 , 2)- 3(1 , 2) & 218.440  &   45.5           & N & Y & N & N & N & N & N & N & N & Y & N  \\
H$_2$CO         & 3( 2, 2)- 2( 2, 1)& 218.476  &   68.1           & N & Y & N & N & N & N & N & N & N & Y & N \\
H$_2$CO         & 3( 2, 1)- 2( 2, 0)& 218.760  &   68.1           & Y & Y & N & N & Y & N & Y & N & Y & Y & N  \\
OCS             & (18 - 17)         & 218.903  &  99.8            & N & Y & N & N & Y & N & N & N & N & N & N  \\
C$^{18}$O       &  (2 - 1)          & 219.560  &  15.8            & N & Y & N & Y & N & N & Y & Y & Y & Y & N \\
SO              &  6(5)-5(4)        & 219.949  &  35.0            & N & Y & Y & N & Y & N & Y & N & Y & Y & N \\
CH$_3$OH        & 8(0 , 8)- 7(1 , 6) E & 220.078  &   96.6        & N & Y & N & N & N & N & N & N & N & N & N\\
CH$_3$CN        & (12$_4$-11$_4$)    & 220.679 &   183.3          & N & N & Y$^d$ & N & N & N & N & N & N & N & N\\
CH$_3$CN        & (12$_3$-11$_3$)    & 220.709  &  133.3          & N & Y$^d$ & Y$^d$ & N & N & N & N & N & N & N & N 	\\
CH$_3$CN        & (12$_2$-11$_2$)    & 220.730  &  97.4           & N & Y$^d$ & Y$^d$ & N & N & N & N & N & N & N & N \\
CH$_3$CN $^e$   & (12$_1$-11$_1$)    & 220.743  &  76.0           & N & Y & Y & N & N & N & N & N & N & N & N \\
CH$_3$CN $^e$   & (12$_0$-11$_0$)    & 220.747  &  68.8           & N & Y & Y & N & N & N & N & N & N & N & N \\
\hline   
\multicolumn{14}{c}{USB}\\
\hline
CH$_3$OH        & 8( -1, 8)- 7(0 , 7) E  & 229.759  &   89.1       & N & Y & N & N & N & N & N & N & N & Y & N \\
CH$_3$OH        &  3( -2, 2)- 4( -1, 4) E & 230.027  &   39.8      & N & Y & N & N & N & N & N & N & N & N & N  \\
OCS             & (19 - 18)             & 231.061  &  110.8        & N & Y & N & N & N & N & N & N & N & N & N \\
$^{13}$CS       &  (5 - 4)              & 231.221  &  33.29        & N & Y & N & N & N & N & N & N & Y & N & N \\
\hline
\end{tabular} 
\end{center}
\end{small}
{\bf Notes.}\\
$^a$ {From Splatalogue (http://www.splatalogue.net/), \citet{Muller2005}.}\\
$^b$ {Y: Detected at $\ge$5$\sigma$, N: Undetected (no emission at $\ge$5$\sigma$).  All measurements were made from 	image cubes corrected for the primary beam response, and a detection was determined from the spectra extracted from those 	beam corrected image cubes. Thus, for sources far from the pointing centre such as C-MM3, the rms noise is higher.}\\
$^c$ {These components are blended in the spectra and cannot be separated.}\\
$^d$ {CH$_3$CN components where emission is $<$5$\sigma$ but $>$3$\sigma$.}\\
$^e$ {These CH$_3$CN components are blended in the spectra: 'Y' indicates emission $>$5$\sigma$ for the blended 	line.  See Table 3 for line fits.}\\
\end{landscape}

\begin{table*}
\begin{minipage}{180mm}
\begin{small}
\begin{center}
\caption{Gaussian Fits to Molecular Lines Detected at Millimetre Continuum Peaks \label{table:molecularlinefits}}
\centering
\begin{tabular}{l c c c c c c c}
\hline\hline
 Species        & Transition  & Frequency& E$_{upper}$ & \multicolumn{4}{c}{Fitted Line Parameters}\\[0.3ex]
\cline{5-8}

                &             &          &    &   Peak Intensity$^a$  &  V$_{centre}$$^a$  &  Linewidth  $^a$    &   $\int$ S $dv$ $^a$   \\
                &             &  (GHz)  & (K) & (Jy beam$^{-1}$) & (km\,s$^{-1}$)&  (km\,s$^{-1}$) & (Jy beam$^{-1}$\,km\,s$^{-1}$)   \\
\hline

                &             &          &                 &   C-MM3   &   &   &    \\
\hline
H$_2$CO  $^b$   & 3(0, 3) - 2(0, 2) & 218.222  &   21.0    &   0.91 (0.11) & 6.60 (0.22) & 3.93 (0.54) & 3.80 (0.49)\\
H$_2$CO  $^b$   & 3(0, 3) - 2(0, 2) & 218.222  &   21.0    &   0.44 (0.08) & -6.50 (0.33) & 4.72 (0.98) & 2.22 (0.40) \\
H$_2$CO         & 3(2, 1) - 2(2, 0) & 218.760  &   68.1    &   0.58 (0.11) & 8.22 (0.25) & 2.68 (0.62) & 1.66 (0.36) \\ 
\hline
                &             &          &                 &   C-MM4   &   &   &    \\
\hline
CH$_3$OH        & 5(1 , 4) - 4(2, 2) & 216.946  &   55.9    &   0.25 (0.03) &  8.88 (0.15) &  3.70 (0.35) & 0.98 (0.12) \\
SiO $^c$        &   (5 - 4)   & 217.105  &  31.3           &   0.22 (0.03)  &  15.04 (0.41)&  5.33 (0.97) & 1.26 (0.30)  \\
DCN             &   (3 - 2)         & 217.239  &  20.9     &   0.87 (0.04)   &  8.53 (0.08) & 3.54 (0.19) & 3.29 (0.23) \\
{\it c}-C$_3$H$_2$ & 6$_{16}$ - 5$_{05}$& 217.822&38.6 &   0.55 (0.03) &  7.77 (0.06) & 2.54 (0.15) & 1.50 (0.08) \\
{\it c}-C$_3$H$_2$&  5$_{14}$ - 5$_{23}$  & 217.940& 35.4  &   0.24 (0.03)  & 8.70 (0.19) & 2.92 (0.46) & 0.74 (0.11) \\
H$_2$CO         & 3(0, 3) - 2(0, 2)& 218.222  &   21.0    &   1.75 (0.03) & 7.62 (0.03) & 3.3 (0.07) & 6.14 (0.11) \\
HC$_3$N         & (24 - 23)         & 218.324  &   131.0   &   0.21 (0.03) & 8.11 (0.20) & 3.88 (0.49) & 0.85 (0.14) \\
CH$_3$OH        & 4(2, 2) - 3(1, 2) & 218.440  &   45.5    &   0.51 (0.03) & 8.72 (0.11) & 3.39 (0.26) & 1.83 (0.13) \\
H$_2$CO         & 3(2, 2)- 2(2, 1)& 218.476  &   68.1    &   0.55 (0.03) & 7.82 (0.11) & 3.85 (0.26) & 2.24 (0.14) \\
H$_2$CO         & 3(2, 1)- 2(2, 0)& 218.760  &   68.1    &   0.80 (0.03) & 8.11 (0.06) & 3.57 (0.14) & 3.06 (0.11) \\
OCS             & (18 - 17)         & 218.903  &  99.8     &   0.21 (0.02) & 8.84 (0.33) & 6.29 (0.89) & 1.43 (0.18) \\
C$^{18}$O       &  (2 - 1)          & 219.560  &  15.8     &   2.59 (0.05) & 8.03 (0.18) & 2.38 (0.04) & 6.57 (0.11) \\
SO              &  6(5)-5(4)        & 219.949  &  35.0     &   1.32 (0.05) & 8.16 (0.05) & 3.04 (0.13) & 4.28 (0.17) \\
CH$_3$CN        & (12$_3$-11$_3$)    & 220.709  &  133.3   &   0.10 (0.01) & 5.77 (0.51) & 2.76 (1.08) & 0.28 (0.11)\\
CH$_3$CN        & (12$_2$-11$_2$)    & 220.730  &  97.4    &   0.14 (0.01) & 5.32 (0.4) & 2.46 (0.80) & 0.37 (0.10)\\
CH$_3$CN $^d$  & (12$_1$-11$_1$)    & 220.743  &  76.0    &   0.21 (0.01) & 7.69 (0.28) & 1.61 (0.53) & 0.37 (0.10) \\
CH$_3$CN $^d$   & (12$_0$-11$_0$)    & 220.747  &  68.8    &   0.16 (0.01) & 7.34 (0.29) & 2.34 (0.59) & 0.40 (0.10) \\ 
CH$_3$OH        & 8(0 , 8)- 7(1 , 6) E & 220.078  &   96.6 &   0.21 (0.03)& 7.96 (0.30) & 4.70 (0.73) & 1.07 (0.15) \\
CH$_3$OH        & 8($-$1, 8)- 7(0 , 7) E  & 229.759  & 89.1 &   0.36 (0.03) & 8.29 (0.15) & 4.06 (0.37) & 1.54 (0.13) \\
CH$_3$OH        &  3($-$2, 2)- 4($-$1, 4) E & 230.027  &39.8 &   0.22 (0.03) & 8.32 (0.28) & 4.35 (0.68) & 1.02 (0.15) \\
OCS             & (19 - 18)             & 231.061  &  110.8&   0.27 (0.03) & 8.42 (0.23) & 4.07 (0.55) & 1.19 (0.15) \\
$^{13}$CS       &  (5 - 4)           & 231.221  &  33.29   &   0.26 (0.04) & 7.61 (0.30) & 4.28 (0.74) & 1.16 (0.19) \\
\hline
                &             &          &                 &   C-MM5   &   &   &    \\
\hline
DCN             &   (3 - 2)         & 217.239  &  20.9     &   0.13 (0.02) & 9.54 (0.22) & 2.09 (0.56) & 0.28 (0.10) \\
HC$_3$N         & (24 - 23)         & 218.324  &   131.0   &   0.25 (0.02) & 9.51 (0.08) & 2.65 (0.28) & 0.70 (0.06) \\
SO $^b$         &  6(5)-5(4)        & 219.949  &  35.0     &   0.55 (0.05) & 7.22 (0.70) & 6.13 (1.73) & 3.57 (1.06) \\
SO $^b$         &  6(5)-5(4)        & 219.949  &  35.0     &   0.33 (0.01) & 13.2 (0.08) & 3.89 (0.21) & 1.35 (0.09) \\
CH$_3$CN        & (12$_4$-11$_4$)    & 220.679 &   183.3   &   0.12 (0.03)& 8.82 (0.30)& 2.68 (0.69) & 0.34 (0.11) \\
CH$_3$CN        & (12$_3$-11$_3$)    & 220.709  &  133.3   &   0.10 (0.02)& 5.59 (0.46) & 2.83 (1.20) & 0.30 (0.10) \\
CH$_3$CN        & (12$_2$-11$_2$)    & 220.730  &  97.4    &   0.11 (0.02)& 7.26 (0.48) & 4.05 (0.98) & 0.49 (0.11) \\
CH$_3$CN $^d$       & (12$_1$-11$_1$)    & 220.743  &  76.0    &   0.12 (0.02)& 7.51 (0.64) & 2.78 (1.26) & 0.35 (0.18) \\
CH$_3$CN $^d$    & (12$_0$-11$_0$)    & 220.747  &  68.8    &   0.20 (0.02)& 8.51 (0.27) & 3.61 (0.50) & 0.78 (0.12) \\

\hline
                &             &          &                 &   C-MM10   &   &   &    \\
\hline
DCN             &   (3 - 2)         & 217.239  &  20.9     &   0.38 (0.05) & 7.80 (0.15) & 2.07 (0.33) & 0.84 (0.17) \\
H$_2$CO         & 3(0, 3) - 2( 0, 2)& 218.222  &   21.0    &   1.09 (0.11) & 7.71 (0.09) & 1.59 (0.19) & 1.85 (0.30) \\
C$^{18}$O       &  (2 - 1)          & 219.560  &  15.8     &   0.25 (0.04) & 7.83 (0.26) & 3.56 (0.61) & 0.96 (0.22)\\
\hline
                &             &          &                 &   C-MM12   &   &   &    \\
\hline
SiO $^c$        &   (5 - 4)   & 217.105  &  31.3           &   0.65 (0.05) &  10.93 (0.16) & 4.36 (0.47) & 3.03 (0.28) \\
{\it c}-C$_3$H$_2$& 6$_{16}$ - 5$_{05}$& 217.822&38.6 &   0.31 (0.06) &  9.34 (0.20)  & 1.86 (0.38) & 0.61 (0.12) \\
H$_2$CO         & 3(0, 3)- 2(0, 2) & 218.222  &   21.0    &   0.57 (0.09) & 11.18 (0.28) & 3.55 (0.66) & 2.17 (0.54) \\
H$_2$CO         & 3(2, 1)- 2(2, 0) & 218.760  &   68.1    &   0.20 (0.02) & 10.86 (0.14) & 2.97 (0.33) & 0.64 (0.10) \\
OCS             & (18 - 17)         & 218.903  &  99.8     &   0.22 (0.02) & 8.49 (0.18) & 3.96 (0.42) & 0.94 (0.13) \\
SO $^b$         &  6(5)-5(4)        & 219.949  &  35.0     &   0.47 (0.06) & 10.85 (0.29) & 4.51 (0.70) & 2.25 (0.46) \\
SO $^b$         &  6(5)-5(4)        & 219.949  &  35.0     &   0.25 (0.03) & 28.90 (0.28) & 5.85 (0.67) & 1.62 (0.24) \\
\hline
\end{tabular} 
\end{center}
\end{small}
\end{minipage}
\end{table*}

\begin{table*}
\begin{minipage}{180mm}
\begin{small}
\begin{center}
\contcaption{}
\centering
\begin{tabular}{l c c c c c c c}
\hline\hline
 Species        & Transition  & Frequency& E$_{upper}$ & \multicolumn{4}{c}{Fitted Line Parameters}\\[0.5ex]
\cline{5-8}

                &             &          &    &   Intensity$^a$  &  V$_{centre}$$^a$  &  Width  $^a$    &   $\int$ S $dv$ $^a$   \\
                &             &  (GHz)  & (K) & (Jy beam$^{-1}$) & (km\,s$^{-1}$)&  (km\,s$^{-1}$) & (Jy beam$^{-1}$\,km\,s$^{-1}$)   \\

\hline
                &             &          &                 &   SMA1   &   &   &    \\
\hline
C$^{18}$O       &  (2 - 1)          & 219.560  &  15.8     &   0.67 (0.08) & 9.92 (0.11) & 2.19 (0.29) & 1.59 (0.28) \\
\hline
                &             &          &                 &   SMA2   &   &   &    \\
\hline
DCN             &   (3 - 2)         & 217.239  &  20.9     &   0.36 (0.04) & 9.95 (0.18) & 3.23 (0.42) & 1.23 (0.21) \\
{\it c}-C$_3$H$_2$ $^e$ & 6$_{16}$ - 5$_{05}$& 217.822&38.6 &   0.27 (0.03) & 9.42 (0.13) & 2.78 (0.31) & 0.81 (0.12) \\
{\it c}-C$_3$H$_2$&  5$_{14}$ - 5$_{23}$  & 217.940& 35.4  &   0.25 (0.03)  & 10.38 (0.15) & 2.54 (0.34) & 0.67 (0.12) \\
H$_2$CO         & 3(0, 3) - 2( 0, 2)& 218.222  &   21.0    &   0.56 (0.04) & 9.72 (0.10) & 3.31 (0.25) & 2.00 (0.20) \\
H$_2$CO         &3(2, 1)- 2(2, 0)   & 218.760  &   68.1   &    0.19 (0.03) & 9.88 (0.23) & 2.89 (0.53) & 0.58 (0.14) \\
C$^{18}$O       &  (2 - 1)          & 219.560  &  15.8     &   1.73 (0.06) & 9.68 (0.05) & 2.70 (0.11) & 4.97 (0.27) \\
SO              &  6(5)-5(4)        & 219.949  &  35.0     &   0.80 (0.02) & 9.52 (0.02) & 2.16 (0.06) & 1.82 (0.06) \\
$^{13}$CS       &  (5 - 4)           & 231.221  &  33.29   &   0.39 (0.05) & 9.52 (0.12) & 2.06 (0.28) & 0.85 (0.15) \\
\hline
                &             &          &                 &   SMA3   &   &   &    \\
\hline
H$_2$CO         & 3(0, 3)- 2(0, 2)& 218.222  &   21.0    &   2.48 (0.08) & 8.96 (0.08) & 4.59 (0.18) & 12.08 (0.63) \\
CH$_3$OH        & 4(2 , 2)- 3(1 , 2) & 218.440  &   45.5    &   1.08 (0.07) & 8.55 (0.08) & 2.56 (0.20) & 2.94 (0.30) \\
H$_2$CO         & 3(2, 2)- 2(2, 1)& 218.476  &   68.1    &   1.30 (0.04) & 8.83 (0.05) & 3.47 (0.12) & 4.80 (0.21) \\
H$_2$CO         & 3(2, 1)- 2(2, 0)& 218.760  &   68.1    &   1.20 (0.04) & 8.81 (0.05) & 3.56 (0.12) & 4.58 (0.12) \\
C$^{18}$O       &  (2 - 1)          & 219.560  &  15.8     &   0.39 (0.04) & 8.83 (0.23) & 4.84 (0.54) & 2.01 (0.30) \\
SO  $^e$            &  6(5)-5(4)        & 219.949  &  35.0     &   0.43 (0.06) & 9.93 (0.41) & 6.26 (0.96) & 2.84 (0.57) \\
CH$_3$OH        & 8($-$1, 8) - 7(0, 7) E  & 229.759  & 89.1 &   0.97 (0.04) & 8.60 (0.07) & 3.40 (0.17) & 3.52 (0.23) \\
\hline
                &             &          &                 &   C-MM5b   &   &   &    \\
\hline
SiO             &   (5 - 4)   & 217.105  &  31.3           &   0.34 (0.03) & 14.61 (0.16) & 4.15 (0.37) & 1.49 (0.17) \\
DCN             &   (3 - 2)         & 217.239  &  20.9     &   0.94 (0.05) & 9.31 (0.07) & 2.52 (0.17) & 2.51 (0.22) \\
H$_2$CO         & 3(0, 3) - 2(0, 2)& 218.222  &   21.0    &   0.68 (0.06) & 8.93 (0.09) & 2.22 (0.22) & 1.60 (0.21) \\
HC$_3$N         & (24 - 23)         & 218.324  &   131.0   &   0.44 (0.03) & 8.97 (0.07) & 2.11 (0.17) & 0.99 (0.11) \\
H$_2$CO         & 3(2, 1) - 2(2, 0)& 218.760  &   68.1    &   0.12 (0.02) & 9.63 (0.22) & 4.11 (0.52) & 0.54 (0.09) \\
C$^{18}$O       &  (2 - 1)          & 219.560  &  15.8     &   1.54 (0.08) & 9.28 (0.06) & 2.39 (0.13) & 3.92 (0.29) \\
SO $^b$         &  6(5)-5(4)        & 219.949  &  35.0     &   0.39 (0.07) & 9.83 (0.17) & 2.18 (0.47) & 0.90 (0.24) \\
SO $^b$         &  6(5)-5(4)        & 219.949  &  35.0     &   0.14 (0.03) & 15.32 (0.21) & 3.12 (0.51) & 0.48 (0.10) \\
\hline
\end{tabular}
\end{center}
\end{small}
{\bf Notes.}\\
$^a$ {Formal errors from the single Gaussian fits are given in the brackets.}\\
$^b$ {Two distinct velocity components are present in the spectrum; a single Gaussian fit is reported for each component.}\\
$^c$ {Two velocity components appear to be present in the spectrum, but are not sufficiently well-separated in velocity to be fit separately.  The reported fit is for the stronger component.}\\
$^d$ {The k=0 and k=1 CH$_3$CN components are blended in the spectra.  The reported parameters are from multiple-Gaussian fits using the GILDAS CLASS package.}\\
$^e$ {Complex line profile, not well-fit by a single Gaussian.}\\

\end{minipage} 
\end{table*}

\begin{table}
\begin{small} 
\begin{center}
\caption{Systemic Velocity Estimates for 1.3\,mm Continuum Sources}
\centering
\begin{tabular}{c c c c c}
\hline
\hline 
    Source    &  \multicolumn{4}{c}{$v_{\rm LSR}$\,(km\,s$^{-1}$)}\\
\cline{2-5}
	      &    Compact$^a$  &     Diffuse$^b$ &    N$_2$H$^+$$^c$ & Adopted$^d$\\
\hline
C-MM3         &     --      &      --           &     7.1        &  7.1 \\
C-MM4         &  8.2 (0.5)  &   8.3 (0.4)       &     8.9        &  8.3\\
C-MM5         &  8.5 (1.0)  &   9.5             &     --         &  8.9\\
C-MM10        &  --         &   7.8 (0.02)      &     --         &  8.9\\
C-MM12        &   8.5       &   9.3             &     --       &  8.9\\
C-MM13        &   --        &   --              &     8.2      & 8.2\\
SMA1          &   --        &   9.9            &     --       & 8.9\\
SMA2          &   --        &   9.8 (0.4)       &     --       & 8.9\\
SMA3          &   --        &   8.8             &     --       &8.9\\
SMA4          &   --        &   --              &     --       &8.9\\
C-MM5b        &   9.0       &   9.3 (0.02)      &     --       &8.9 \\
\hline
\end{tabular}
\label{table:vlsr}
\end{center}
\end{small}

{\bf Notes.}\\
$^a$ {Average v$_{\rm LSR}$ estimates from molecular lines categorised as exhibiting ``compact'' emission (see Section \ref{section:results_lines}) and includes emission from CH$_3$CN (note only CH$_3$CN transitions with $>$5$\sigma$ detections are included), OCS, HC$_3$N, and CH$_3$OH 5(1 , 4)- 4(2 , 2), CH$_3$OH 8(0 , 8)- 7(1 , 6) E and CH$_3$OH 3( -2, 2)- 4( -1, 4) E. Note not all of these lines are detected at $>$5$\sigma$ towards each core; see Table 2. Furthermore, $^{13}$CS is not used in the v$_{\rm LSR}$ estimate for ``compact'' emission as it also displays diffuse emission. The standard deviation is given in the brackets. If no value is given, emission was detected only in one transition.}\\
$^b$ {Average v$_{\rm LSR}$ estimates from  molecular lines categorised as exhibiting ``diffuse'' emission (see Section \ref{section:results_lines}) and includes emission from C$^{18}$O, DCN and c-C$_3$H$_2$. Again $^{13}$CS is not used in the v$_{\rm LSR}$ estimate for ``diffuse'' tracers as it also appears to trace compact emission. Note not all of these lines are detected towards each core; see Table 2. The standard deviation is given in the brackets. If no value is given, emission was detected only in one transition.}\\
$^c$ {Taken from \citet{Peretto2007}}\\
$^d$ {v$_{\rm LSR}$ adopted for each core for the remainder of the analysis}\\
\end{table}

Towards C-MM3, C-MM13, SMA1 and SMA4 no emission is detected from
lines that exhibit either ``compact'' or ``diffuse'' morphology.  As a
result, we cannot estimate the $v_{\rm LSR}$'s of these sources from
our SMA data. $v_{\rm LSR}$ estimates for C-MM3 and C-MM13 are of
particular interest, given the outflow activity near C-MM3
(Section~\ref{sec:outflows}) and the velocity gradient reported by
\citet{Peretto2007} in NGC 2264-C (spanning C-MM2, C-MM3, C-MM13, and
C-MM4).  Since we are unable to estimate $v_{\rm
  LSR}$'s for several continuum sources from our SMA data (and
have robust estimates based on compact molecular line emission for
only a small minority), we adopt the following approach.  For C-MM4, where we have the most detected compact emission,
we use our estimate of the $v_{\rm LSR}$ of 8.3\,km\,s$^{-1}$ for the
remaining analysis.  For C-MM3
and C-MM13, we adopt the $v_{\rm LSR}$'s reported by
\citet{Peretto2007} based on N$_2$H$^{+}$(1-0) emission
(Table~\ref{table:vlsr}).  
For the remaining millimetre continuum sources, we adopt the average
$v_{\rm LSR}$ from our SMA data (averaged across all cores, for both
compact and diffuse tracers) of 8.9\,km\,s$^{-1}$.  
We note that for C-MM10 and C-MM5b, we have multiple, consistent
velocity measurements from our SMA data (though for C-MM10 only from
``diffuse'' tracers); however, as these cores show no evidence of
outflow activity, we adopt the three $v_{\rm LSR}$'s described above
for simplicity in plotting composite maps of red and blueshifted
emission (e.g. Figure~\ref{image:redblue}).
The $v_{\rm LSR}$ adopted for each
continuum peak, and used throughout the remainder of this paper, is
given in Table \ref{table:vlsr}.


\subsubsection{Extended and High Velocity Molecular Line Emission: Molecular Outflows\label{sec:outflows}}

\begin{figure*}
\includegraphics[width=0.95\textwidth]{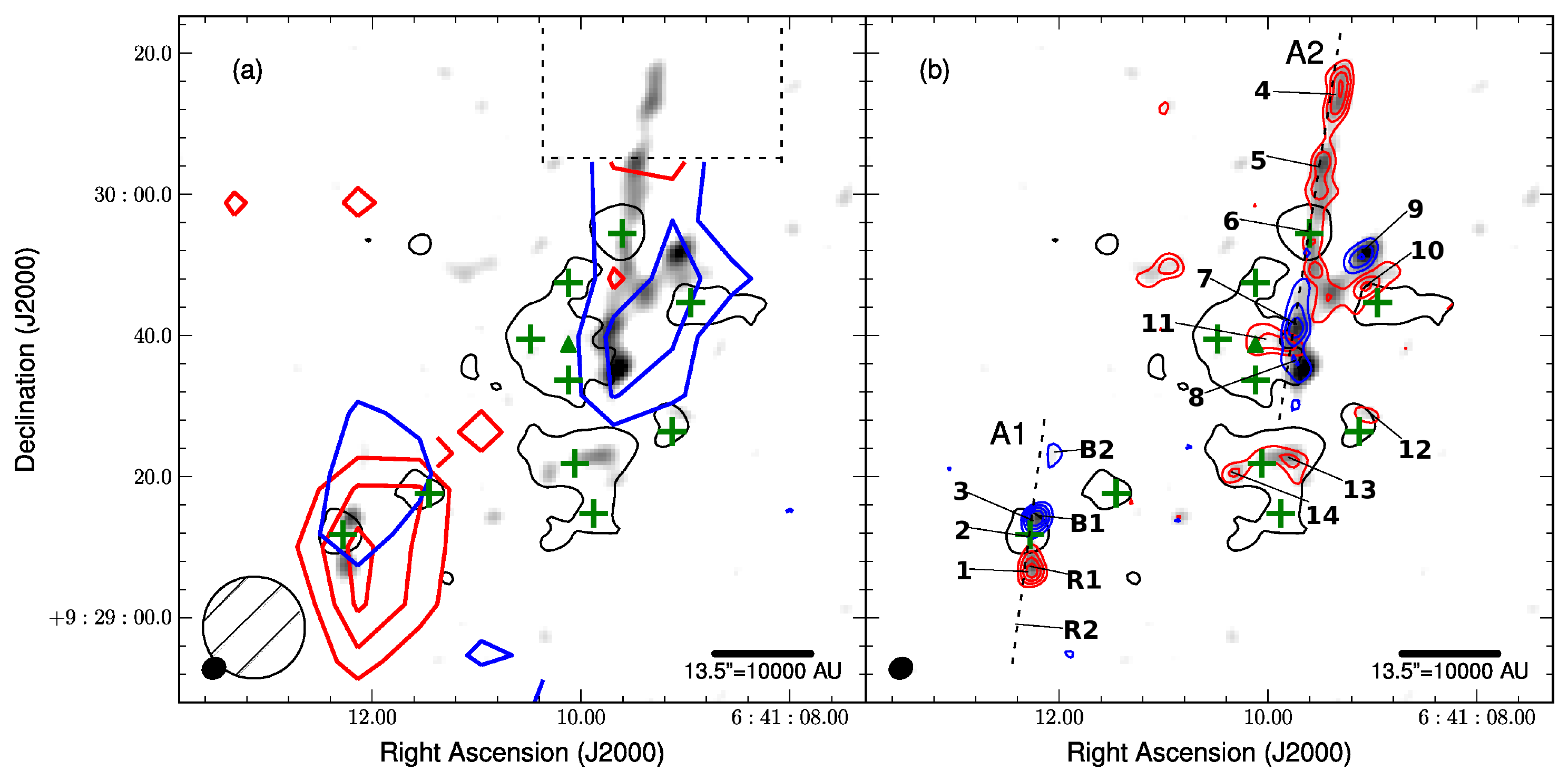}
\caption{SiO kinematics. SMA SiO integrated intensity map (greyscale, integrated over the velocity range $-$20 to 37\,km\,s$^{-1}$) overlaid with contours of 1.3 mm continuum emission (black contour, level 3$\sigma$$=$6\,mJy beam$^{-1}$) and blue/redshifted SiO emission from {\bf (a)} the JCMT (SiO (8-7)) and {\bf (b)} the SMA (SiO (5-4)). The velocity intervals for the blue/redshifted SiO begin 3\,km\,s$^{-1}$ from the $v_{\rm LSR}$ ($v_{\rm LSR}$=7.1 km\,s$^{-1}$ for C-MM3, 8.3\,km\,s$^{-1}$ for C-MM4 and 8.9 km\,s$^{-1}$ for C-MM12, see Section~\ref{results:vlsr}) and extend to the maximum outflow velocity.  The velocity intervals are the same for the JCMT and SMA data and are: Blue: $-$20 to $+$4.1\, km\,s$^{-1}$ for C-MM3 (and C-MM13), $-$1.0 to $+$5.3\,km\,s$^{-1}$ for C-MM4, and $-$20 to $+$5.9\,km\,s$^{-1}$ for C-MM12 (and the other millimetre continuum cores); Red: $+$10.1 to $+$37\,km\,s$^{-1}$ (C-MM3 and C-MM13), $+$11.3 to $+$20\,km\,s$^{-1}$ (C-MM4), and $+$11.9 to $+$32\,km\,s$^{-1}$ (C-MM12 and other cores).  In both panels, green pluses (+) mark the positions of the 10 millimetre continuum peaks from the dendrogram analysis and the green triangle marks the position of C-MM5b. {\bf(a)} JCMT SiO\,(8-7) contour levels: (3,5,7,9) $\times$ $\sigma$\,$=$\,0.2\,K km\,s$^{-1}$. The 14.5$\arcsec$ JCMT beam (hatched circle) and the SMA synthesised beam (filled ellipse) are shown at lower left.  The dashed rectangle shows the area of missing data from the HARP receiver element H14.  {\bf (b)} SMA SiO\,(5-4) contour levels: (3,5,7,9) $\times$ $\sigma$\,$=$\,0.36 Jy beam$^{-1}$ km\,s$^{-1}$. The dashed black lines represent the possible outflow axes A1 and A2. The numbered positions 1-14 mark the components discussed in Section~\ref{sec:outflows}, and mark the positions at which the spectra shown in Figure \ref{image:spectra} were extracted. Positions 1-3 are part of the outflow axis A1, positions 4-8 are part of outflow axis A2, and the remaining positions mark locations of ambiguous SiO emission that is not obviously associated with an outflow. The labels R1, R2, B1 and B2 mark the components of the redshifted and blueshifted outflow lobes of C-MM3 named by \citet{Saruwatari2011}. A colour version of this figure is available online.\label{image:sio_mom0} }
\end{figure*}

As shown in Figure~\ref{image:chemistry}, several species are
characterised by collimated, extended emission: SiO, SO, CH$_3$OH,
H$_2$CO, and $^{12}$CO.  The two most prominent collimated emission
features in the integrated intensity maps (both elongated N-S) appear
to be centred on C-MM3 and C-MM12.  To investigate the nature of the
extended molecular line emission, and its relation to outflows
previously reported based on lower-angular-resolution data, we analyse
its velocity structure.  To identify potential outflow axes, we use
our high-resolution SMA observations of the well-known outflow tracer
SiO \citep[e.g.][and references therein]{Gusdorf2008b,Gusdorf2008a,Guillet2009,Leurini2014}.

Figure \ref{image:sio_mom0} presents integrated intensity maps of the red-
and blueshifted SiO emission observed with the SMA (SiO\,(5-4), resolution
$\sim$3$^{''}$) and the JCMT (SiO\,(8-7),
resolution $\sim$15$^{''}$).
We compare the SMA and JCMT data to test whether the spatial filtering
of the SMA misses any large-scale active outflows (which would be seen
in SiO emission with the JCMT).
The JCMT observations (Figure~\ref{image:sio_mom0}(a)) reveal
two potential outflow systems: one encompassing both C-MM3 and
C-MM13 (with the peak emission nearest to C-MM3), and a second
that intersects both C-MM10 and C-MM12. In the
second system, only blueshifted emission is detected with the JCMT; however,
receiver H14 was not operational during our observations,
resulting in a gap in the JCMT map north of C-MM12.  
Figure \ref{image:sio_mom0}(b) presents the higher spatial
resolution SMA SiO\,(5-4) data.  With the SMA, the SiO emission
near C-MM3 and C-MM13 is clearly resolved, revealing a single
bipolar outflow centred on C-MM3 (outflow axis A1, Figure
\ref{image:sio_mom0}(b)).  The red and blueshifted lobes are
spatially well-separated and are centred on the continuum source,
indicating that the high-velocity SiO\,(5-4) emission traces a bipolar
molecular outflow.
The second potential outflow system
identified in the JCMT data, towards C-MM12 and C-MM10, splits into multiple
components at the higher
spatial resolution of the SMA observations.  As shown in Figure
\ref{image:sio_mom0}(b), the SMA data reveal a collimated, bipolar outflow centred on
C-MM12 (axis A2, Figure~\ref{image:sio_mom0}(b)): as in the C-MM3
outflow, the red and
blueshifted lobes are spatially well-separated and centred on the
millimetre continuum source.  For both outflows, the sense of the
velocity gradient is consistent in the SMA and JCMT observations
(e.g.\ the blueshifted lobe is north of C-MM3, and the redshifted
lobe south of C-MM3, in both the SMA and JCMT maps).
The fact that our SMA SiO data recover (and resolve) both of the SiO
outflows identified with the JCMT is strong evidence that the SMA is
not ``missing'' large-scale active outflows.

In addition to the two clear outflows driven by C-MM3 and
  C-MM12 along axes A1 and A2, there are several additional components of SiO
  emission present in the field, labelled 9-14 in Figure
  \ref{image:sio_mom0}. To the north of C-MM10, red- and
  blueshifted emission are present (components 9 and 10).  Redshifted emission also extends through the blueshifted axis of
  A2 to C-MM5b (component 11). In addition, extended redshifted SiO emission
  is observed towards C-MM4 (components 13 and 14) and redshifted emission
  is found coincident with SMA3 (component 12). The nature of this
  additional SiO emission is unclear.

\begin{figure*}

\includegraphics[width=0.83\textwidth]{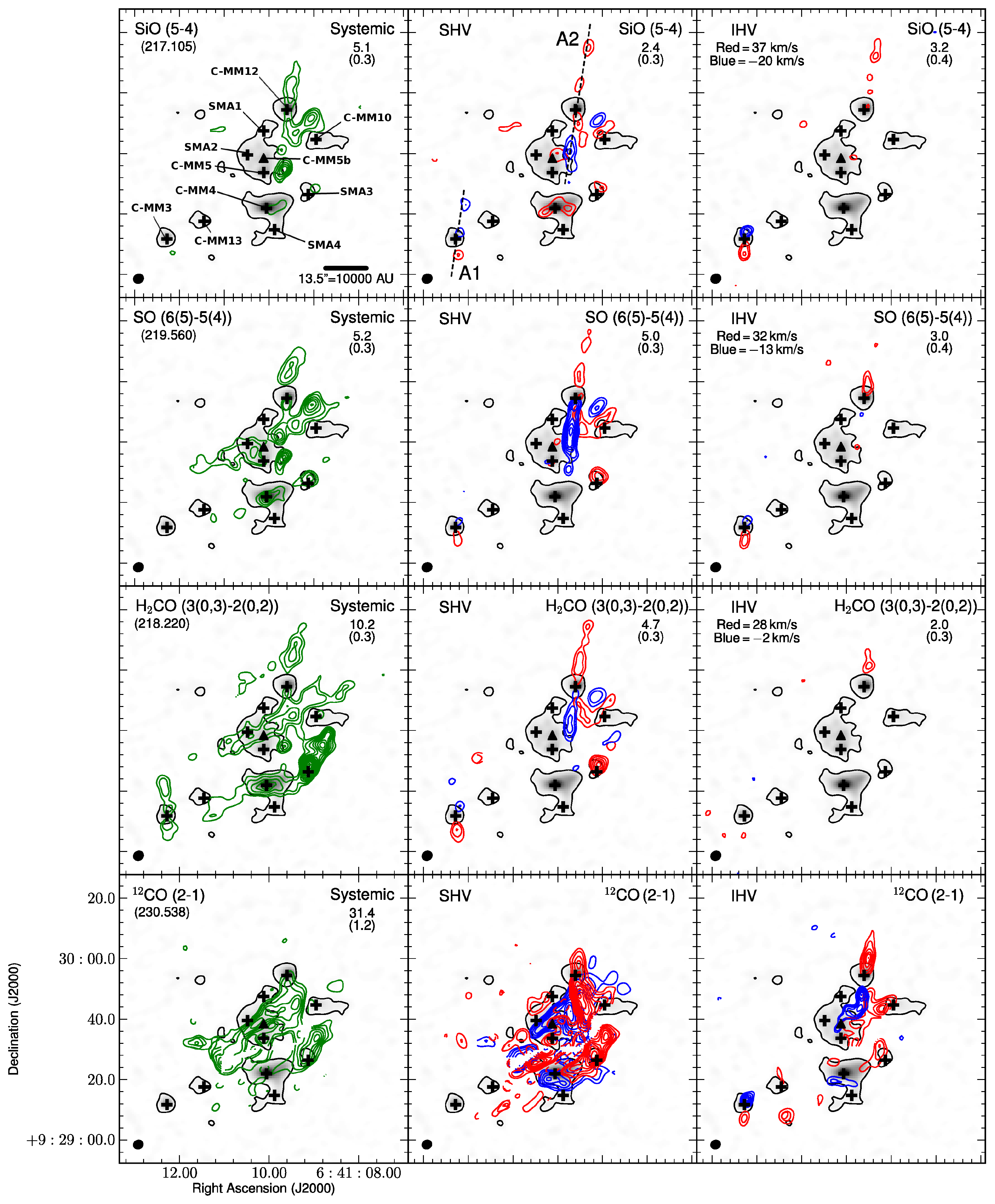}
\caption{Integrated intensity maps (coloured contours) of emission at systemic (left column), SHV (middle column) and IHV (right column) velocities for the indicated transitions, overlaid on the SMA 1.3\,mm continuum image (greyscale and black contour, contour level 3$\sigma$\,$=$\,6\,mJy/beam). ``Systemic'' is defined as $v_{\rm LSR}\pm$3\,km\,s$^{-1}$ (green contours), SHV as $v_{\rm LSR}\pm$3-10\,km\,s$^{-1}$, and IHV as $v_{\rm LSR}\pm$$>$10\,km\,s$^{-1}$ (where red and blue contours represent the red and blueshifted emission respectively). The maps shown are composites, with an adopted $v_{\rm LSR}$ of 7.1\,km\,s$^{-1}$ for C-MM3, 8.3\,km\,s$^{-1}$ for C-MM4 (including the vicinity of C-MM13), and 8.9\,km\,s$^{-1}$ for the remainder of the map (Section~\ref{results:vlsr}). The maximum velocity extent of the emission is given at top left in the IHV panels, except for $^{12}$CO (2-1), where the maximum velocity differs significantly from outflow to outflow across the map (see Section~\ref{sec:outflows}). Contour levels are shown as 3\,$\sigma$, 5\,$\sigma$  and then to peak, increasing in steps of 3\,$\sigma$. The peak and rms (in Jy beam$^{-1}$ km\,s$^{-1}$) are given at upper right in each panel, except for SHV and IHV $^{12}$CO. For $^{12}$CO\,(2-1), the contoured rms levels are: SHV, red: 1.9 Jy\,beam$^{-1}$ km\,s$^{-1}$ (C-MM3/C-MM13), 1.9  Jy\,beam$^{-1}$ km\,s$^{-1}$  (other sources); SHV, blue: 1.4 Jy\,beam$^{-1}$ km\,s$^{-1}$ (C-MM3/C-MM13), 1.9 Jy\,beam$^{-1}$ km\,s$^{-1}$  (other sources); IHV, red: 1.0 Jy\,beam$^{-1}$ km\,s$^{-1}$ (C-MM3/C-MM13), 1.4  Jy\,beam$^{-1}$ km\,s$^{-1}$ (other sources); IHV, blue: 0.4  Jy\,beam$^{-1}$ km\,s$^{-1}$ (C-MM3/C-MM13), 0.6 Jy\,beam$^{-1}$ km\,s$^{-1}$ (other sources). In all panels, black pluses (+) mark the positions of the ten millimetre continuum peaks from the dendrogram analysis, and the black triangle marks the position of C-MM5b. The outflow axes from Figure \ref{image:sio_mom0}b are overlaid for reference in the top middle panel (SHV SiO). The $^{12}$CO emission in the systemic and SHV panels shows a discontinuity in the emission to the north of C-MM4, this is due to the composite of the different velocity ranges used when integrating the emission towards the continuum peaks and given the CO emission is more abundant over the field than the other molecular transitions. A colour version of this figure is available online.\label{image:redblue}}
\end{figure*}

To further explore the nature and structure of the high velocity
outflow emission in NGC 2264-C, we consider four velocity regimes \citep[e.g.][]{Santiago2009}:
systemic ($v_{\rm LSR}\pm$3\,km\,s$^{-1}$), standard high velocity (SHV,
$v_{\rm LSR}\pm$3 to
10\,km\,s$^{-1}$), intermediate high velocity (IHV, $v_{\rm
  LSR}\pm$10-30km\,s$^{-1}$), and extremely high velocity (EHV, $v_{\rm
  LSR}\pm$30-50km\,s$^{-1}$).  
Figure~\ref{image:redblue} presents SMA integrated intensity maps for the
systemic, SHV, and IHV regimes for SiO (5-4), SO (6(5)-5(4)), H$_2$CO (3(0, 3)-
2(0, 2)) and $^{12}$CO (2-1).  These are the only four transitions in
which emission is detected at velocities $>$10 km\,s$^{-1}$ from the
$v_{\rm LSR}$ (e.g.\ in the IHV regime).  We do not present EHV maps
because very little EHV emission is detected: the only $>$3$\sigma$ emission with
$\left| v-v_{\rm LSR} \right| >$30\,km\,s$^{-1}$ is $^{12}$CO\,(2-1) at $\sim$31\,km\,s$^{-1}$, towards the
redshifted lobe of the C-MM3 outflow.
As described in Section~\ref{results:vlsr}, above, we adopt a $v_{\rm LSR}$ of 7.1 km
s$^{-1}$ for C-MM3 and a $v_{\rm LSR}$ of 8.3\,km\,s$^{-1}$ for the region covering C-MM4 (including C-MM13).  For the rest of the region, the velocity
ranges are calculated with respect to a $v_{\rm LSR}=$8.9\,kms$^{-1}$ (adopted for the other continuum cores, Section~\ref{results:vlsr}). 

Towards the outflow axis A2, low-velocity (systemic) emission is traced by SiO, SO, and H$_2$CO (as shown in Figure~\ref{image:redblue}). In contrast, along axis A1 only H$_2$CO is detected at low velocities. In addition, two CH$_3$OH lines (CH$_3$OH 4(2,2)$-$3(1,2) and CH$_3$OH 8(-1,8)-7(0,7)E) exhibit extended emission along the outflow axes (see Figure~\ref{image:chemistry}) only in the systemic and SHV regimes, and so are not shown in Figure~\ref{image:redblue}. There is also a noticeable extension in the systemic emission from SO and H$_2$CO running from north west to south east coincident with the components 9-11.

In the SHV regime, collimated red- and blueshifted emission centred
on C-MM3 and C-MM12 is detected in SiO, SO, and H$_2$CO. 
The IHV $^{12}$CO emission displays a similar morphology, consistent
with the SHV and IHV molecular line emission tracing bipolar molecular
outflows driven by these two millimetre continuum cores.  (The
$^{12}$CO emission in the SHV regime is affected by poorly-imaged
extended structure and confusion with the surrounding cloud, making it
difficult to identify outflows in $^{12}$CO in this velocity range). 
Both lobes of the outflow associated with C-MM3 are also detected in
the IHV regime in SiO and SO, while only the redshifted lobe of the
outflow associated with C-MM12 is detected in IHV SiO, SO, and H$_2$CO
emission (Figure~\ref{image:redblue}).

The additional SiO components (numbered 9-14 in Figure~\ref{image:sio_mom0}) are also associated with SHV SO and
  H$_2$CO emission,  with the exception of the redshifted emission towards C-MM4.  Redshifted SHV SiO emission is observed towards both C-MM4 and SMA3 (components 12-14), while (redshifted) SO and H$_2$CO emission are only observed towards SMA3 (component 12).  In the IHV regime, only $^{12}$CO emission is detected towards these components.  While both redshifted (near SMA3) and blueshifted (south of C-MM4) emission are present, these potential lobes are not centred on a millimetre continuum source and it is unclear if they are related to a single outflow.
Compared to the two bipolar outflows (along A1 and A2), the velocity extent of the emission towards the additional, ambiguous SiO components is also more modest. For $^{12}$CO\,(2-1), the maximum redshifted velocity is $\sim$38\,km\,s$^{-1}$, 32\,km\,s$^{-1}$, 28\,km\,s$^{-1}$ and 24\,km\,s$^{-1}$ for A1 (C-MM3 outflow, components 1-3), A2 (C-MM12 outflow, components 4-8), towards C-MM10/C-MM5b (components 9-11) and towards C-MM4/SMA3 (components 12-14), respectively. The minimum velocity for blueshifted $^{12}$CO emission is $-$17.2\,km\,s$^{-1}$ for A1 (C-MM3 outflow) and $-$2.8\,km\,s$^{-1}$ for A2 (C-MM12 outflow) and towards C-MM4. 

\begin{figure*}
\centering
\includegraphics[width=0.485\textwidth]{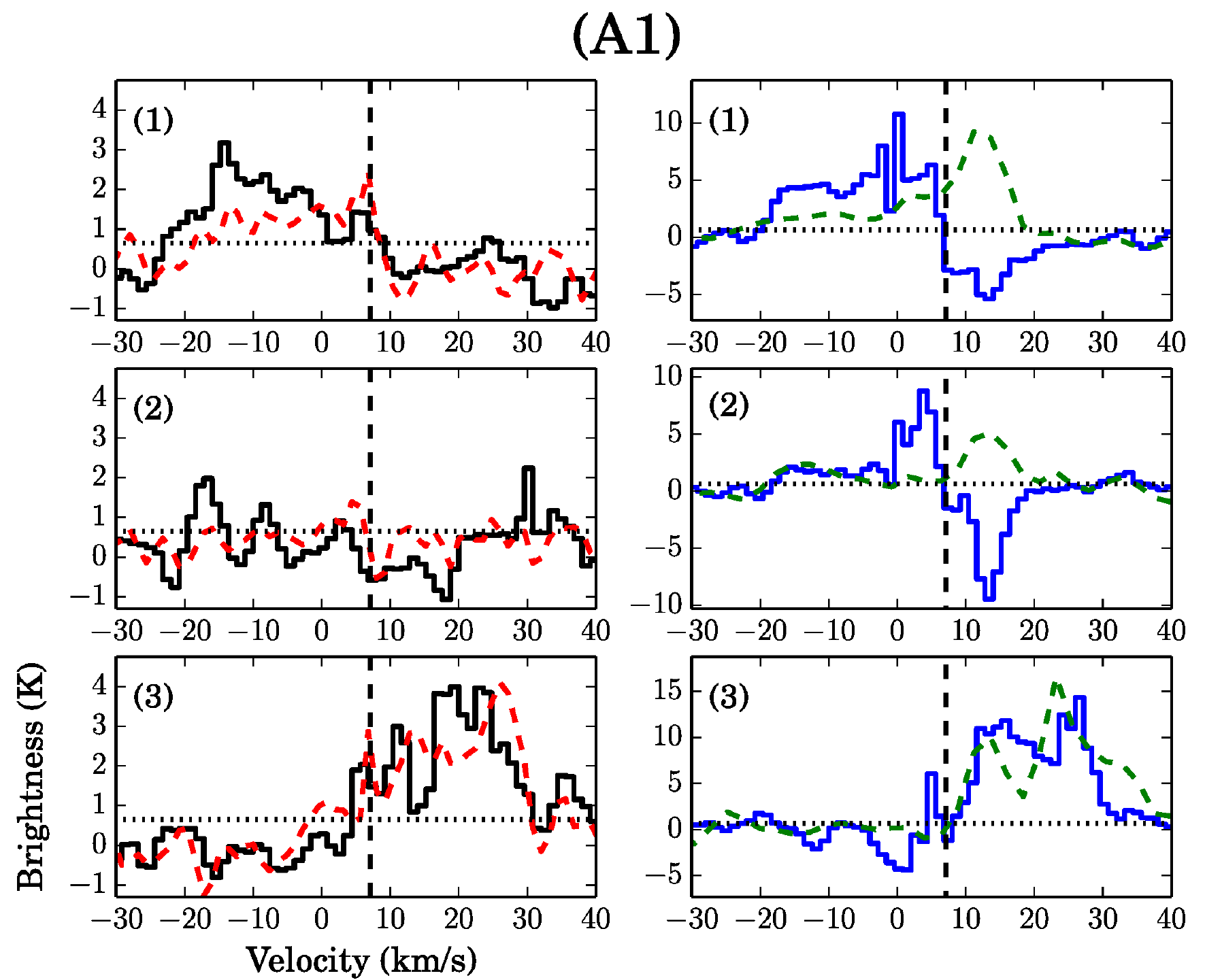}
\includegraphics[width=0.485\textwidth]{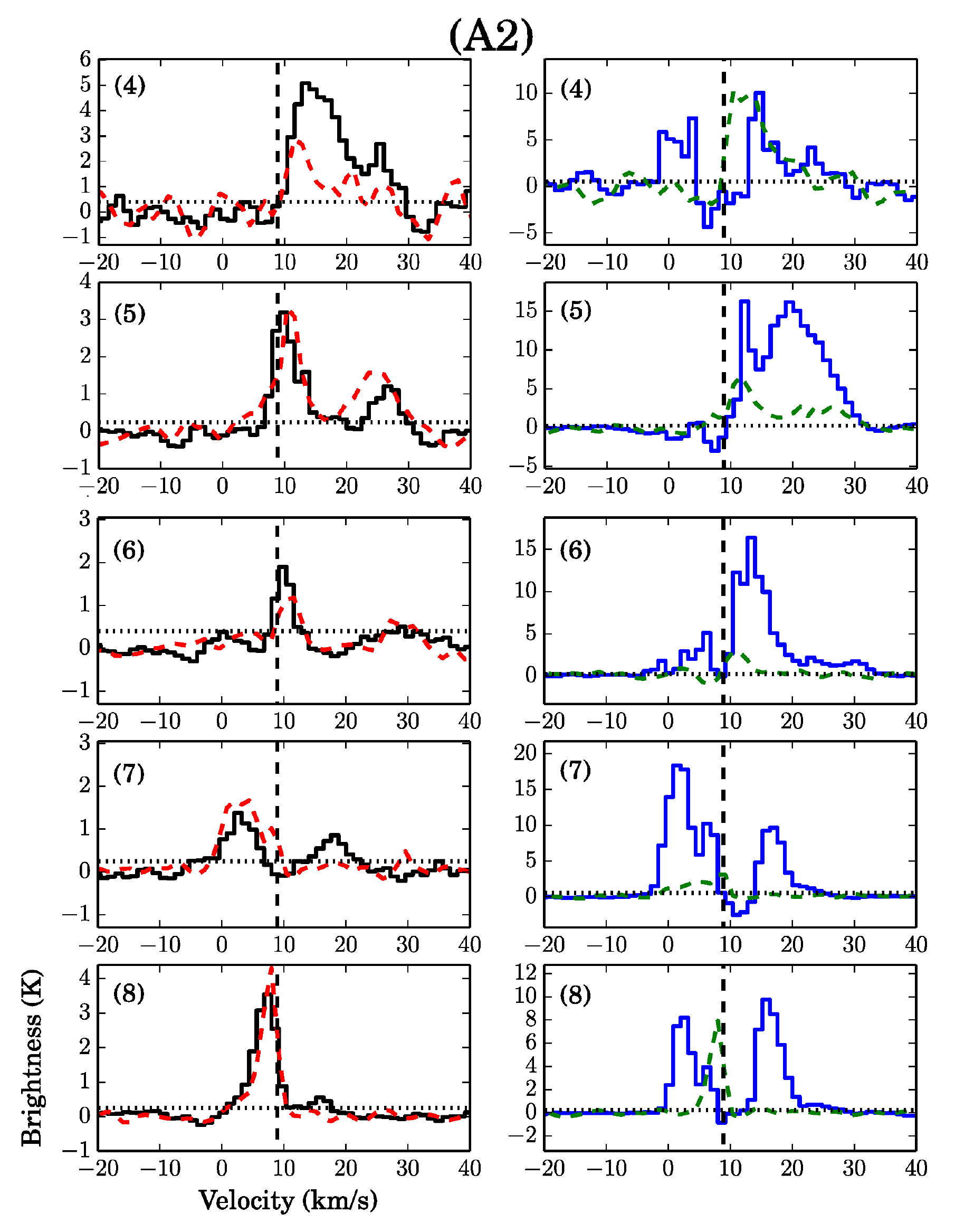}
\includegraphics[width=0.485\textwidth]{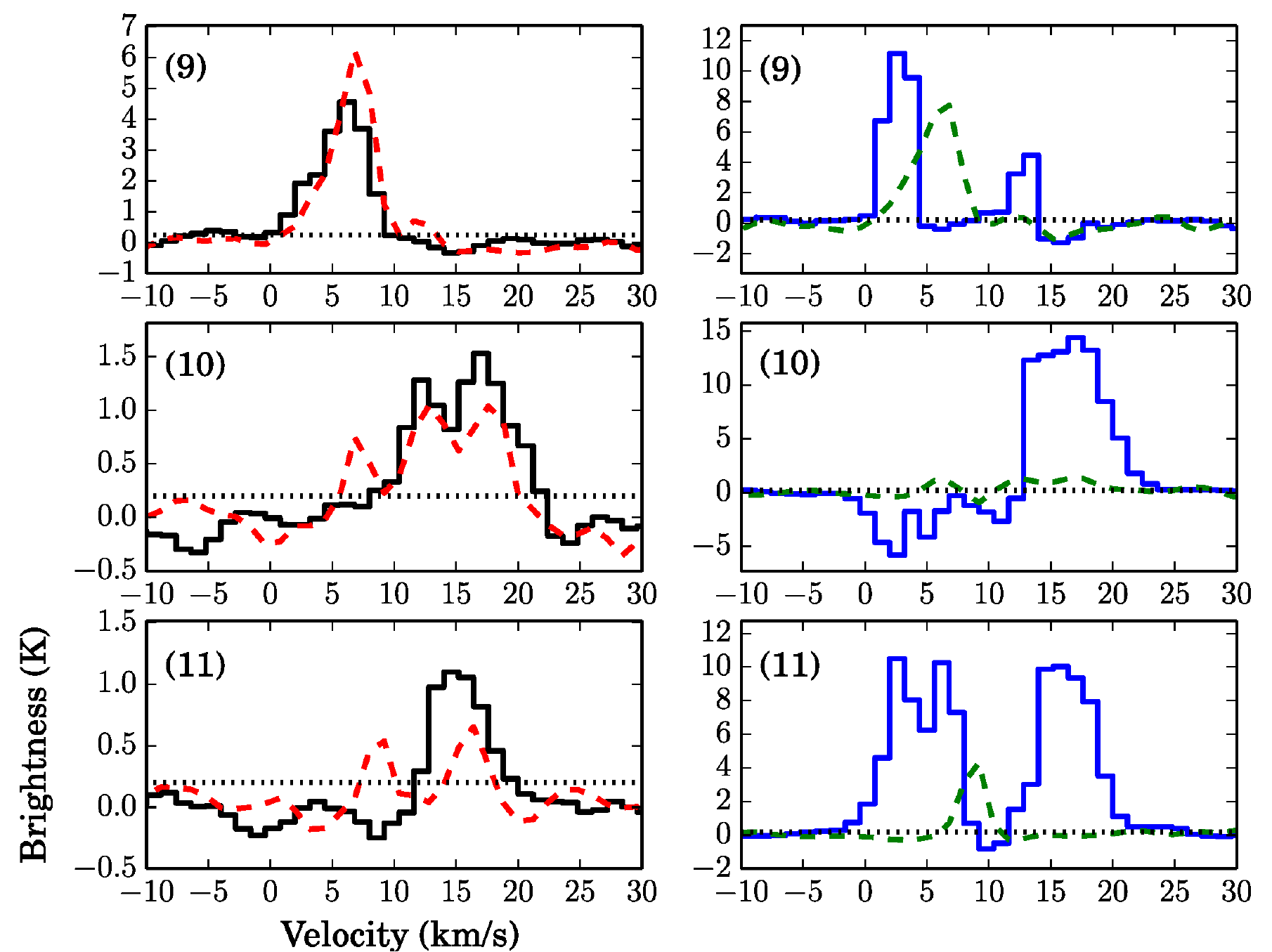}
\includegraphics[width=0.485\textwidth]{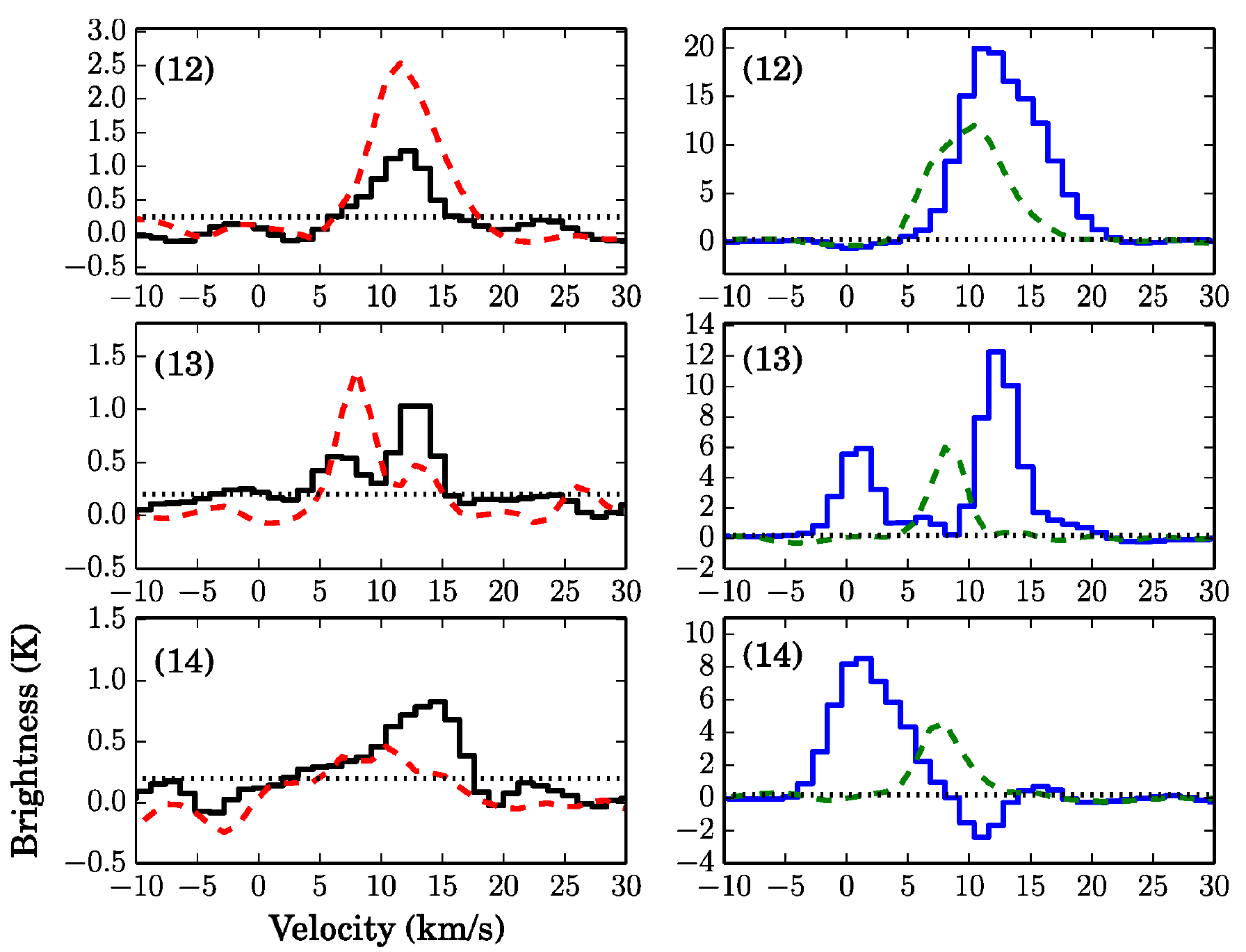}
\caption{Spectra of SiO (5-4), SO 6(5)-5(4), H$_2$CO 3(0, 3)- 2(0, 2) and $^{12}$CO (2-1) extracted from a single pixel at the positions/components labelled in Figure \ref{image:sio_mom0}.  For each position, the left panel shows SiO (5-4) (solid black line) and SO 6(5)-5(4) (dashed red line) and the right panel shows $^{12}$CO (2-1) (solid blue line) and  H$_2$CO 3(0, 3)- 2(0, 2) (dashed green line). Panels labelled 1-3 represent emission extracted from positions along A1 and the vertical dashed black line represents the $v_{\rm LSR}$ of C-MM3 at 7.1\,kms$^{-1}$. Panels labelled 4-8 represent emission extracted from positions along A2 and the vertical black dashed line represents the  $v_{\rm LSR}$ of C-MM12 of 8.9\,kms$^{-1}$. Panels with a number 9 or greater represent emission extracted from positions where SiO emission is present but not obviously associated to an outflow. The H$_2$CO 3(0, 3)- 2(0, 2) emission has been scaled up by a factor of two in all spectra. The horizontal black dashed lines are the rms values from the line free chanels.  \label{image:spectra}}
\end{figure*}

To examine the outflow kinematics in greater detail, Figure \ref{image:spectra} presents SiO, SO, H$_2$CO and $^{12}$CO spectra at the positions labelled in Figure \ref{image:sio_mom0}, along the potential outflow axes A1 and A2 (numbered 1-8), and towards the additional components (numbered 9-14). Works by \citet{Codella2014}, \citet{Tafalla2010}, and \citet{CFLee2010} identified similarities between SiO and SO emission at high velocities. As shown in Figure \ref{image:redblue}, SiO and SO exhibit similar morphologies in NGC 2264-C; however, the SiO emission extends to higher absolute velocities compared with the SO emission (e.g. for SiO the maximum red- and blueshifted velocities are 37\,km\,s$^{-1}$ and $-$20\,km\,s$^{-1}$, respectively, compared with 32\,km\,s$^{-1}$ and $-$13\,km\,s$^{-1}$ for SO).
Close examination of the line profiles also reveals a velocity gradient in $^{12}$CO towards C-MM4 and SMA3, from redshifted (component 12) to blueshifted (component 14).  This $^{12}$CO emission is lower-velocity than observed towards axes A1 and A2.  The SiO emission is also considerably narrower towards C-MM4 and SMA3 than along axes A1 and A2.

We note that SiO\,(5-4) emission associated with outflows from low-mass (proto)stars is unlikely to be detected as extended, collimated structure in our SMA observations. Scaling the results of \citet{GomezRuiz2013} and \citet{Codella2014} to the distance of NGC 2264-C and our SMA beam, we would detect only the strongest SiO\,(5-4) emission, and that at the $\sim$4-8\,$\sigma$ level. We also note that in both the JCMT and SMA observations, the width of the collimated emission appears to be limited by the size of the beam.

\subsubsection{Candidate Millimetre CH$_3$OH Masers\label{section:maseridge}}

A notable feature of Figure~\ref{image:chemistry} is the very strong 229.759 GHz CH$_{3}$OH emission associated with the ``ridge''. The 229.759\,GHz CH$_3$OH 8(-1, 8)-7(0, 7)E transition is a known Class I methanol maser, first reported towards DR21 (OH) and DR21 West by \citet{Slysh2002} based on observations with the IRAM 30-m telescope. Probable maser emission in this transition is often seen in SMA observations of massive star-forming regions \citep[e.g.][and references therein]{QiuandZhang2009,Fontani2009,Fish2011,Cyganowski2011,Cyganowski2012}. Most recently, \citet{Hunter2014} directly demonstrated the maser nature of 229.759 GHz CH$_{3}$OH emission in NGC6334I(N), by showing that the observed line brightness temperature ($T_{B}$) is greater than the upper energy of the transition ($E_{\rm upper}$) in very high-resolution SMA observations.

\begin{figure*}
\includegraphics[width=0.95\textwidth]{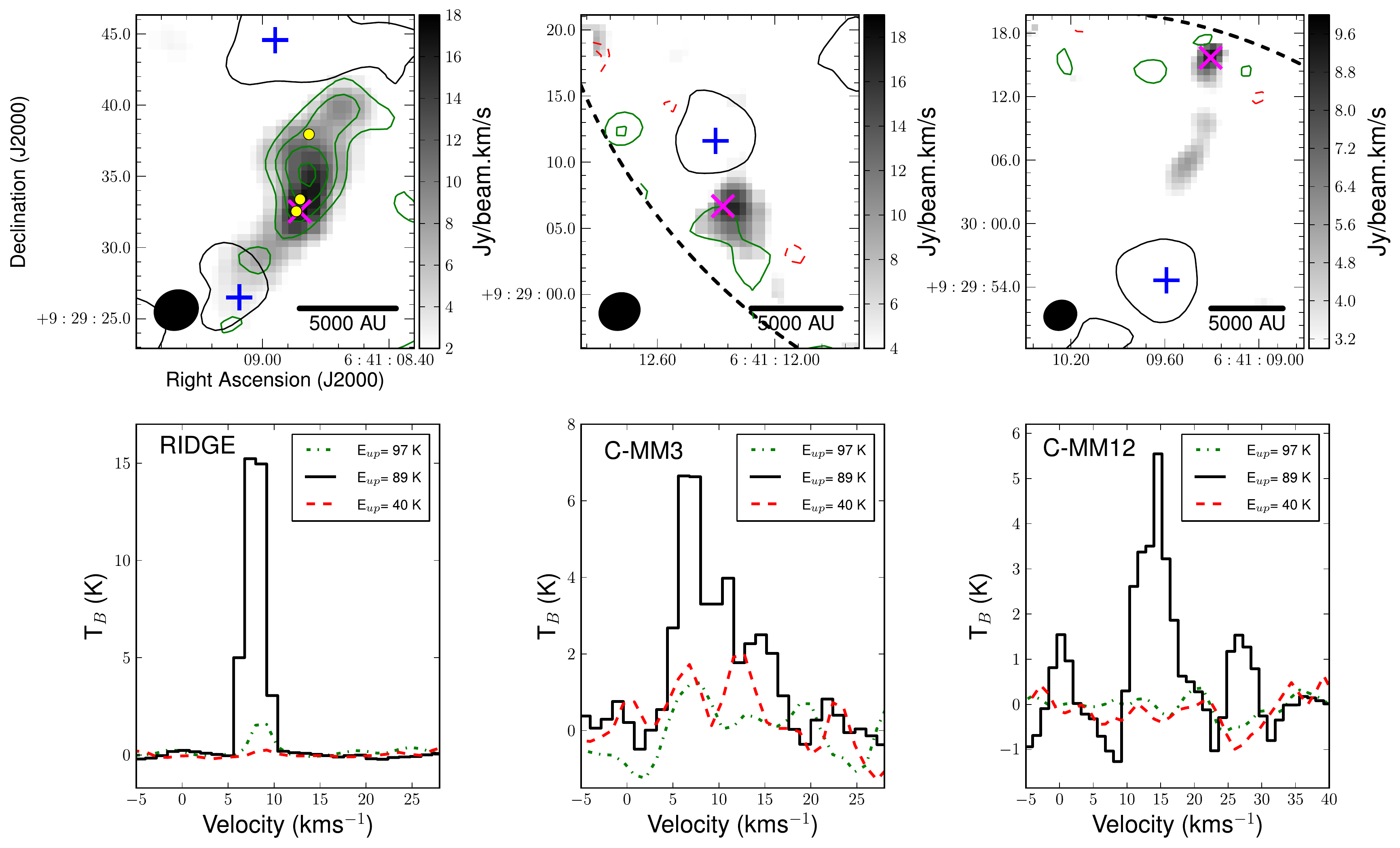}
\caption{Top: Zoom views of integrated intensity maps, corrected for the primary beam response, of CH$_3$OH 8( -1, 8)- 7(0 , 7) E  (229.759\,GHz, greyscale), CH$_3$OH 8( 0, 8)- 7(1, 6) E (220.078\,GHz, solid green contours) and CH$_3$OH 3( -2, 2)- 4( -1, 4) E(230.027\,GHz, dashed red contours) towards the ridge feature and the redshifted outflow lobes from C-MM3 and C-MM12. The integrated velocity ranges are the same as in Figure \ref{image:chemistry}.  Contour levels: CH$_3$OH 8(0, 8)- 7(1 ,6) E 
  (220.078\,GHz): 3, 6, 9, 12\,$\sigma$, for 1\,$\sigma$ values of 0.2, 0.6 and 0.6 Jy beam$^{-1}$ km\,s$^{-1}$ towards the ridge, C-MM3 and C-MM12, respectively.  CH$_3$OH 3( -2, 2)- 4( -1, 4) E (230.027\,GHz): 3, 5\,$\sigma$ for 1\,$\sigma$ values of 0.2, 0.4 and 0.5 Jy beam$^{-1}$ km\,s$^{-1}$  towards the ridge, C-MM3 and C-MM12, respectively. The synthesized beam
 is shown at lower left in each panel. The blue pluses (+) mark the positions of the millimetre continuum peaks; the solid black line is the 3$\sigma$ contour of the 1.3\,mm continuum emission. The thick dashed line in
 the C-MM3 and C-MM12 images is the 10\% level of the primary beam response. The yellow circles show the positions of the three 44\,GHz CH$_3$OH maser spots detected by \citet{Slysh2009}. Bottom: Spectra of the three CH$_{3}$OH transitions. The spectra are extracted at the positions of peak CH$_3$OH 8( -1, 8)- 7(0 , 7) E (229.759\,GHz) emission towards the ridge and the outflows of C-MM3 and C-MM12.  These positions are marked by magenta crosses(x) (see also Table~\ref{table:maser}). \label{image:maser}}
\end{figure*}

To investigate the nature of the 229.759 GHz CH$_3$OH emission in NGC
2264-C, Figure \ref{image:maser} presents integrated intensity maps
and corresponding line profiles for the CH$_3$OH 8( -1, 8)- 7(0 , 7)
E, CH$_3$OH 8(0, 8)- 7(1 , 6) E and CH$_3$OH 3(-2, 2)- 4( -1, 4) E
transitions at three locations where the CH$_3$OH 8( -1, 8)- 7(0 , 7)
E 229.759\,GHz emission is strongest. These are the ridge, 
the redshifted outflow lobe of C-MM3, and the redshifted outflow lobe of C-MM12. As shown in
Figure~\ref{image:maser}, the 229.759\,GHz CH$_3$OH emission in the
ridge is more than twice as strong as that towards the redshifted
outflow lobe of either C-MM3 or C-MM12.

Like most previous 229 GHz studies \citep[with the notable exception of][]{Hunter2014}, our SMA observations do not have sufficient angular resolution to establish masing in the 229.759\,GHz line based on its brightness temperature. \citet{Slysh2002} proposed the ratio of the 229.759\,GHz and 230.027\,GHz CH$_3$OH lines as a diagnostic of maser emission, with values of 229.759/230.027$>$\,3 indicating nonthermal 229.759 GHz emission. We note that the 230.027 GHz line is undetected at all three positions, and the 3$\sigma$ limits are used to calculate the line ratios (see Table \ref{table:maser} for the 3$\sigma$ limits towards each position). At two positions in our field, this line ratio is $\ge$\,3: towards the redshifted lobe of the C-MM3 outflow, where the ratio is $\sim$3, and towards the strongest 229.759\,GHz emission in the ridge, where the ratio is considerably higher ($\sim$ 47). Furthermore, throughout the ridge the line ratio is consistently $>$8. While the line ratios towards the ridge are considerably higher than towards the C-MM3 outflow, the ridge line ratios fall within the range of line ratios, $\sim$7-100, found towards the outflow lobes of two extended green objects by \citet{Cyganowski2011}. By comparison, the line ratio towards the C-MM4 continuum peak is $<$\,2, consistent with thermal emission from warm gas (Section~\ref{dis:temp_est}).

Table~\ref{table:maser} presents fits to the 220.078, 229.759, and 230.027 GHz CH$_{3}$OH lines at the three positions shown in Figure~\ref{image:maser}. Emission from 229 GHz CH$_{3}$OH masers often coincides spatially and spectrally with emission in lower-frequency Class I CH$_{3}$OH maser transitions \citep[e.g.][]{Cyganowski2011,Cyganowski2012,Fish2011}. Class I CH$_3$OH maser emission, in the form of the 44\,GHz transition, was initially observed to the west of IRS1 in the direction of the ridge feature by \citet{Haschick1990}. More recently, \citet{Slysh2009} identified three 44\,GHz maser spots with the VLA (resolution 0.15$\arcsec$) that coincide spatially with the ridge (positions shown in Figure \ref{image:maser}). The strongest of the three 44\,GHz maser spots is coincident with the position of the strongest 229\,GHz candidate maser emission. Furthermore, the 229/230\,GHz line ratio is found to be $>$15 at the positions of all three 44\,GHz maser spots. The $v_{\rm LSR}$ velocity of the 229\,GHz emission at the positions of the three 44 GHz maser spots is 8.53, 8.68, and 8.48\,km\,s$^{-1}$, respectively (for increasing declination in Figure \ref{image:maser}), within $\sim$1\,km\,s$^{-1}$ of the 44\,GHz $v_{\rm LSR}$ measurements from \citet{Slysh2009}, of 7.22, 7.59, and 7.72\,km\,s$^{-1}$. The coincidence of the Class I 44\,GHz CH$_3$OH maser spots with the 229.759\,GHz emission in the ridge, both spatially and spectrally, supports the interpretation that the 229.759\,GHz transition is masing in the ridge.

\begin{table*}
\begin{minipage}{180mm}
\begin{small}
\label{table:maser}
\begin{center}
\caption{Methanol Line Fits for Candidate 229.759 GHz CH$_{3}$OH Masers in NGC 2264-C \label{table:maser}}
\centering
\begin{tabular}{c c c c c c c c c c}
\hline\hline
&&&&&&\multicolumn{4}{c}{Fitted Line Parameters}\\
\cline{7-10}
 Species   & Transition                & $\nu$ & E$_{upper}$ &  RA     &   DEC   &  Intensity$^{a}$     &  V$_{centre}$$^{a}$  &  Width$^{a}$      &   $\int$S$dv$$^{a}$    \\ [0.5ex]
           &                           & (GHz) &    (K)      & (J2000) & (J2000) & (Jy beam$^{-1}$) & (km\,s$^{-1}$)  &  km\,s$^{-1}$ & Jy beam$^{-1}$\,km\,s$^{-1}$   \\
\hline
\multicolumn{10}{c}{Ridge}\\     
\hline
CH$_3$OH &8(0 , 8)- 7(1 , 6) E          & 220.078 & 96.6  &  06 41 08.8 & $+$09 29 32.5  & 0.66 (0.03) & 8.63 (0.04) & 2.29 (0.11) & 1.63 (0.10)\\
CH$_3$OH & 8( -1, 8)- 7(0 , 7) E        & 229.759 & 89.1  &  06 41 08.8 & $+$09 29 32.5  & 6.64 (0.14) & 8.53 (0.03) & 2.45 (0.06) & 17.31 (0.58)\\
CH$_3$OH &  3( -2, 2)- 4( -1, 4) E      & 230.027 & 39.8  &  06 41 08.8 & $+$09 29 32.5  & $<$0.14$^*$ & -- & -- & --\\
\hline
\multicolumn{10}{c}{C-MM3 Outflow}\\
\hline
CH$_3$OH & 8(0 , 8)- 7(1 , 6) E    & 220.078 & 96.6  & 06 41 12.3 & $+$09 29 06.7 & $<$0.55$^*$ &--  &-- & -- \\
CH$_3$OH & 8( -1, 8)- 7(0 , 7) E   & 229.759 & 89.1  & 06 41 12.3 & $+$09 29 06.7 & 2.24 (0.4) & 8.16 (0.58) & 5.79 (1.63) & 13.84 (4.81) \\
CH$_3$OH &  3( -2, 2)- 4( -1, 4) E & 230.027 & 39.8  & 06 41 12.3 & $+$09 29 06.7 & $<$0.74$^*$ & --& -- & -- \\
\hline
\multicolumn{10}{c}{C-MM12 Outflow}\\
\hline
CH$_3$OH & 8(0 , 8)- 7(1 , 6) E    & 220.078 & 96.6  & 06 41 09.3 & $+$09 30 15.7  & $<$0.50$^*$ & -- & -- & --\\
CH$_3$OH & 8( -1, 8)- 7(0 , 7) E   & 229.759 & 89.1  & 06 41 09.3 & $+$09 30 15.7  & 1.74 (0.21) &14.70 (0.30) & 4.85 (0.70) & 8.99 (1.70) \\
CH$_3$OH &  3( -2, 2)- 4( -1, 4) E & 230.027 & 39.8  & 06 41 09.3 & $+$09 30 15.7  & $<$0.90$^*$ & -- & -- & -- \\
\hline

\end{tabular}
\end{center}
\end{small}

{\bf Notes.}\\
$^a$ {The formal errors from the single Gaussian fits are given in the brackets.}\\
$^*$ {Non detection.  The value given is the 3$\sigma$ limit, calculated from the rms in the spectrum extracted from the primary-beam-corrected image cube for each position.}\\
\end{minipage}
\end{table*}

\section{Discussion}

\subsection{Nature of the Millimetre Continuum Peaks}
\begin{figure}
\includegraphics[width=0.4\textwidth]{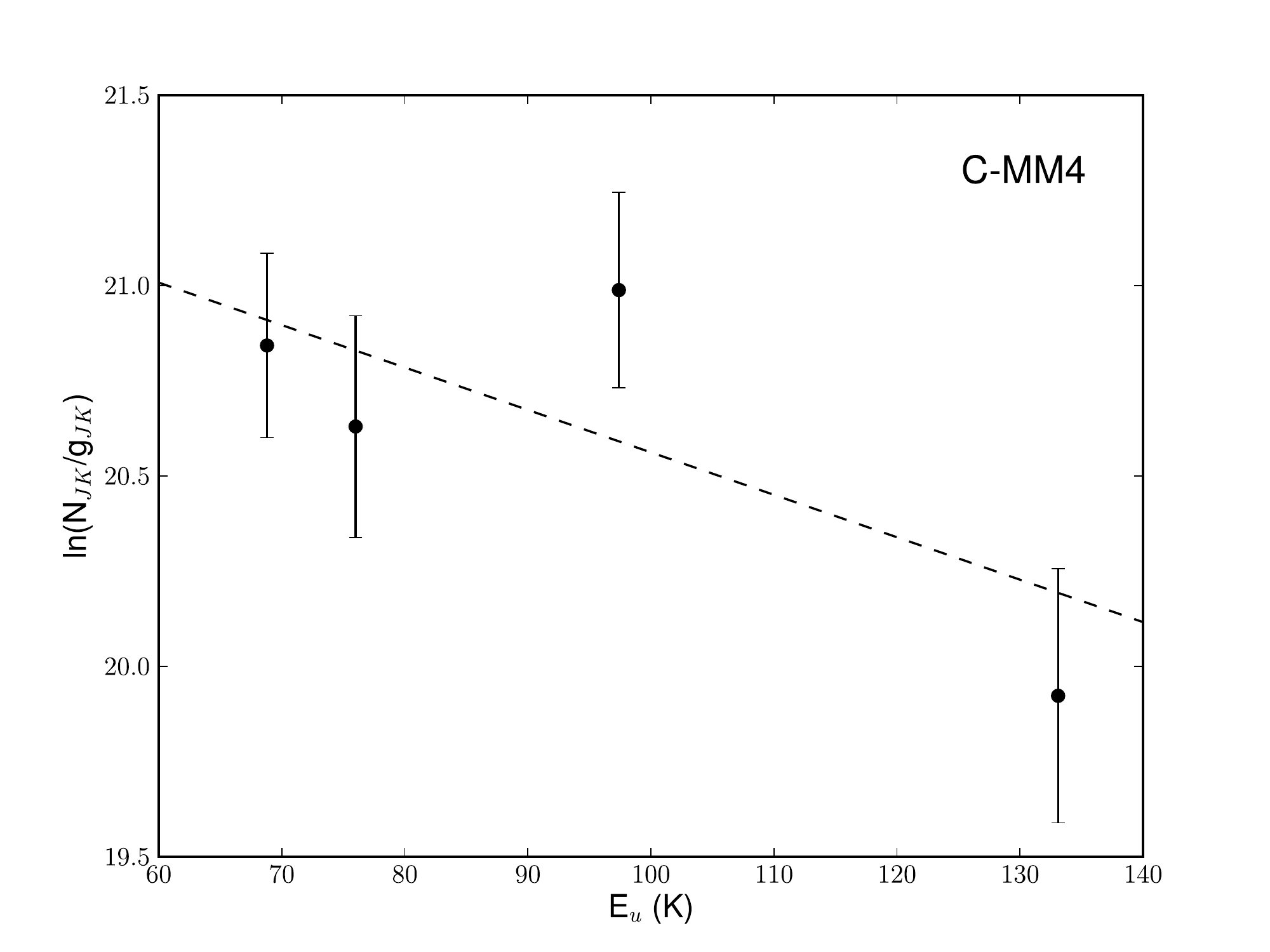}
\includegraphics[width=0.4\textwidth]{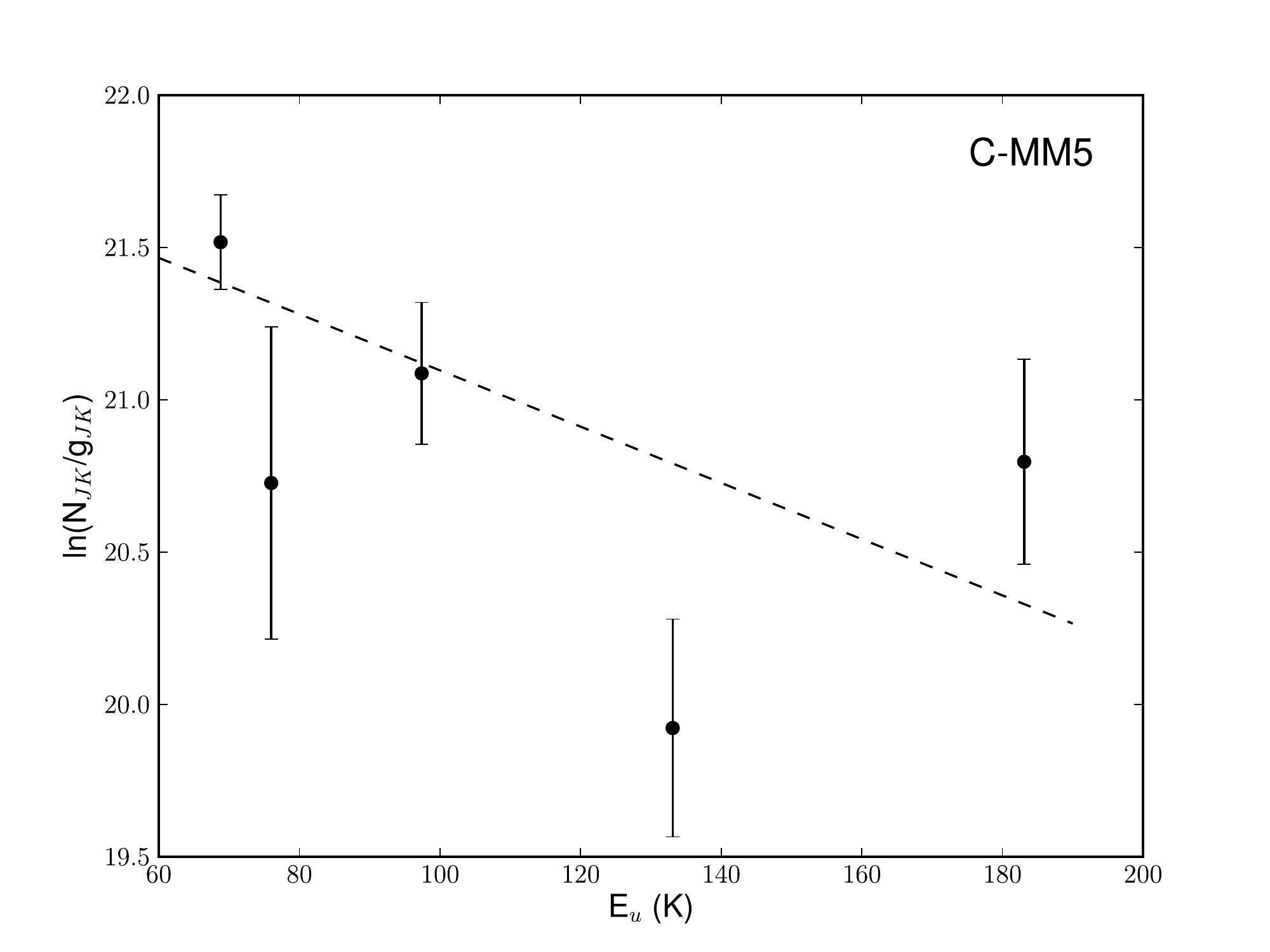}
\includegraphics[width=0.4\textwidth]{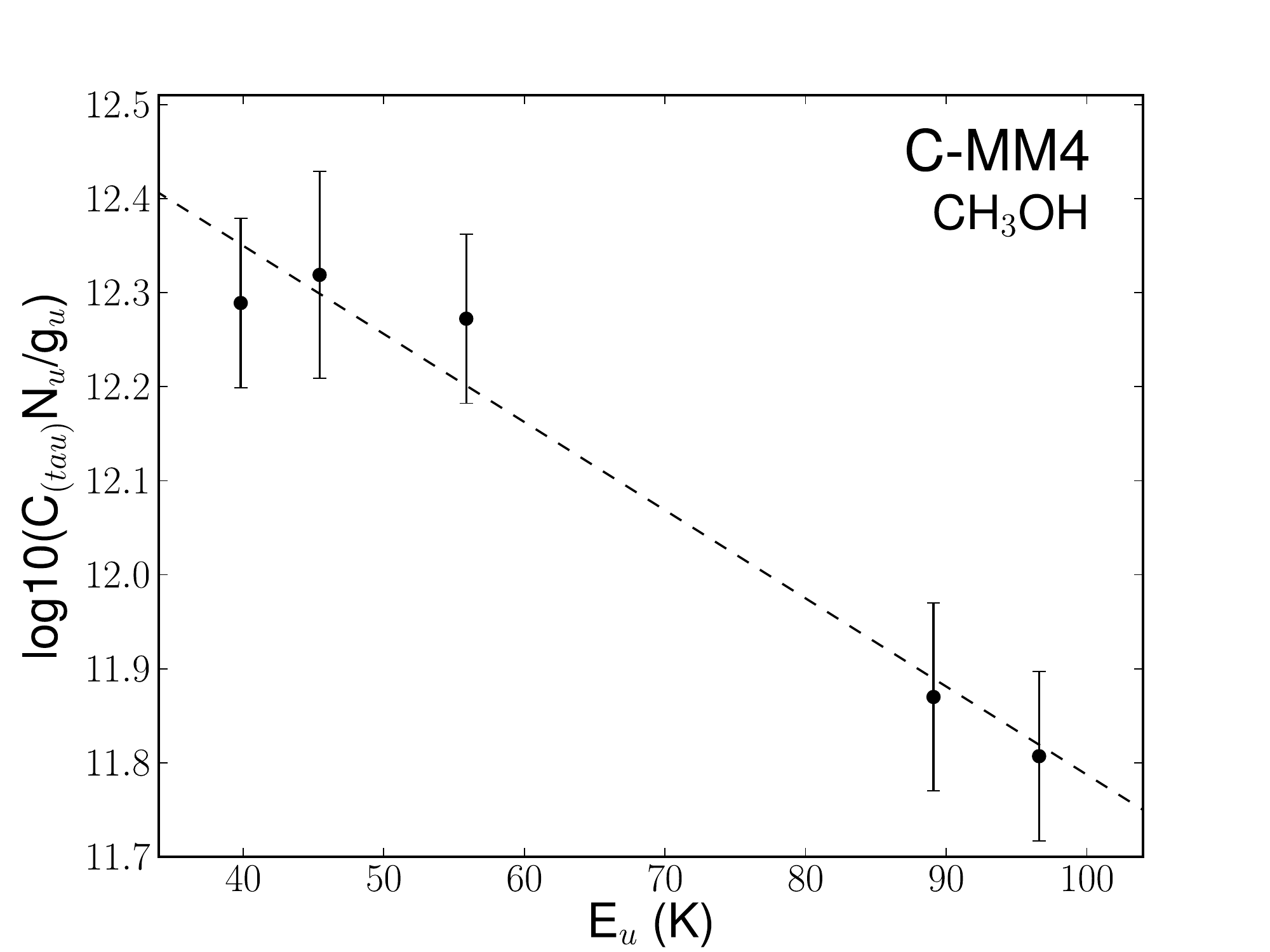}
\caption{Rotation diagrams for CH$_3$CN J$=$12$_{K}$-11$_{K}$ observed towards the continuum peaks of CMM4 and C-MM5 (top two panels) and for CH$_3$OH towards the continuum peak of C-MM4 (bottom panel). The  CH$_3$OH rotation diagram has been corrected for optical depth  (Section~\ref{dis:temp_est}).  \label{image:ch3cn}}
\end{figure}
\subsubsection{Temperature Estimates from Molecular Line Emission \label{dis:temp_est}} Methyl cyanide (CH$_3$CN) is commonly used to estimate the temperature
of warm/hot gas in star forming regions. Using the method of rotational
diagrams and assuming optically thin emission and LTE
(e.g. \citealt{LorenandMundy1984}, \citealt{Zhang1998}), gas temperatures may
be derived from the relative integrated intensities of the K-ladder
components. Compact CH$_3$CN emission is identified towards two 1.3\,mm
continuum peaks, C-MM4 and C-MM5. The highest-energy K component detected (at
$>$3$\sigma$) is K\,$=$\,3 towards C-MM4 and K\,$=$\,4 towards C-MM5, with
upper-level energies of E$_{upper}=$133\,K and E$_{upper}=$183\,K
respectively. Following \citet{Araya2005}, we estimate temperatures of
$\sim$90\,$\pm$48\,K and $\sim$108\,$\pm$36\,K and CH$_{3}$CN column densities of $\sim$8.6$\times$10$^{12}$\,cm$^{-2}$ and $\sim$1.6$\times$10$^{13}$\,cm$^{-2}$  for C-MM4 and C-MM5, respectively, from a linear regression fit (accounting for the uncertainties in the line integrated intensities) to the rotational diagrams, shown in Figure \ref{image:ch3cn}.
For C-MM4, the temperature is also estimated by applying the rotation diagram method \citep[e.g.][]{GoldsmithandLanger1999} to the five detected CH$_{3}$OH transitions, which have E$_{upper}$ ranging from $\sim$40-100\,K. As noted by \citet{GoldsmithandLanger1999}, the optical depth can alter the derived rotation temperature. We iteratively solve for the optical depth and rotational temperature following e.g. \citet{Brogan2009} and \citet{Cyganowski2011}. The best fit gives a rotational temperature of $\sim$46$\pm$11\,K for C-MM4 (see Figure \ref{image:ch3cn}). We note that both the A and E CH$_3$OH transitions are included in the rotational diagram analysis.

\subsubsection{Mass Estimates from the SMA 1.3\,mm Continuum Emission \label{dis:mass}}

Thermal emission from dust and free-free emission from ionised gas can
both contribute to the observed continuum flux at millimetre wavelengths.
To estimate the contribution from ionised gas,
we extrapolate the 3.6\,cm VLA flux densities from \citet{Reipurth2004} to 1.3\,mm, assuming
S$_{\nu}$\,$\propto\nu^{-0.1}$. VLA2 (see Table 2 from \citealt{Reipurth2004}) is coincident with the millimetre continuum peak C-MM4, while VLA3 and VLA4 both fall within the 15$\sigma$ continuum contour level for C-MM5, where VLA4 is coincident with the peak of the millimetre continuum emission. If we extrapolate the 3.6cm emission from VLA2 for C-MM4, and the total emission from both VLA3 and VLA4 for C-MM5, we find that the estimated ionised contribution is negligible for both C-MM4 and C-MM5 ($<$1$\%$ at 1.3\,mm from 
free-free emission).  We thus conclude that the thermal contribution from the dust
dominates the emission at 1.3\,mm and estimate the gas mass
assuming a simple isothermal model
of optically thin dust emission, 
\begin{equation}\label{equation:mass}
M_{gas}=\frac{RF_{\nu}D^2}{\kappa_{m}(\nu)B_{\nu}(T_d)},
\end{equation}
where R is the gas to dust mass ratio (assumed to be 100), F$_{\nu}$
is the flux density, D is the distance to the region, $\kappa_m(\nu)$ is the dust opacity, B$_{\nu}$(T$_d$) is the Planck function and
T$_d$ is the (assumed) dust temperature.
We
adopt $\kappa_m=$1.0\,cm$^2$g$^{-1}$ at 1.3\,mm, for
grains with ice mantles in high density regions \citep[10$^{8}$cm$^{-3}$;][]{Henning1994}.

Table~\ref{table:mass} presents mass estimates for a range of assumed
dust temperatures based on greybody fits to the large-scale dust
emission \citep{WardThompson2000}.  Two temperature components--of
17\,K and 38\,K--were required to fit the SED
\citep{WardThompson2000}, and we adopt these as lower and upper-limit
temperatures for our mass estimates.
We adopt this approach because, for the majority of the dust continuum sources
(C-MM4 and C-MM5 excepted, Section~\ref{dis:temp_est}),
gas temperatures cannot be estimated from our SMA data.     
For C-MM4 and C-MM5, Table~\ref{table:mass} also presents mass
estimates assuming T$_{dust}=$T$_{gas}$ and the gas temperatures from
Section ~\ref{dis:temp_est}.
We note that the dust temperatures at the scales probed by our SMA
observations are likely to be higher than those measured by
\citet{WardThompson2000} from single-dish data.
As a result, the mass estimates obtained assuming the single-dish dust
temperatures (which range from
$\sim$0.1$-$7\,M$_{\odot}$)
are likely upper limits for the SMA cores.

Despite the (considerable) uncertainties introduced by the dust
temperature, C-MM3, C-MM4, and C-MM12 appear as the three
most massive 1.3 mm continuum sources: they are the only sources with
estimated masses $>$1 M$_{\odot}$ for all assumed temperatures.
Notably, the outflow-driving sources C-MM3 and C-MM12 are both
spatially compact, in contrast to C-MM4 (Table~
\ref{table:Continuum}, Figure~\ref{image:continuum}). Furthermore, unlike C-MM4, both C-MM3 and C-MM12 are individual separate structures with no indication of further substructure.
We note that the mass estimate for C-MM5 in Table~\ref{table:mass}
encompasses both C-MM5 and C-MM5b, and so may be an overestimate for
the mass of C-MM5 alone.
The newly identified millimetre continuum sources (e.g.\ SMA1, SMA2,
SMA3, and SMA4) have the lowest estimated
masses, $\leq$0.5\,M$_{\odot}$. 
The 5\,$\sigma$ sensitivity limit of the observations corresponds to 
$\sim$\,0.14\,M$_{\odot}$ and $\sim$\,0.05\,M$_{\odot}$ for 17 and
38\,K, respectively.

\begin{table}
\begin{small}
\begin{center}
\caption{Mass estimates for 1.3\,mm continuum sources in NGC 2264-C. \label{table:mass}}
\centering
\begin{tabular}{l c c c c}
\hline\hline

Source & Mass        &Mass         &Mass  $^{a}$    &  Mass $^{b}$ \\
       &17\,K        &   38\,K     & CH$_3$CN & CH$_3$OH\\
       & (M$_{\odot}$) & (M$_{\odot}$)&   (M$_{\odot}$)       &   (M$_{\odot}$)      \\
\hline
C-MM3  &  5.4 & 2.0 & --&--\\
C-MM4  &  6.8 & 2.5 &1.0$^{+1.3}_{-0.4}$& 2.0$^{+0.8}_{-0.4}$\\
C-MM5  &  1.8 & 0.7 &0.21$^{+0.1}_{-0.1}$&--\\
C-MM10 &  1.6 & 0.6 & --&--\\
C-MM12 &  3.8 & 1.4 & --&--\\
C-MM13 &  0.7 & 0.3 & --&--\\
SMA1   &  0.3 & 0.1 & --&--\\
SMA2   &  0.4 & 0.1 & --&--\\
SMA3   &  0.4 & 0.1 & --&--\\
SMA4   &  0.5 & 0.2 & --&--\\

\hline
\end{tabular}
\end{center}
{\bf Notes.}\\
$^{a}$ {Mass estimated using T$_{\rm gas}=$T$_{\rm dust}$ and the rotation temperature estimate for CH$_3$CN (Section~\ref{dis:temp_est}).}\\
$^{b}$ {Mass estimated using T$_{\rm gas}=$T$_{\rm dust}$ and the rotation temperature estimate for CH$_3$OH (Section~\ref{dis:temp_est}).}\\
\end{small}
\end{table}

\subsubsection{Notes on Individual Millimetre Continuum Sources \label{dis:individualsource}}

The 1.3\,mm continuum sources in NGC 2264-C display a wide range of associated molecular line emission in our SMA data.  C-MM4 exhibits the richest line emission, with a hot-core-like spectrum, while C-MM3, C-MM13, and SMA-4 are not associated with any molecular line emission that can be directly associated with the millimetre continuum cores (as opposed to e.g.\ outflow emission, Section~\ref{results:vlsr}). The presence/absence and relative strengths of molecular lines detected in wide-bandwidth interferometric observations may be indicative of evolutionary state; however, there is a degeneracy with core mass \citep[e.g.][]{Zhang2007a,GalvanMadrid2010,Cyganowski2012,Wang2014,Immer2014}. We discuss the nature of the individual 1.3\,mm continuum sources in NGC 2264-C, with reference to their molecular line emission (or lack thereof) and their 24/70$\mu$m properties \citep[also indicative of evolutionary state, e.g.][]{Battersby2011}, below.

{\bf CMM-3:} The molecular line emission observed in the vicinity of CMM-3 with the SMA is associated with the outflow driven by this millimetre continuum core \citep[Section~\ref{sec:outflows}, see also][]{Saruwatari2011}.  No compact molecular line emission is detected coincident with the millimetre continuum source; this may, however, be due to sensitivity limitations, as CMM-3 is located at the $\sim$20\% power level of our SMA primary beam.  No Herschel PACS 70$\mu$m or Spitzer MIPSGAL 24$\mu$m emission is detected towards CMM-3\footnote{Strong 70$\mu$m, 24$\mu$m, and 2.2$\mu$m emission are observed to the south-east of C-MM3. This emission is not directly associated with C-MM3, but is from a separate star with a weak, diffuse nebulosity surrounding it.}, indicating that CMM-3 is likely young; the detection of N$_2$H$^{+}$\,(1-0) \citep{Peretto2007} and the enhancement of SCUBA 450\,$\mu$m emission are also consistent with youth.

C-MM3 is one of the most massive cores in NGC 2264-C (based on our SMA mass estimates, Section~\ref{dis:mass}). Our SMA observations
recover $\sim$30\% of the 1.3\,mm single dish flux density (e.g. \citealt{Peretto2006}; \citealt{WardThompson2000}). We note that our mass estimate for C-MM3 is lower than those of \citet{Saruwatari2011}\footnote{Our measured flux density for C-MM3 agrees within $\sim$20\% with that reported by \citet{Saruwatari2011} based on SMA observations with similar spatial resolution.}, \citet{Peretto2006}, \citet{Peretto2007}, and \citet{WardThompson2000}, due to differences in spatial filtering (when comparing interferometric and single-dish data), and in assumed opacities and dust temperatures (along with the recently revised distance to the region). If we adopt the temperature, opacity (scaled to 1.3mm), and distance assumed
by \citet{Peretto2007}, we obtain a mass estimate from our SMA data comparable to their estimate based on their PdBI observations. Interestingly, \citet{Peretto2007} suggest that C-MM3 harbours a disc of mass 1.1\,M$_{\odot}$, which would be a substantial fraction of our estimated mass of $\sim$2-5 M$_{\odot}$ (Table~\ref{table:mass}).

\begin{figure*}
\includegraphics[width=0.95\textwidth]{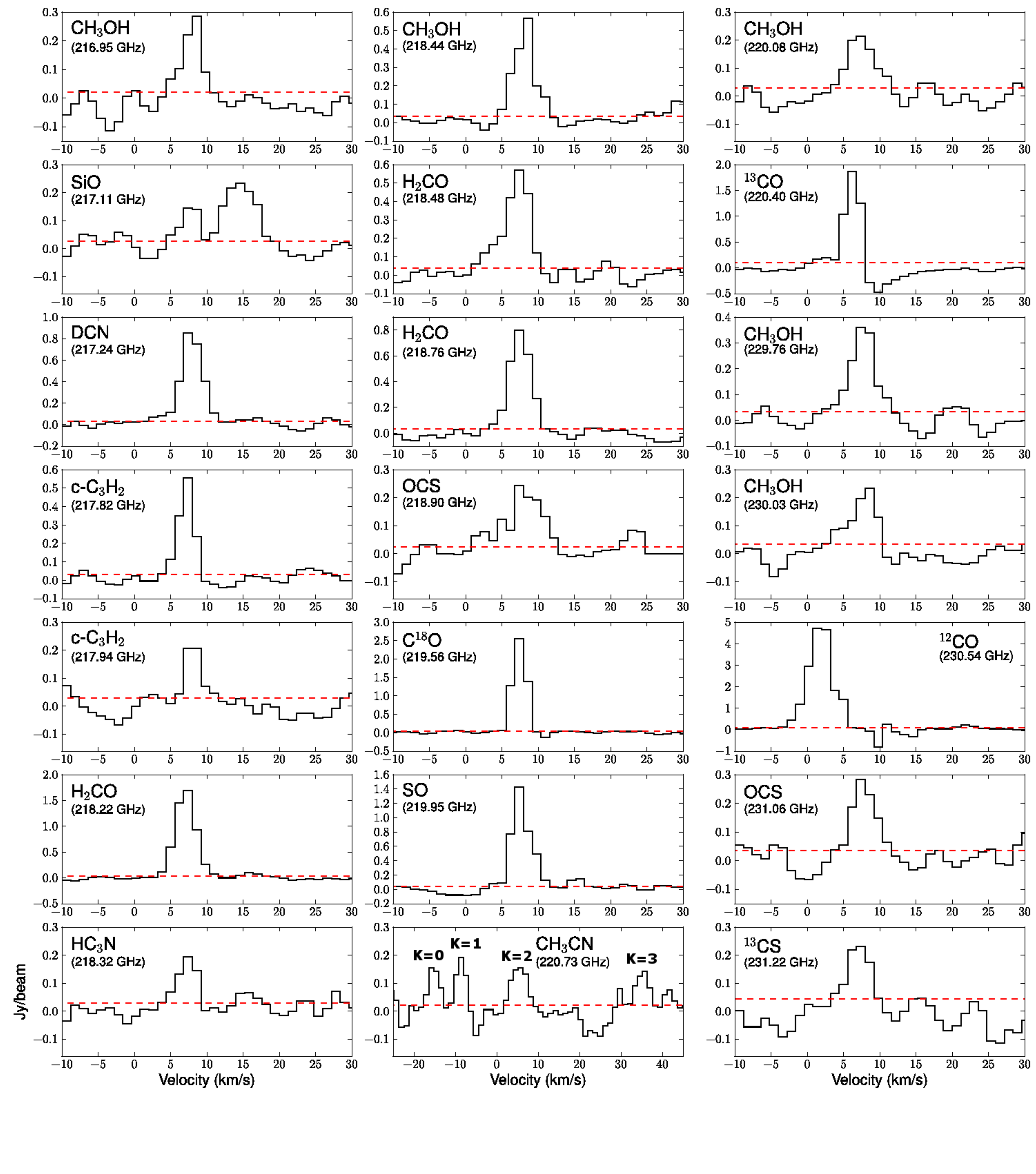}
\caption{Spectra of transitions detected at the continuum peak of C-MM4, extracted from image cubes corrected for the primary beam response. Each transition detected above 5$\sigma$ is shown. The CH$_3$CN ladder is shown in a single panel, centred on the K=2 transition (rest frequency$=$220.730\,GHz): the velocity range shown includes all K ladder transitions detected at $>$3$\sigma$ with the transitions labelled in the panel. In each panel, the horizontal red dashed line is the spectral rms noise and the transition and rest frequency are given in the top left corner. \label{image:cmm4lines}}
\end{figure*}

\begin{figure*}
\includegraphics[width=0.95\textwidth]{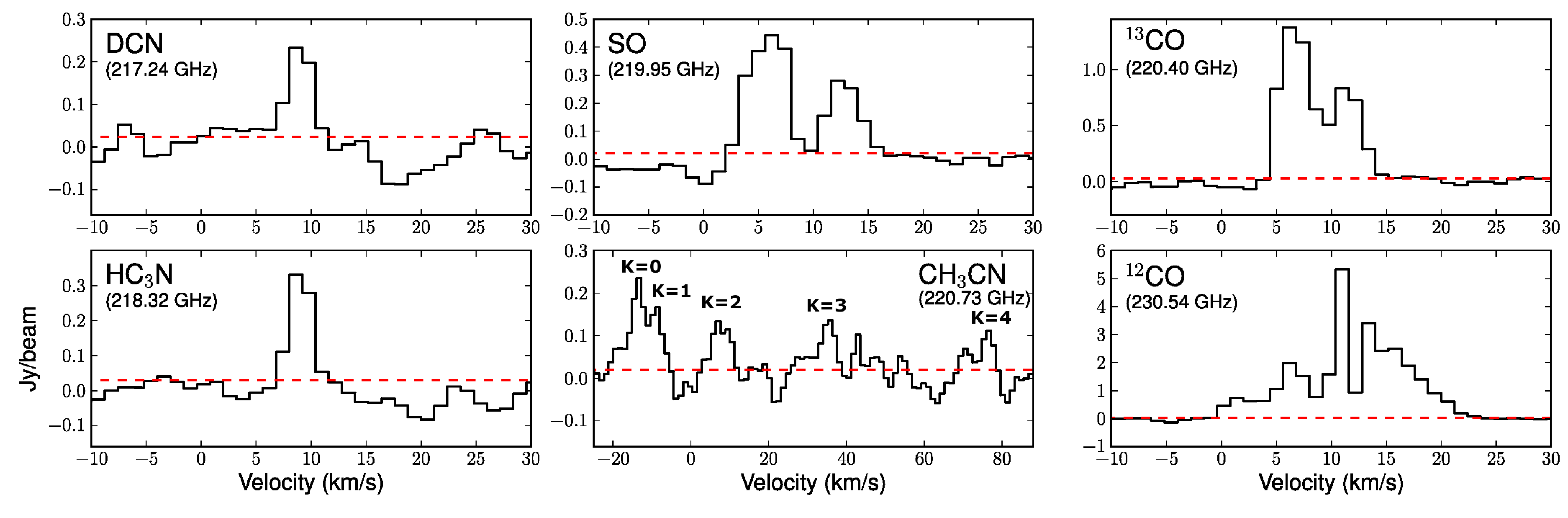}
\caption{Spectra of transitions detected at the continuum peak of C-MM5, extracted from image cubes corrected for the primary beam response. Each transition detected above 5$\sigma$ is shown. The CH$_3$CN ladder is shown in a single panel, centred on the K=2 transition (rest frequency$=$220.730\,GHz): the velocity range shown includes all K ladder transitions detected at $>$3$\sigma$ with the transitions labelled in the panel. In each panel, the horizontal red dashed line is the spectral rms noise and the transition and rest frequency are given in the top left corner.\label{image:cmm5lines}}
\end{figure*}

{\bf C-MM4}, the most molecular line rich core, displays a hot core like spectrum. Compact molecular line emission with E$_{upper}$ temperatures of 55-133\,K from typical hot core tracers such as OCS, CH$_3$CN, CH$_3$OH and HC$_3$N is coincident with the continuum peak on scales of $\sim$0.01\,pc. Figure \ref{image:cmm4lines} displays the line profiles for all transitions observed towards the continuum peak of C-MM4. In addition to the compact emission, several transitions tracing more diffuse emission are also identified such as N$_2$H$^{+}$\,(1-0) \citep{Peretto2007}, DCN, and C$^{18}$O. No clear outflow is identified in the molecular line emission towards this core. The FIR 70$\mu$m Herschel emission (see Figure \ref{image:infrared1}) is dominated by the bright infrared source AFGL 989-IRS1 in this region. However, at the continuum peak of C-MM4 the 70$\mu$m flux is approximately 50\% brighter than at other positions at similar radii from IRS1, suggesting that C-MM4 is contributing to the 70\,$\mu$m flux. In addition, 3.6\,cm emission (VLA2 in \citealt{Reipurth2004}) is detected coincident with the continuum peak. No 24\,$\mu$m emission is detected towards the continuum peak of C-MM4.

We find a 1.3\,mm flux density of 496\,mJy for this source, which recovers approximately half of the single dish flux (e.g. \citealt{Peretto2006}; \citealt{WardThompson2000}). \citet{Peretto2006} estimate a mass of 35\,M$_{\odot}$ from their single dish observations, which is comparable to their single dish mass estimate for C-MM3 of $\sim$40M$_{\odot}$. However, the morphology of C-MM4 is noticeably more extended  than C-MM3. The molecular line emission also suggests that at least the central region of C-MM4 is likely warmer than C-MM3. Our mass estimates for C-MM4 using the CH$_3$CN and CH$_3$OH rotational temperatures are $\sim$1-2\,M$_{\odot}$. These results indicate C-MM4 is potentially lower in mass and more evolved than C-MM3.

{\bf C-MM5} coincides with the position of the IR-bright B2 2000\,L$_{\odot}$ star AFGL 989-IRS1. This well studied source dominates both the 24$\mu$m and 70$\mu$m emission in this region. We detect comparatively little molecular line emission towards this continuum peak. The molecular line emission detected is shown in Figure \ref{image:cmm5lines}. Furthermore, we find no obvious indication that an outflow is driven by this source in any of the observed molecular transitions, despite the detection of a twisted jet feature at 1.65$\mu$m and 2.2$\mu$m by \citet{Schreyer1997} to the north-west of C-MM5. This source also drives a dense stellar wind \citep{Bunn1995} that has excavated a low density cavity around AFGL 989-IRS1 \citep{Schreyer2003}. That may account for the lack of molecular line emission detected towards this source, if a low density cavity created by a wind/jet has removed part or all of the outer envelope. While molecular line emission from the lower energy transitions is lacking, compact emission from CH$_3$CN and HC$_3$N is coincident with the peak of the continuum emission. \citet{Grellmann2011} have previously identified a disc in this source. As emission from CH$_3$CN is known to trace discs (e.g. \citet{Cesaroni2014} and references therein), we suggest the emission from CH$_3$CN and HC$_3$N may be related to the presence of a disc in AFGL989-IRS1. On the other hand, as C-MM5, unlike C-MM4, does not display a typical hot core-like spectrum, and given CH$_3$CN has also been previously detected in low velocity shocks and in outflows (e.g. \citealt{Bell2014}; \citealt{Csengeri2011}), the emission in C-MM5 from these compact tracers could also be due to shocks. An additional indication that AFGL\,989-IRS1/C-MM5 is comparatively evolved is the detection of 3.6\,cm emission (VLA sources 3 and 4, \citealt{Reipurth2004}).

For C-MM5, our SMA observations only recover $\sim$25\% of the single dish flux. However, we only consider the 1.3\,mm flux density estimated
for C-MM5 and not the entire parent structure, which also includes the new detections SMA1 and SMA2. \citet{Peretto2006} estimate a mass of $\sim$18M$_{\odot}$ from their single dish observations. We estimate a gas mass, using the temperature estimate from CH$_3$CN, of $\sim$0.2\,M$_{\odot}$ for C-MM5, which is lower than for C-MM3 or C-MM4. Furthermore, this mass estimate potentially incorporates an additional source, C-MM5b, and so the mass of C-MM5 may be lower still. As discussed in Section \ref{section:continuum}, we suggest C-MM5b is likely a
separate source that at the current resolution we are unable to spatially
resolve. The presence of the strongest 70$\mu$m, 24$\mu$m, 1.65$\mu$m, and
2.2$\mu$m emission, along with the previous detections of a disc and 3.6\,cm radio continuum, suggest that C-MM5 is likely the most evolved source in our field.

{\bf C-MM10} displays very little molecular line emission. However, 70$\mu$m emission is observed towards the continuum peak (Figure \ref{image:infrared1}). This source is notably extended (Figure \ref{image:continuum}) compared with the compact morphology of cores such as C-MM3 and C-MM12. Furthermore, only $\sim$10\% of the single dish flux reported in \citet{Peretto2006} is recovered with the SMA, which may suggest that the bulk of the 1.3\,mm emission associated with this source is from an extended region that we currently resolve out. C-MM10 is likely a low mass source, but its exact nature is unclear.

{\bf CMM-12} lacks emission from most typical hot core tracers. Only compact emission from OCS (E$_{upper}=$99.8\,K) is detected towards the continuum peak. The majority of the molecular line emission in the vicinity of this source (such as SiO, H$_2$CO and SO) is from the outflow driven by this millimetre continuum core. As emission from OCS has previously been suggested to be tracing the inner regions of the outflow (e.g. \citealt{Cyganowski2011}), the OCS emission is likely also tracing the outflow, rather than the continuum peak. C-MM12 also lacks any obvious 70$\mu$m or 24$\mu$m emission. Our estimated mass for this source is 1.4-3.8\,M$_{\odot}$, approximately half of the estimated mass for C-MM3. We recover a notably high proportion of the single dish flux, $\sim$90\%. The single dish mass is estimated to be $\sim$9\,M$_{\odot}$ \citep{Peretto2006}. These results suggest C-MM12 is likely a low/intermediate mass protostar.

{\bf C-MM13} displays one of the most molecular line poor spectra. Furthermore, we find no enhancement of 450\,$\mu$m, 70\,$\mu$m or 24\,$\mu$m emission towards C-MM13. The N$_2$H$^{+}$(1-0) line, however, displays the strongest emission coincident with this peak \citep{Peretto2007}. Our estimated mass for this source is in the range 0.3-0.7\,M$_{\odot}$, compared with $\sim$8\,M$_{\odot}$ from \citet{Peretto2007}. Scaling our mass assuming the parameters from \citet{Peretto2007}, we would still only estimate a mass of $\sim$2\,M$_{\odot}$ for C-MM13.  Furthermore, C-MM13 is not detected in the single dish observations. These results suggest C-MM13 is a low mass, likely young protostar. However, its nature, as with the continuum peak C-MM10, is unclear.

{\bf SMA1, SMA2, SMA3, and SMA4} are all new 1.3\,mm detections\footnote{SMA4 was marginally detected at 3.2mm with the PdBI by \citet{Peretto2007}, but was not considered as an unambiguous detection.}. None of these newly detected 1.3\,mm continuum peaks display enhancements
in 450$\mu$m, 70\,$\mu$m or 24$\mu$m emission, and only towards SMA4 is there N$_2$H$^{+}$(1-0) emission \citep{Peretto2007}. SMA2 and SMA3 are the most molecular line rich of the four. No emission from the ``compact'' tracers (Section~\ref{section:results_lines}) is detected towards any of these four continuum peaks. The mass estimates are $<$0.5\,M$_{\odot}$ for all four new detections, for all assumed gas temperatures. Three of the four new detections are nested within a parent tree structure. SMA1 and SMA2 are part of the same parent structure as AFGL 989-IRS1/C-MM5, where SMA1 is spatially coincident with the twisted jet feature to the north of this source, and SMA4 forms part of the same parent structure as C-MM4. Furthermore, \citet{Kamezaki2014} identified an associated water maser coincident with SMA4, and given the association of N$_2$H$^+$ and 3.2\,mm emission in \citet{Peretto2007} they designate this source as C-MM4S. In addition, they note the association of an X-ray source (FMS2-1269) \citep{Flaccomio2006} with this continuum peak, suggesting that it is an X-ray emitting class 0 source. SMA3 is the only new mm detection that is found as an independent structure. It is spatially coincident with the southern edge of the ridge feature. 

\subsection{Nature of the Ridge \label{discussion:ridge}}

As noted in section \ref{section:maseridge}, a prominent feature in our SMA molecular line observations is the ``ridge''. This feature is detected in all three H$_2$CO transitions, and three CH$_3$OH transitions: CH$_3$OH 4(2 , 2)- 3(1 , 2) (218.440\,GHz), CH$_3$OH 8(0 , 8)- 7(1 , 6) E (220.078\,GHz) and CH$_3$OH 8($-$1, 8)- 7(0 , 7) E (229.759\,GHz). In all of these transitions the ridge feature displays the strongest emission compared with any other location in the region. The emission from these transitions typically extends over a spatial scale of $\sim$15000\,AU in length (taken along the PA$=$150 degrees), with a perpendicular width of $\sim$2000-5000\,AU. We note that along the length of the ridge the width varies and is not always resolved by our beam. The ridge is not clearly associated with a millimetre continuum peak: only the southern edge of the ridge spatially coincides with the continuum peak SMA3. At this position, we find compact redshifted emission traced by SiO, SO and H$_2$CO in the SHV regime, and $^{12}$CO emission in the IHV regime. While no blueshifted counterpart is found towards the ridge, we do find a possible blueshifted component traced by $^{12}$CO coincident with C-MM4. Thus, the redshifted emission at the southern edge of the ridge may be from an outflow running northwest-southeast between C-MM4 and SMA3 where the associated driving source is not obvious.

The ratio of the 229\,GHz and 230\,GHz CH$_3$OH lines (discussed in Section \ref{section:maseridge}) is an indication of the non-thermal nature of the 229\,GHz emission towards the ridge. In addition, \citet{Slysh2009} detect 44 \,GHz Class I CH$_3$OH masers towards the ridge. While Class I CH$_{3}$OH masers have typically been associated with outflows (e.g. \citealt{PlambeckandMenten1990}; \citealt{Kurtz2004}; \citealt{Cyganowski2009}), \citet{Voronkonov2010,Voronkonov2014} suggest that Class I masers can also be excited in the shocks formed by expanding HII regions.  The maser emission in the ridge could therefore be attributable to shocks from an outflow (although we do not detect a clear powering source) or from a wind impacting the ambient medium. As the ridge is located at the edge of a low density cavity driven by AFGL 989-IRS1 (e.g. \citealt{Schreyer2003}; \citealt{Nakano2003}), the maser emission in the ridge could be a result of the dense stellar wind from IRS1 \citep{Bunn1995} colliding with a region of high density ambient material.

\subsection{Outflow Properties}

\begin{table*}
\begin{minipage}{180mm}
\begin{center}
\caption{SiO (5-4) and $^{12}$CO (2-1) Outflow Properties\label{table:SiO}}
\centering
\begin{tabular}{l c c c c c c c c}
\hline\hline
Outflow  $^{a}$       &  L$^{b}$& $v_{max}$ $^{c}$&$|v_{max}-v_{\rm LSR}|$ &Inclination$^{d}$&$l_{flow}$  $^{e}$ &$|v_{max}-v_{\rm LSR}|$ $^{f}$  &T$_{dyn}$ $^{g}$ &  Integrated  $^{h}$\\
                &                &                 &   &        & $_{(corrected)}$         & $_{(corrected)}$       &     &  Flux Density  \\
                & ($\arcsec$)         &  (km\,s$^{-1}$)   &(km\,s$^{-1}$) &      (Degrees)  & (AU)    &  (km\,s$^{-1}$)    &   (years)    & (Jy.km\,s$^{-1}$)                  \\
\hline
\multicolumn{9}{c}{SiO\,(5-4)}\\ 
\hline
C-MM3 R1 &   6                &  36.8      &  29.7     &     45-60        &      6200-5100    &       42-60      &   700-410      &  21               \\
C-MM3 B1 &  4                 &  -19.6     &  26.7     &     45-60        &      4200-3400    &       38-53      &   520-300      &  13             \\
C-MM3 B2 &  13                &  -5.2      &  12.3     &     45-60        &      13600-11000  &       17-25      &   3700-2100    &  6              \\

C-MM12 red lobe &  22                &  32.0      &  23.0     &     45-60        &     23000-18700   &       32-46      &   3400-2000    &  67               \\
C-MM12 blue lobe&  22                &  -1.6      &  10.6     &     45-60        &     23000-18700   &       15-22      &   7300-4200    &  20               \\
\hline
\multicolumn{9}{c}{$^{12}$CO\,(2-1)}\\ 
\hline
C-MM3 R1  &   6                & 38.0      &  30.9      &     45-60        &   6200-5100        &      44-62      &   680-390    & 68   \\
C-MM3 B1 &   3                & -17.2     &  24.3      &     45-60        &   3100-2500        &      34-49      &   430-240    & 30  \\
C-MM3 B2 &   11               & -5.2      &  12.3      &     45-60        &   11500-9300       &      17-25      &   3100-1800  & 1   \\
C-MM12 red lobe &   19               & 32        &  23.0      &     45-60        &   19800-16200      &      33-46      &   2900-1700  & 179   \\
C-MM12 blue lobe&   17               & -2.8      &  11.8      &     45-60        &   17700-14500      &      17-24      &   5000-2900  & 79   \\
\hline

\end{tabular}
\end{center}
{\bf Notes.}\\
$^{a}$ {The blueshifted lobe of C-MM3 has been treated as two separate components, B1 and B2. As we do not detect emission from the redshifted R2 component of C-MM3 identified by \citet{Saruwatari2011}, all properties for the C-MM3 red lobe are calculated from the R1 component only (see Figure \ref{image:sio_mom0}).}\\
$^{b}$ {Angular distance between the continuum peak and the farthest positional offset for the SiO (5-4) and $^{12}$CO emission in arcseconds .}\\
$^{c}$ {The absolute velocity of the first/last channel in the image cubes (corrected for the primary beam response) where emission is detected above 3\,$\sigma$.}\\
$^{d}$ {The outflow length, maximum velocity, and dynamical timescale are influenced by the assumed inclination. \citet{Saruwatari2011} assume an inclination of 60$^{\circ}$ for C-MM3; as we are unable to determine the inclination, we report outflow properties assuming both an inclination of 45$^{\circ}$ and 60$^{\circ}$ from the line of sight.} \\
$^{e}$ {Calculated from the maximum lobe length 
  and corrected for inclination using $l_{flow(corrected)}$=$l_{flow}$/sin($i$); converted
  to AU assuming a distance of 738\,pc \citep{Kamezaki2014}. }\\
$^{f}$ {Calculated from $|v_{max}-v_{\rm LSR}|$ corrected for inclination using ($|v_{max}-v_{\rm LSR}|$)/cos($i$).} \\
$^{g}$ {Dynamical timescale calculated using T$_{dyn}$$=$$l_{flow(corrected)}$/$v_{max(corrected)}$}.\\
$^{h}$ {Total integrated intensity calculated from images corrected for the primary beam response. For SiO, the emission is integrated from the $v_{\rm LSR}$ (7.2\,km\,s$^{-1}$ and 8.9\,km\,s$^{-1}$ for C-MM3 and C-MM12 respectively) to the maximum velocity of each lobe. For $^{12}$CO\,(2-1), the emission is integrated over the velocity ranges stated in Table \ref{table:co}, which exclude emission at low velocities close to the systemic velocity due to confusion.}\\
\end{minipage}
\end{table*}

\begin{table*}
\begin{minipage}{180mm}
\begin{center}
\caption{Derived Outflow Properties Estimated from the $^{12}$CO (2-1) Emission. \label{table:co}}
\centering
\begin{tabular}{l c c c c c c c c}
\hline\hline
Outflow         &  Inclination $^{a}$&$\Delta\,v$$^{b}$&M$_{out}$             &  P$_{out}$$^{c}$               &  E$_{out}$$^{c}$  & $\dot{M}_{out}$$^{c}$        & $\dot{P}_{out}$$^{c}$  \\
                &  (Degrees)         & (km\,s$^{-1}$)&(10$^{-3}$M$_{\odot}$)& (M$_{\odot}$ km\,s$^{-1}$) & (erg)       &  M$_{\odot}$yr$^{-1}$) &(M$_{\odot}$ km\,s$^{-}$yr$^{-1}$) \\
                &                    &          &                      &                          & ($\times$10$^{43}$) &    ($\times$10$^{-6}$)   & ($\times$10$^{-5}$)\\
\hline
C-MM3 red lobe $^{d}$ &  45-60             & 15.2-38.0  &   2.5             &  0.06-0.08               &  1.53-3.0         &    3.6-6.4          & 8.8-20.5\\
C-MM3 blue lobe\,$^{d}$ &  45-60        &$-$17.2-0.8 &   1.1             &  0.02-0.03               &  0.5-1.0          &    2.6-4.6          & 4.6-12.5\\
 
C-MM12 red lobe &  45-60             & 16.4-32.0  &   6.5             &  0.12-0.17               &  2.5-4.9          &    2.2-3.8          & 4.1-10.0\\
C-MM12 blue lobe&  45-60             & -2.8-0.8   &   2.8             &  0.04-0.05               &  0.5-0.9          &    0.6-1.0          & 0.8-1.7\\
\hline
\end{tabular}
\end{center}
{\bf Notes.}\\
$^a$ {Inclination from line of sight: the outflow properties are estimated for assumed inclinations of 45$^{\circ}$ and 60$^{\circ}$.}\\
$^b$ {The velocity range used to estimate the outflow properties, taken from the absolute velocities in the image cube, the maximum velocity is taken from the velocity of the first/last channel in the image cubes (corrected for the primary beam response) where emission is detected above 3\,$\sigma$.}\\
$^c$ {Values are calculated from the dynamical timescale estimates from the $^{12}$CO\,(2-1) emission presented in Table \ref{table:SiO}.}\\
$^d$ {The values presented for the blueshifted outflow lobe of C-MM3 are the sum of the B1 and B2 outflow components shown in Figure
    \ref{image:sio_mom0}; $\dot{M}_{out}$ and $\dot{P}_{out}$ are calculated from the dynamical timescale of the B1 component. The R2 component reported by \citet{Saruwatari2011} is not detected in our observations so all values for the redshifted lobe are derived only from the R1 component (see Figure \ref{image:sio_mom0}).}\\
\end{minipage}
\end{table*}

We identify two unambiguous, high velocity bipolar outflows, traced
by SiO, SO, H$_2$CO, CH$_3$OH and $^{12}$CO emission, driven by C-MM3 and
C-MM12. 
In Table \ref{table:SiO}, we present the outflow parameters (e.g. outflow length, maximum velocity and dynamical timescale) estimated from both the SiO\,(5-4) and $^{12}$CO\,(2-1) emission for the two clearly defined bipolar outflows. The dynamical timescale is given as T$_{dyn}$$=$$l_{flow}$/($|v_{max}-v_{\rm LSR}|$), where $l_{flow}$ is the observed maximum outflow length on the sky, and $v_{max}$ is the absolute velocity of the maximum/minimum channel in the image cube with a 3$\sigma$ detection. The inclination ({\it i}) of the outflow to the line of sight affects the estimates of both the outflow length ($l_{flow(corrected)}$=$l_{flow}$/sin($i$)) and the maximum velocity with respect to the v$_{\rm LSR}$ e.g. (($|v_{max}-v_{\rm LSR}|$)/cos($i$)), and thus the dynamical timescale. We present estimates for inclinations of both 45 and 60 degrees in Table \ref{table:SiO}. Towards C-MM3, we treat the two spatially separated blueshifted components, B1 and B2 (identified by \citet{Saruwatari2011}, see Figure \ref{image:sio_mom0}), independently and provide dynamical timescale estimates for both. For the blueshifted component B1 and the redshifted lobe, we find similar maximum lengths, and absolute velocity estimates from the SiO\,(5-4) emission. We note that towards the redshifted lobe, only emission from component R1 is detected. \citet{Saruwatari2011} also identified a second redshifted component R2 (see their Figure 2 and Figure \ref{image:sio_mom0} presented here); however, R2 is outside of the 10\% power level of our primary beam. Thus, we only use emission from R1 to estimate the outflow parameters for the redshifted lobe of C-MM3.

The dynamical timescales estimated from SiO\,(5-4) are $\sim$300-500 yrs and $\sim$400-700\,yrs for B1 and the redshifted lobe of C-MM3, respectively. Additionally, we find similar estimates of the outflow properties from the $^{12}$CO\,(2-1) emission. We note that the blueshifted emission from B1 is at a slightly lower absolute maximum velocity than the redshifted emission, and is approximately 3\,km\,s$^{-1}$ and 6\,km\,s$^{-1}$ slower for SiO\,(5-4) and $^{12}$CO\,(2-1) respectively.
At the position of the more northern blueshifted component B2, an even lower absolute maximum velocity with respect to the v$_{\rm LSR}$ of $\sim$12km\,s$^{-1}$ and dynamical timescale estimate of $\sim$2000-3500\,yrs is obtained for both the SiO\,(5-4) and $^{12}$CO\,(2-1) emission. In both tracers, B2 is approximately an order of magnitude older compared with either the higher velocity blueshifted component B1 or the redshifted lobe. We suggest the outflow driven by C-MM3 may be impacting with dense material at this position, which is causing it to slow further from the central core. In addition to the blueshifted components, B1 and B2, \citet{Saruwatari2011} also identified a second weaker, lower velocity redshifted component, R2, which is similarly offset and has a similar maximum velocity offset with respect to the v$_{\rm LSR}$ as B2. If both the red and blueshifted lobes are similarly slowed due to an impact with ambient material, then C-MM3 may be centrally located in a dense envelope.

The red- and blueshifted outflow lobes driven by C-MM12 have a similar maximum length on the sky of $\sim$20$\arcsec$ ($\sim$15000\,AU) in both the SiO\,(5-4) and $^{12}$CO\,(2-1) emission. However, the blueshifted lobe has a considerably lower maximum velocity with respect to the v$_{\rm LSR}$ compared with the
redshifted lobe in both tracers. This results in noticeably longer dynamical timescale estimates for the blueshifted outflow lobe, which may indicate that the blueshifted lobe is slowed, possibly due to an impact with ambient material. Moreover, the integrated SiO\,(5-4) flux density of the lower velocity blueshifted lobe is significantly lower than the integrated SiO\,(5-4) flux density towards the higher velocity, redshifted lobe (see Table~\ref{table:SiO}). This result again points towards SiO\,(5-4) being a more effective tracer of high velocity shocks.

Assuming the $^{12}$CO\,(2-1) emission is optically thin at high velocities, we follow the procedure outlined in \citet[][and references therein]{Cyganowski2011} and estimate the outflow mass from the $^{12}$CO\,(2-1) emission using
\begin{equation}
M_{out} = \frac{1.186\times10^{-4}\,Q(T_{ex})\,e^{\frac{E_{upper}}{T_{ex}}}D^2\int{S_{\nu}d\nu}}{\nu^3\mu^2S\chi}, 
\end{equation}
where $M_{out}$ is the outflow gas mass given in units of $M_{\odot}$, T$_{ex}$ is the excitation temperature in Kelvin, Q(T$_{ex}$) is the partition function, $\nu$ is the frequency in GHz, $\chi$ is the abundance relative to H$_2$, D is the distance to NGC 2264-C in kpc, and S$_\nu$ is the $^{12}$CO\,(2-1) flux in Jy. We do not estimate an outflow mass using the SiO emission due to the uncertainty in the SiO to H$_2$ abundance ratio. If we adopt the same excitation temperature used by \citet{Maury2009} of 20\,K, then $Q$(T$_{ex}$) is estimated to be 7.56 following \citet{Mangum2015}\footnote{Assuming a rotational constant estimated using Splatalogue (http://www.splatalogue.net/), \citet{Muller2005}} for $^{12}$CO. We use a standard value of 10$^{-4}$ for the $^{12}$CO to H$_2$ abundance ratio \citep{Frerking1982}. Following the procedure from \citet{Qiu2009}, we estimate the outflow momentum and energy from
\begin{equation}
P_{out} = \sum\,M_{out}(\Delta\,v)\Delta\,v
\end{equation}
and
\begin{equation}
E_{out} = \frac{1}{2}\sum\,M_{out}(\Delta\,v)(\Delta\,v)^2,
\end{equation}
where M$_{out}$($\Delta\,v$) is the outflow mass calculated for a given channel of velocity $\Delta\,v$\,$=$\,$|v_{channel}$\,$-$\,$v_{\rm LSR}|$. In Table \ref{table:co}, we provide estimates of M$_{out}$, P$_{out}$ and E$_{out}$ for the red- and blueshifted outflow lobes driven by C-MM3 and C-MM12. The values presented for the blueshifted lobe driven by C-MM3 are taken from the total contribution of the B1 and B2 components. For the redshifted lobe we do not detect emission from R2 and only emission from the R1 component is included in the derived outflow properties. We only consider the high velocity gas in the estimates of the outflow mass, momentum and energy and the respective velocity ranges assumed for each lobe are presented in Table \ref{table:co}. In addition, we estimate the mass and momentum outflow rates, $\dot{M}_{out}$\,$=$\,$M_{out}$/t$_{dyn}$ and $\dot{P}_{out}$\,$=$\,$P_{out}$/t$_{dyn}$, using the dynamical timescale estimates presented in Table \ref{table:SiO}. Given that the main blueshifted component, B1, contains more than 90\% of the total mass contained in both the B1 and B2 components, we use the dynamical timescale estimated for B1 to derive estimates of the mass and momentum outflow rates presented in Table \ref{table:co} for C-MM3. All properties are corrected for inclinations of {\it i} = 45$^{\circ}$ and 60$^{\circ}$. The estimated outflow masses for the blue- and redshifted lobes are 1.1$\times$10$^{-3}$\,M$_{\odot}$ and 2.5$\times$10$^{-3}$\,M$_{\odot}$ respectively, compared to previous outflow mass estimates using $^{12}$CO\,(2-1) by \citet{Saruwatari2011} of 1.6$\times$10$^{-3}$\,M$_{\odot}$ and 2.6$\times$10$^{-3}$\,M$_{\odot}$ for the blue- and redshifted lobes respectively. The lower estimates found in this work may be due the R2 component in our data being located outside of the 10\% power level and thus not included in the mass estimate and in the blueshifted lobe we detect $^{12}$CO\,(2-1) emission down to an absolute velocity of -17.2\,kms$^{-1}$ compared with -20\,kms$^{-1}$ in \citet{Saruwatari2011}.
We find that the outflow driven by C-MM12 is more than twice as massive and contains approximately twice the momentum and energy compared with the outflow from C-MM3. However, the mass and momentum outflow rates estimated for the outflow driven by C-MM3 are more than double those found towards the outflow from C-MM12. Given these results and the fact that the dynamical time is found to be shorter for C-MM3 compared with C-MM12, we speculate that the more massive core C-MM3 may be driving a younger, less massive outflow compared with the lower mass core C-MM12. We note, however, that we have not fully accounted for possible inclination differences between the outflows which can significantly affect the estimation of the dynamical timescale. In both outflows we observe an apparent asymmetry between the outflow properties of the higher velocity, redshifted, and lower velocity, blueshifted, outflow lobes. In both cases, the redshifted lobes are twice as massive massive and contain approximately three times more momentum and energy compared with their respective blueshifted lobes.

\subsubsection{Comparison with Previous Single Dish Observations of Outflows in NGC 2264-C}

One of our main goals was to unambiguously identify the driving sources of the several high velocity outflow lobes observed in NGC 2264-C by \citet{Maury2009} using single dish $^{12}$CO\,(2-1) at $\sim$11\,$\arcsec$ resolution, through higher resolution SMA observations ($\sim$3\,$\arcsec$). \citet{Maury2009} identified five outflow lobes (F1, F2, F5, F7, F11; see their Figure 2) in our SMA field of view, suggesting either C-MM3 or C-MM13 as the most likely candidate to be driving the collimated red- (to the north) and blueshifted (to the south) outflow lobes F1 and F2 respectively. While we identify no obvious outflow emission, from any of our high velocity outflow tracers, along either F1 or F2, we do identify high velocity compact bi-polar outflow driven by C-MM3. However, in contrast to the direction of the red- and blueshifted F1 and F2 lobes shown in \citet{Maury2009} (north to south respectively), we observe the redshifted lobe of C-MM3 to the south and the blueshifted lobe to the north (i.e. in the opposite direction). Similarly, in our single dish SiO (8-7) observation towards C-MM3 (see Figure \ref{image:sio_mom0}) the red- and blueshifted emission is observed to the south and north of C-MM3 respectively. These results suggest neither C-MM3 nor C-MM13 is driving the F1 and F2 lobes observed by \citet{Maury2009}. They do find an extended, poorly collimated redshifted lobe (F11) to the south of C-MM3. Thus, the redshifted high velocity outflow from C-MM3 may then be contributing to the redshifted emission from F11. However, F11 is significantly extended compared with the compact redshifted emission we observe towards C-MM3.

The red and blueshifted lobes F7 and F5 extend over several continuum peaks at the lower spatial resolution of the single dish $^{12}$CO observations (see Figure 2, \citet{Maury2009}), and it is not obvious which continuum peak(s) are responsible for the emission. With the higher spatial resolution  provided by the SMA, we now identify a collimated bipolar outflow driven by C-MM12. Thus, we can associate the lobes F7 and F5 to the outflow driven by C-MM12. Furthermore, as previously mentioned, the fact that our SMA SiO data recover (and resolve) both of the SiO outflows identified with the JCMT is strong evidence that the SMA is not ``missing'' any large-scale active outflows.

We estimate outflow masses for the blue- and redshifted lobes driven by C-MM12 of 2.8$\times$10$^{-3}$\,M$_{\odot}$ and 6.5$\times$10$^{-3}$\,M$_{\odot}$ respectively compared to the minimum\footnote{\citet{Maury2009} estimate both a minimum and maximum mass for each lobe depending on velocity range used, see Section 4 of \citet{Maury2009} for a detailed explanation.} mass estimates from \citet{Maury2009} of 0.9$\times$10$^{-2}$M$_{\odot}$ and 6.0$\times$10$^{-2}$M$_{\odot}$ for the blue- (F5) and redshifted (F7) outflow lobes respectively. Thus, we only recover $\sim$10\% and $\sim$30\% of the mass of the red and blueshifted lobes F7 and F5, respectively. However, we have only included redshifted emission situated to the north of C-MM12 and do not include redshifted emission to the south of C-MM12. As shown previously, to the south of C-MM12 there is redshifted emission running from northwest to southeast that may be from another potential outflow axis (components 9 to 11, see Figures \ref{image:sio_mom0} and \ref{image:redblue}), which at the lower spatial resolution is blended with the redshifted emission associated with the outflow driven by C-MM12. Furthermore, we integrate both the red- and blueshifted emission over a smaller velocity range compared with \citet{Maury2009}. If we now include the redshifted emission to the south of C-MM12 and extend the velocity ranges, we now estimate a mass of $\sim$20$\times$10$^{-3}$ M$_{\odot}$ which is $\sim$30\% of the mass estimated by \citet{Maury2009} and a similar blueshifted estimate as previously found. Our beam size is $\sim$4 times greater than \citet{Maury2009}, thus the difference in mass estimates is likely due in part to spatial filtering, but also due to the ability to more accurately separate emission not directly associated with the outflow at higher resolution.

\subsubsection{Relative Evolutionary State}

The two dominant bi-polar outflows are being driven by the IR-dark, compact,
millimetre bright, and potentially youngest sources, C-MM3 and C-MM12. The lack
of 70$\mu$m and 24$\mu$m emission towards either source and the associated
outflow emission indicates that both of these sources are likely at a very
early stage of evolution. In comparison, no obvious active outflow emission
from SiO\,(8-7), SiO \,(5-4), or high velocity $^{12}$CO\,(2-1) is
associated with C-MM5, which is coincident with the most evolved star in the
region, AFGL989-IRS1, a 9.5\,M$_{\odot}$ star. Thus, the presence of an SiO
outflow appears to decline with evolution for this region. \citet{Klaassen2012}
found an opposite trend in their survey of SiO\,(8-7) emission towards massive star forming regions, finding an increase of
the SiO luminosity with evolution, which was also weakly observed by
\citet{Leurini2014} in their sample. \citet{Davies2011} predict the lifetime
for YSOs to dramatically drop for sources with a bolometric luminosity of 
$\sim$10$^{5.5}$L$_{\odot}$ or greater. The typical bolometric luminosity of the
sources in \citet{Klaassen2012} is of the order 10$^{4}$-$10^5$L$_{\odot}$,
with several objects at a few $\times$10$^{5}$L$_{\odot}$. In comparison,
AFGL989-IRS1 has a bolometric luminosity of $\sim$10$^{3}$L$_{\odot}$. Thus, it
is likely that some possible compact HII regions in \citet{Klaassen2012} will
be the same age or possibly younger than the most evolved source in our region,
AFGL 989-IRS1, and so the presence of SiO may decline with age, as seen by
\citet{SanchezMonge2013}, consistent with the low mass scenario of the jet
declining with age. In addition, the most evolved sources observed
  by \citet{Klaassen2012} and \citet{Leurini2014} may also host companion younger objects given the coarse angular resolution of the datasets (JCMT 15$\arcsec$, and APEX 18$\arcsec$ respectively).
\begin{figure*}
\includegraphics[width=0.95\textwidth]{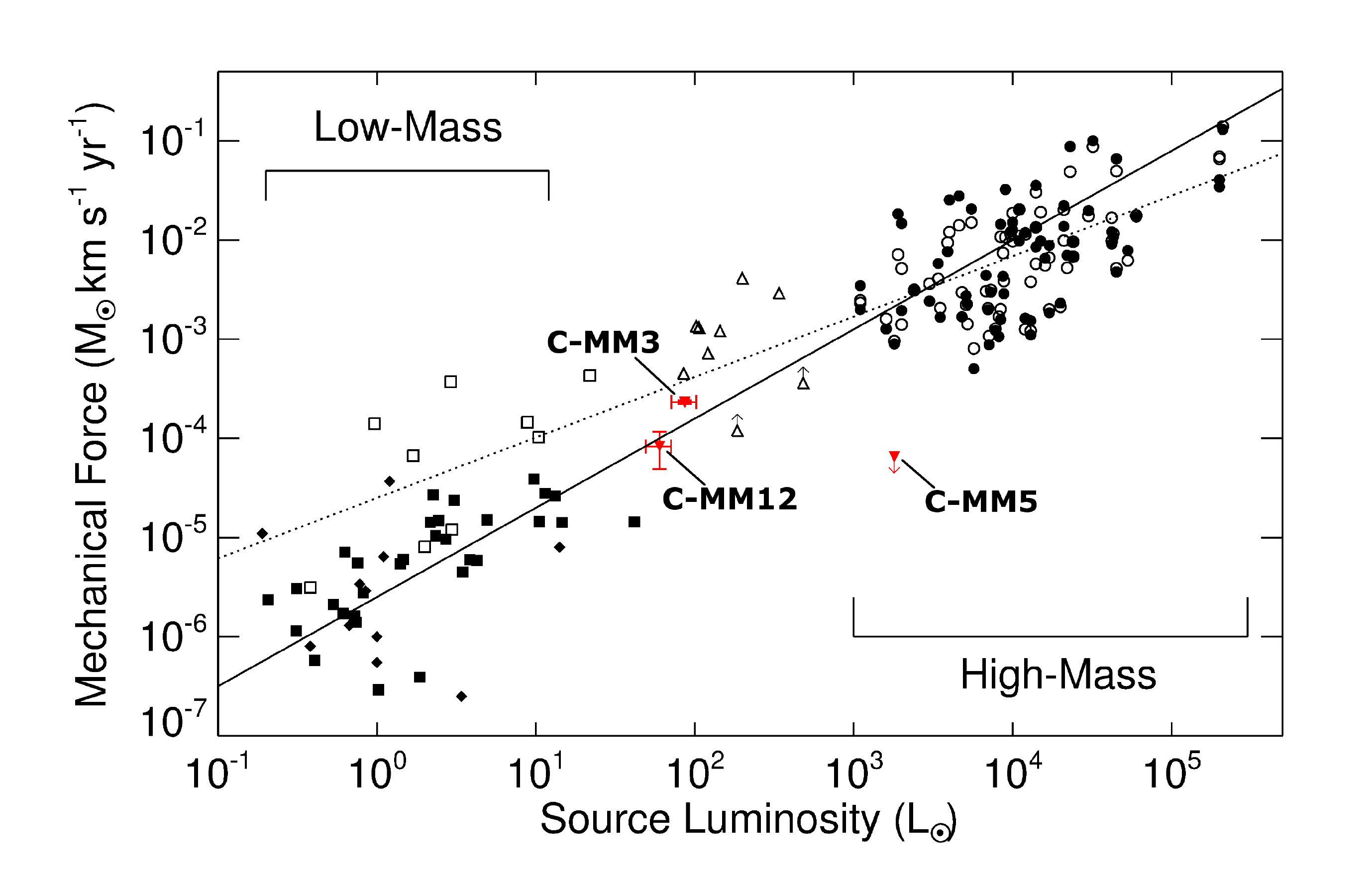}
\caption{Outflow force versus bolometric luminosity adapted from Figure 7 of \citet{Maud2015outflows}. The filled and open circles are from \citet{Maud2015outflows} and represent outflow forces calculated using a fixed dynamical timescale (t$_{dyn}$$=$$R_{max}$/$v_{max}$, the same method adopted in this paper) and those summed from maps where a position-variable t$_{dyn}(x,y)$ is used, respectively (see \citet{Maud2015outflows} for more details). The filled diamonds are the $\sim$ Class I sources from \citet{Vandermarel2013}, the open and filled squares are Class 0 and Class I low-mass outflow sources from \citet{Bontemps1996}, and the open triangles are intermediate-mass Class 0 analogues from \citet{DuarteCabral2013}. The red inverted triangles represent C-MM3, C-MM12, and the upper limit of the estimated force for AFGL989-IRS1/C-MM5, as labelled. The error bars on the outflow force represent the maximum and minimum values (summing both the red- and blueshifted components) estimated for inclinations of 45 and 60 degrees, with the main point representing the average value. The uncertainties on the luminosities for C-MM3 and C-MM12 come from the range in the luminosity estimated from the SED fit for C-MM3, with the position of the marker representing the average. For AFGL989-IRS1/C-MM5 the bolometric luminosity is taken from the RMS survey database \citep{Lumsden2013}. The black dotted line represents the linear fit from the \citet{Maud2015outflows} sample (open and filled circles) extended to the lower luminosity regime. The black solid line is the linear fit to the Class I sources from \citet{Bontemps1996} extended to the high mass regime. \label{image:literature}}
\end{figure*}

The IR radiation of NGC 2264-C is dominated by AFGL989-IRS1, however,
C-MM3 is sufficiently spatially offset ($\sim$40$\arcsec$) from this
source that it is possible to obtain an estimate of the bolometric luminosity of C-MM3. We employ the online SED fitting tool of
\citet{Robitaille2007} to fit an SED for C-MM3, using data at five wavelengths:
70$\mu$m and 160$\mu$m Herschel PACS archival data (ID 1342205056,
P.I. F.\ Motte), 450$\mu$m and 850$\mu$m SCUBA data from
\citet{DiFrancesco2008}, and the IRAM 1.3mm single dish flux from
\citet{WardThompson2000}. The 70$\mu$m emission is not enhanced towards the position of
C-MM3 but does peak nearby (see section
\ref{dis:individualsource}). We therefore take an upper limit of the
70$\mu$m emission towards this source. The best fits to the data are from models 3011153 and 3003259\footnote{http://caravan.astro.wisc.edu/protostars}, giving bolometric luminosity estimates of 71\,L$_{\odot}$ and 102\,L$_{\odot}$ respectively over a range of inclination angles from near edge on to $\sim$50
degrees. These results are consistent with a previous estimate of 50$\pm$10\,L$_{\odot}$ by
\citet{Maury2009} for C-MM3, again using the online SED tool \citet{Robitaille2007} and three flux
measurements (Spitzer/MIPS 70$\mu$m, APEX/ArTeMiS 450$\mu$m and IRAM 1.2mm MAMBO data \citep{Peretto2006} see \citet{Maury2009} for more details). The remaining outflow driving
source, C-MM12, is too close to AFGL-989 to unambiguously identify any
IR emission associated with it. We crudely estimate a
bolometric luminosity for this source assuming that the luminosity
ratio between the two sources
(e.g. L$^{C-MM12}_{bol}$/L$^{C-MM3}_{bol}$) is equal to the 1.3\,mm
flux density ratio. This gives an estimate of the bolometric
luminosity for C-MM12 of between $\sim$50-70\,L$_{\odot}$.

\subsubsection{NGC 2264-C in the Context of Low and High Mass Star Formation Studies}
The correlation between outflow force and source bolometric luminosity has been previously suggested as an indication of a similar outflow driving mechanism across the low and high mass star-forming regimes. Figure \ref{image:literature} presents previous data taken from the literature of the bolometric luminosity as a function of outflow force covering several orders of bolometric luminosity (low mass class 0 and class I sources; \citet{Vandermarel2013}, \citet{Bontemps1996}, intermediate mass class 0 sources; \citet{DuarteCabral2013}, and high mass young stellar objects; \citet{Maud2015outflows}). While linear (log-log space) correlations are found between the two regimes (see Figure \ref{image:literature}), there is an order of magnitude scatter about these fits. The linear fit to the low mass class I sources \citep{Bontemps1996}, when extended to the high mass regime is found to better fit the data compared with the class 0 fit extended to the high mass regime. While both the class 0 and class I fits have a similar gradient \citep{Bontemps1996}, the class 0 fit is, on average, offset higher by an order of magnitude compared with the class I fit.

Sources observed in the RMS sample are by selection IR-bright and thus relatively evolved compared with embedded typically younger class 0 sources. Therefore, it may not be surprising that the fit from the class I low mass protostars has a better correlation with the high mass outflow sources than the fit from the class 0 IR-dark sample. Moreover, as discussed in \citet{Maud2015outflows}, each high mass outflow source, given the observed spatial resolution and distances, likely contains multiple protostars, where the luminosity of the cluster is dominated by the IR-bright RMS source. Thus, the high mass sample of outflow sources plotted in Figure \ref{image:literature} have a force representing the total for the cluster, but the assumption that these are driven by the RMS source, and hence have the correct bolometric luminosity, is not necessarily the case. A combination of low/intermediate-mass stars could be responsible for the observed total outflow force (e.g. \citealt{Maud2015outflows} and references therein), as observed by \citet{Klaassen2015} towards IRAS 17233-3606, and observed here in this data, where the IR-bright RMS source is not found to drive an outflow.

In our SMA observations of NGC 2264-C, we find that the complex outflow emission observed in the single dish CO observations contains two well-collimated, bipolar outflows, neither of which is driven by the IR-bright RMS source. Instead, it is the likely younger, IR-dark sources C-MM3 and C-MM12 that drive the outflows in this region.  The derived bolometric luminosities and outflow forces ($\dot{P}$, see Table \ref{table:co}) for these two sources are plotted in Figure \ref{image:literature}, along with an estimated upper limit for the outflow force of C-MM5/AFGL989-IRS1.  We note that this is a crude estimate, calculated assuming a 20$\arcsec$ aperture centred on C-MM5, with the $^{12}$CO\,(2-1) emission integrated over a similar velocity range as seen in C-MM3 and C-MM12. It is stressed that this gives a very approximate limit for the outflow properties. The derived limit on the outflow force for AFGL989-IRS1 is lower than expected given its bolometric luminosity (1800 L$_{\odot}$, taken from the RMS survey), which may be a result of the more evolved nature of this source. 

The estimated outflow forces for C-MM3 and C-MM12 are below what would be expected given their likely young, class 0-type nature. However, as the class 0 low mass sample of \citet{Bontemps1996} was observed with a single dish telescope compared with the interferometric observations performed here, even though the spatial resolution is similar (given the larger distance to NGC 2264-C, $\sim$738pc) the interferometric observations will likely miss some flux. Furthermore, the outflow forces estimated for C-MM3 and C-MM12 have not been corrected for optical depth effects and are thus likely underestimated by a factor of $\sim$5-10 (e.g. \citealt{Dunham2014}; \citealt{Offner2011}). \cite{Maury2009} estimated a total outflow force for lobes F5 and F7 (which at the higher resolution of our SMA observations are found to be associated with C-MM12) of 4.6$\times$10$^{-4}$M$_{\odot}$\,kms$^{-1}$yr$^{-1}$, compared with 1.1$\times$10$^{-5}$M$_{\odot}$ kms$^{-1}$yr$^{-1}$ estimated here. The true value of the outflow force for C-MM12 likely lies somewhere between these two values, accounting for the spatial filtering but also the ability to separate the outflows themselves at our higher resolution.

There are still major difficulties when trying to link outflow properties between the low and high mass regimes; the scales probed are different, and particularly in the high mass regime it is often not clear if a single or multiple sources are responsible for the emission. In addition, the different methods of estimating properties such as dynamical timescales between samples make it difficult to directly compare. \citet{Richer2000} suggest the perceived correlation may not be directly related to the actual driving mechanism, as the higher masses available in the high mass sample would provide higher values for outflow properties regardless of the underlying mechanism.  This is also addressed by \citet{Maud2015outflows}, who discuss that the core mass could be the main driver of the outflow scaling relation. Taken at face value, the linear (log-log) fit between the low and high mass regimes does indeed suggest a correlation from low to high mass indicative of a similar formation mechanism between the regimes. However, there are significant caveats, as discussed above. Our SMA observations of NGC 2264-C emphasise the need for complementary high resolution observations with the ability to resolve a postulated single outflow and driving core into many components, to fully probe the outflow(s) in intermediate/high mass star forming regions.

\section{Summary}

We have observed the intermediate/high mass star forming region NGC
2264-C at an angular resolution of $\sim$3$\arcsec$ with the SMA at
1.3\,mm and provide the first interferometric observations of the
outflow tracers SiO (5-4) and $^{12}$CO\,(2-1). Of the ten 1.3\,mm
continuum peaks identified, four of which are new detections, only two
are clearly driving bipolar outflows. The SiO (5-4) observations
unambiguously reveal two high velocity bipolar outflows, driven by
C-MM3 and C-MM12. High velocity emission from SO,
$^{12}$CO, and H$_2$CO, along with lower velocity emission from
CH$_3$OH, also traces both outflows. 
Comparing our SMA data to our JCMT SiO\,(8-7) observations, we find that SiO\,(8-7)
is detected only towards
these two bipolar outflows with the JCMT.  In the lower
resolution JCMT observations, the outflow driving sources are ambiguous.  Thus,
the combination of these observations emphasises the power of high-resolution,
multi-line observations to remove ambiguity in identifying
outflow driving sources.

We find a clear evolutionary differentiation among the continuum
sources present in this cluster. Comparison of the molecular
chemistry and of the mid and far IR, millimetre, sub-millimetre, and radio emission reveals that
it is the likely youngest, mm brightest sources, C-MM3 and C-MM12,
that are driving the dominant bipolar outflows. Both of these cores
are IR-dark and molecular line weak and have no detected centimetre-$\lambda$ radio
emission, indicating that they are likely at a very early stage of
evolution. In contrast, the IR-bright, most evolved source in the
region, the RMS source AFGL 989-IRS1, does not drive a molecular
outflow. However, the wind from this source \citep{Bunn1995} is likely
driving a low density cavity \citep{Schreyer2003} that may be
responsible for the molecular line rich ``ridge'' feature we observe, including its Class I CH$_{3}$OH maser emission.

{\bf \it Acknowledgements} We thank the anonymous referee for helpful
comments that have improved the content and overall clarity of this
manuscript.  C.J.C. acknowledges support from the STFC (grant number ST/M001296/1).  C.J.C. was partially supported during this work by a
National Science Foundation Astronomy and Astrophysics Postdoctoral
Fellowship under award AST-1003134. The authors are very grateful to
N. Peretto for providing the reduced PdBI N$_2$H$^+$ (1-0) integrated
intensity maps that were published in \citet{Peretto2007}. This
research made use of APLpy, an open-source plotting package for Python
hosted at http://aplpy.github.com; astrodendro, a dendrogram plotting
routine for Python hosted at http://www.dendrograms.org/ and the
splatalogue molecular database (http:/splatalogue.net/).

\label{here}

\label{lastpage}

\end{document}